\documentclass{article}
\usepackage{graphicx,amssymb,amsfonts,latexsym,amsmath,amsthm,times}
\usepackage[section]{placeins}
\usepackage{subfigure}
\usepackage{epsfig}
\usepackage{MathShorthand}
\usepackage[english]{babel}
\usepackage{setspace}
\usepackage{mathrsfs}
\usepackage{graphicx}
\usepackage{epstopdf}
\usepackage{titlesec}
\usepackage[font=small,labelfont=bf]{caption}
\usepackage[margin=1in]{geometry}

\DeclareMathOperator{\sech}{sech}

\titleformat{\section}{\large\bfseries}{\thesection}{1em}{}
\titleformat{\subsection}{\normalsize\bfseries}{\thesubsection}{1em}{}

\DeclareMathAlphabet{\mathpzc}{OT1}{pzc}{m}{it}

\renewcommand\thesection{\arabic{section}}
\renewcommand\thesubsection{\thesection.\arabic{subsection}}

\renewcommand\theequation{\thesection.\arabic{equation}}
\renewcommand{\thefootnote}{\fnsymbol{footnote}}

\input{tcilatex}

\renewcommand{\thefootnote}{\alph{footnote})}
\begin{document}

\title{Slowly varying control parameters, delayed bifurcations, \\ and
  the stability of spikes in reaction-diffusion systems} \date{\today}

\author{ \renewcommand{\thefootnote}{\alph{footnote})} 
    J.~C.~Tzou, \\ {\small
    \emph{Department of Mathematics and Statistics, Dalhousie University,
      Halifax, Nova Scotia, B3H 3J5 Canada}}\\[3ex]
  \renewcommand{\thefootnote}{\alph{footnote})} M.~J.~Ward, \\ 
   \small \emph{Department of Mathematics,
    University of British Columbia, Vancouver, V6T 1Z2 Canada}\\[3ex]
  \renewcommand{\thefootnote}{\alph{footnote})} 
  T.~Kolokolnikov, \\ {\small
    \emph{Department of Mathematics and Statistics, Dalhousie University,
      Halifax, Nova Scotia, B3H 3J5 Canada}}\\}
\maketitle

\begin{abstract}
We present three examples of delayed bifurcations for spike solutions
of reaction-diffusion systems. The delay effect results as the system
passes slowly from a stable to an unstable regime, and was previously
analysed in the context of ODE's in [P.Mandel and T.Erneux,
  J.Stat.Phys 48(5-6) pp.1059-1070, 1987]. It was found that the
instability would not be fully realized until the system had entered
well into the unstable regime. The bifurcation is said to have been
\textquotedblleft delayed\textquotedblright\ relative to the threshold
value computed directly from a linear stability analysis. In contrast
to the study of Mandel and Erneux, we analyze the delay effect in
systems of \textit{partial} differential equations (PDE's). In
particular, for spike solutions of singularly perturbed generalized
Gierer-Meinhardt and Gray-Scott models, we analyze three examples of
delay resulting from slow passage into regimes of oscillatory and
competition instability. In the first example, for the
Gierer-Meinhardt model on the infinite real line, we analyze the delay
resulting from slowly tuning a control parameter through a Hopf
bifurcation.  In the second example, we consider a Hopf bifurcation of
the Gierer-Meinhardt model on a finite one-dimensional domain. In this
scenario, as opposed to the \textit{extrinsic} tuning of a system
parameter through a bifurcation value, we analyze the delay of a
bifurcation triggered by slow \textit{intrinsic} dynamics of the PDE
system. In the third example, we consider competition instabilities
triggered by the extrinsic tuning of a feed rate parameter. In all
three cases, we find that the system must pass well into the unstable
regime before the onset of instability is fully observed, indicating
delay. We also find that delay has an important effect on the eventual
dynamics of the system in the unstable regime. We give analytic
predictions for the magnitude of the delays as obtained through the
analysis of certain explicitly solvable nonlocal eigenvalue problems
(NLEP's). The theory is confirmed by numerical solutions of the full
PDE systems.
\end{abstract}

\thispagestyle{plain}


\baselineskip=12pt

\vspace{16pt}


\baselineskip=16pt

\setcounter{equation}{0}

\section{Introduction}

\singlespacing

The stability and bifurcation analysis of differential equations is
one of the cornerstones of applied mathematics. In many applications,
the bifurcation parameter is slowly changing, either extrinsically
(e.g.  parameter is experimentally controlled) or intrinsically
(e.g. the bifurcation parameter is actually a slowly-changing
variable). In these situations, the system can exhibit a significant
delay in bifurcation:\ the instability is observed only as the
parameter is increased well past the threshold predicted by the linear
bifurcation theory, if at all. Often referred to as the slow passage
through a bifurcation, and first analyzed in \cite{mandel1987slow,
  baer1989slow}, there is a growing literature on this subject (see
\cite{kuehn2011mathematical} for a recent overview of the subject and
references therein). Some applications of delayed bifurcations include
problems in laser dynamics \cite{baer1989slow}, delayed chemical
reactions \cite{strizhak1996slow}, bursting oscillations in neurons
\cite{bertram1995topological}, and noise-induced delay of the pupil
light reflex \cite{longtin1990noise}, and early-warning signals
\cite{scheffer2009early}.

Delayed bifurcation phenomena is relatively well understood in the context
of ODE's. However much less is known in the context of PDE's. The main goal
of this paper is to study in detail three representative examples of delayed
bifurcations in PDE's, where explicit asymptotic results are obtainable.

In order to present our examples both analytically and numerically, we
focus on slight variants of the Gierer-Meinhardt (GM) and the
Gray-Scott (GS) reaction-diffusion (RD) models. However, the phenomena
that we present in this paper is expected to be representative of a
larger class of RD systems. The specific systems that
we consider are
\begin{equation}
\mbox{GM\ model:\ }\qquad v_{t}=\varepsilon^{2}v_{xx}-v+\frac{v^{p}}{u^{q}}\,,
 \quad \tau u_{t}=Du_{xx}-u+\frac{1}{\varepsilon}\frac{v^{r}}{u^{s}}
\label{gm-intro}
\end{equation}%
and 
\begin{equation}
\mbox{GS\ model:\ }\qquad v_{t}=\varepsilon^{2}v_{xx}-v+Au^{q}v^{p}\,, \quad \tau
u_{t}=Du_{xx}+1-u+\frac{1}{\varepsilon}u^{s}v^{r}\,,  \label{gs-intro}
\end{equation}
for certain choices of the exponents $p$, $q$, $r$, and $s$ (see
below).  In the singular limit $\varepsilon \rightarrow 0,$ both of
these models have equilibria that consist of spike solutions,
characterized by an ${\mathcal O}(\varepsilon)$ width localization
of $v$ as $\varepsilon^{2}$ becomes asymptotically small. The
component $u$ varies over a comparatively long spatial scale and is
independent of $\varepsilon$.  In all three of our examples, we
consider spike solutions that are qualitatively similar to that shown
in Figure \ref{uvefig}.

To illustrate the main complications when generalizing delayed bifurcations
to PDE's, let us first review the following prototypical ODE example \cite%
{baer1989slow}: $\frac{du}{dt}=\left( -1+\varepsilon t\right) u,\ \ \ \ \
u(0)=u_{0}$ where $\varepsilon >0$ is a small parameter. Here, the
equilibrium state is $u=0$ and can be thought of having an \textquotedblleft
eigenvalue\textquotedblright\ $\lambda (\varepsilon t)=-1+\varepsilon t$
which grows slowly in time, and becomes positive as $t$ is increased past $%
t=1/\varepsilon $, at which point the steady state becomes \textquotedblleft
unstable\textquotedblright . On the other hand, the exact solution is given
by $u(t)=u_{0}\exp \left\{ \frac{\left( \varepsilon t-1\right) ^{2}-1}{%
2\varepsilon }\right\} ,$ which starts to grow rapidly only when the term
inside the curly brackets becomes positive, that is at $t=2/\varepsilon $,
well after the bifurcation threshold of $t=1/\varepsilon .$ The difference
between $2/\varepsilon $ and $1/\varepsilon $ is precisely the delay in
bifurcation, and is inversely proportional to the growth rate $\varepsilon $.
More generally, suppose that $u_{e}$ is an equilibrium state of a system of
ODE's that changes slowly in time, so that the standard linearization $%
u=u_{e}+e^{\lambda t}\eta $ yields an eigenvalue $\lambda =\lambda
(\varepsilon t)$ whose real part is slowly growing at a rate $\mathcal{O}%
(\varepsilon )$ and eventually crosses zero. One then replaces the
linearization by a WKB-type anzatz $u=u_{e}+e^{\frac{1}{\varepsilon }\psi
(\varepsilon t)}\eta $ which yields $\psi ^{\prime }(\varepsilon t)=\lambda
(\varepsilon t)$ with $\psi (0)=0.$ The condition $\psi =0$ with $t>0$ then
yields an algebraic expression for the delay.

There are several novel features present in RD systems when compared
to ODE systems. First, the steady state we consider is not constant,
but rather a spike solution such as that shown in Figure
\ref{uvefig}. The stability theory for spike solutions is by now
well-developed; see for example \cite{wei1999single,
iron2001stability, iron2002dynamics, doelman2001large, muratov2002stability, 
kolokolnikov2005existence} and a recent book \cite{wei2013mathematical}. 
One of the key ingredients is the analysis of the so-called nonlocal 
eigenvalue problem (NLEP), first studied in \cite{wei1999single}.

Second, although the instability thresholds $\lambda =0$ are
analytically computable, the location of the unstable eigenvalue
$\lambda $ itself is usually not known explicitly. However, recently,
a sub-family of RD systems has been identified in
\cite{nec2013explicitly} for which a simple asymptotic determination
of this eigenvalue is possible; this is the case when $p=2r-3,$ $r>2$
in (\ref{gm-intro})\ or (\ref{gs-intro}). For this class of RD
systems, we show that an analytic prediction for the delay can be
obtained in ways similar to \cite{mandel1987slow, baer1989slow}.

Third, the bifurcation (and its delay)\ can be triggered
\emph{intrinsically }by the motion of a spike in the system. That is,
a bifurcation may be triggered not by the extrinsic tuning of a
control parameter, but by dynamics intrinsic to the PDE system.

We now summarize our main results. In \S \ref{GMinfline} we study the
slow passage through a Hopf bifurcation. It was previously shown for
both the GM\ model (\cite{ward2003hopf1, ward2003hopf2})\ and
GS\ models (\cite{doelman2001large, muratov2002stability,
  kolokolnikov2005existence})\ that a Hopf bifurcation occurs as the
parameter $\tau $ is increased past some threshold $\tau _{H}>0$. As
$\tau $ is slowly tuned starting from a stable regime past the Hopf
bifurcation threshold $\tau _{H}$ into an unstable regime, the
amplitude of the spike in Figure \ref{uvefig} begins to oscillate
periodically in time while maintaining its shape. The temporal
oscillations of the amplitude are shown in Figure \ref{growingosc}.
However due to the slow change of parameter, there is a significant
delay until the oscillations are fully realized. In \S \ref{GMinfline}
we compute the delay associated with this bifurcation. This is
illustrated in Figure \ref{vm}.

\begin{empty}\begin{figure}[htbp]
  \begin{center}
    \mbox{
    \subfigure[$u_e(x)$ and $v_e(x)$] 
        {\label{uvefig}
        \includegraphics[width=.4\textwidth]{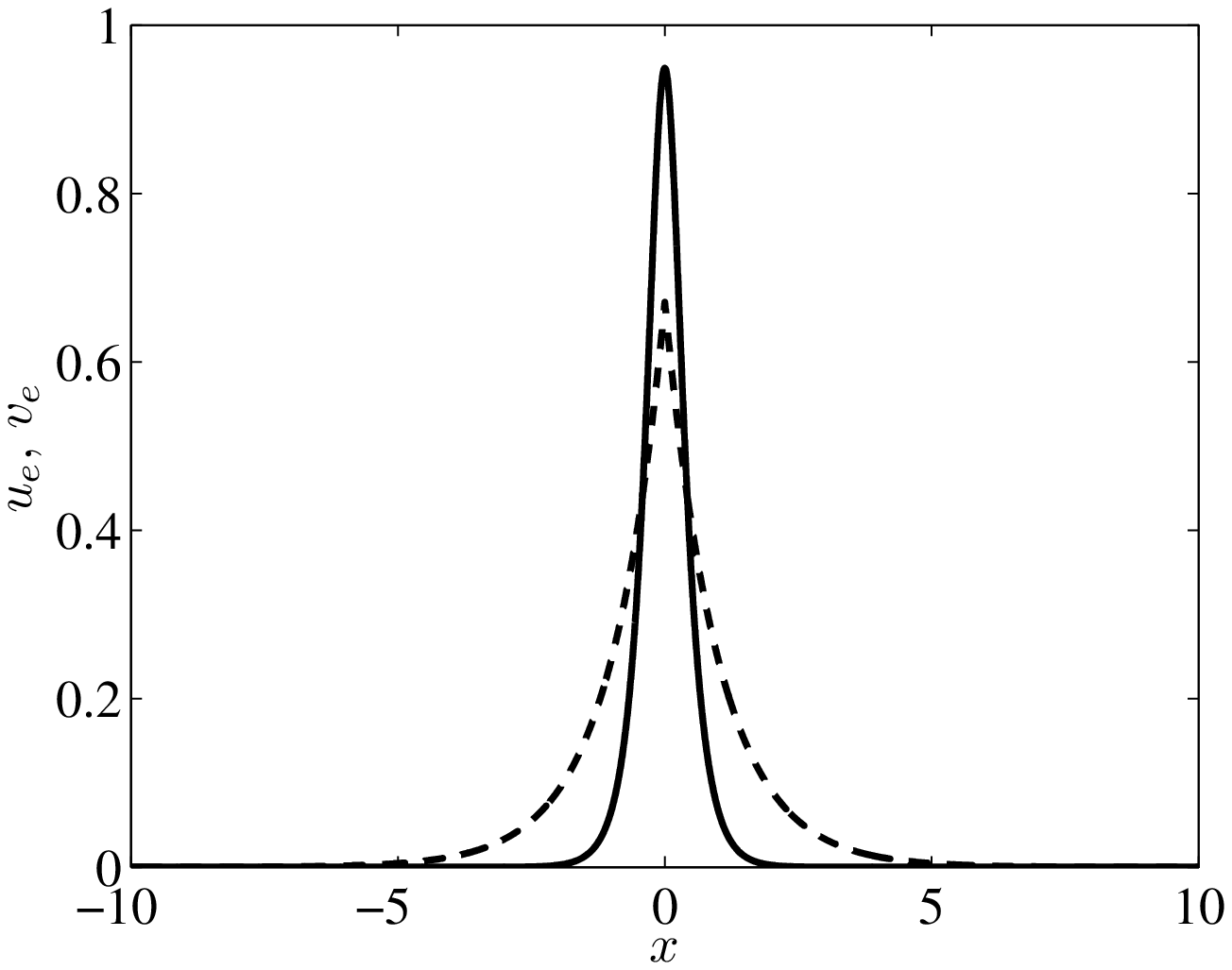}
        }   \hspace{1.5cm}
    \subfigure[amplitude oscillations in time] 
        {\label{growingosc}
        \includegraphics[width=.4\textwidth]{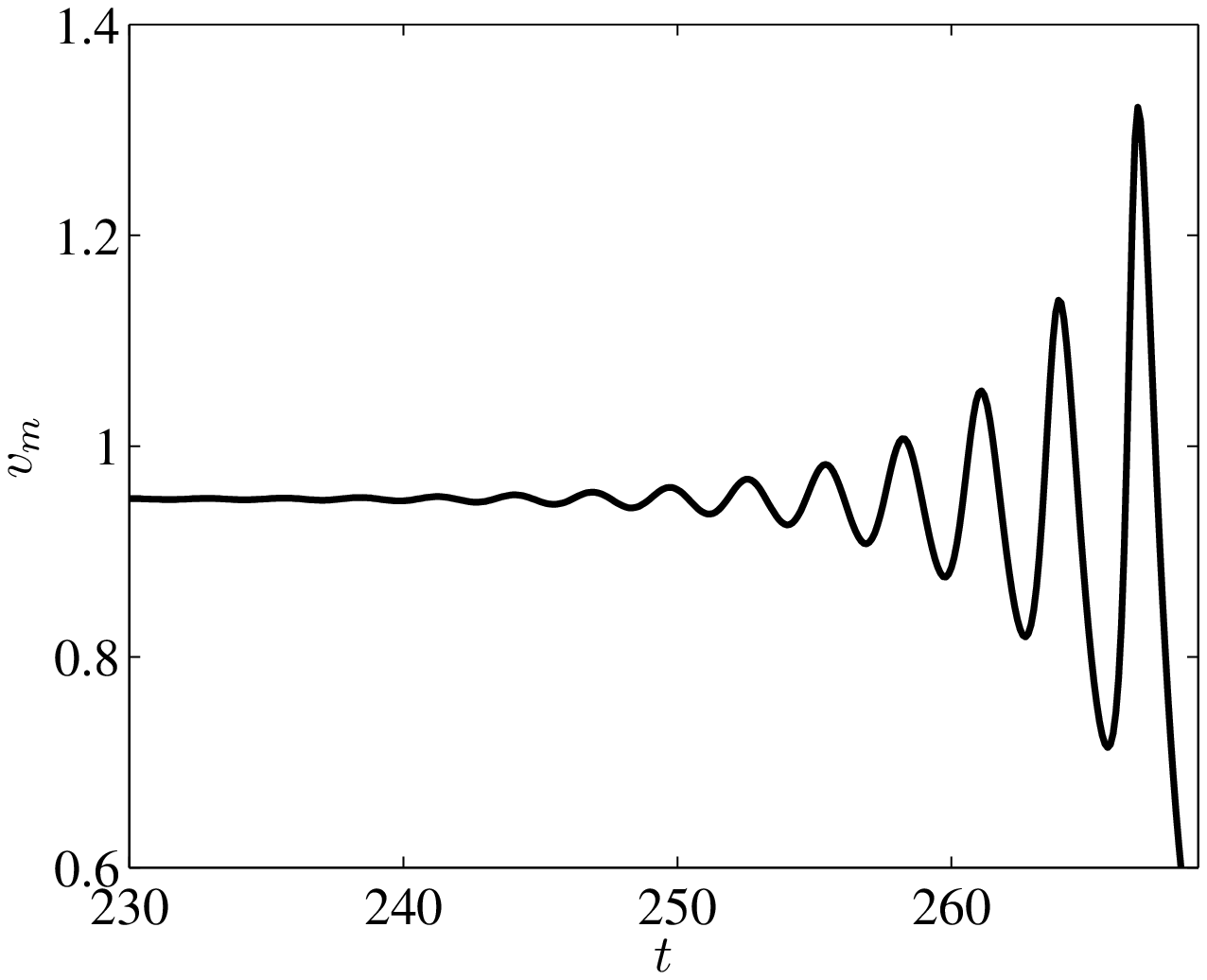}
        }}
    \caption{(a) The asymptotic equilibrium solution of $v$ (solid)
      and $u$ (dashed) for (\ref{GMinf}) with $\varepsilon = 0.3$.  The
      width of the spike in $v_e$ is proportional to $\varepsilon$,
      while $u_e$ is independent of $\varepsilon$. Both $v_e$ and
      $u_e$ are independent of $\tau$. (b) Typical example of
      amplitude oscillations in time when $\tau > \tau_H \approx
      2.114$.  The quantity plotted on the vertical axis is the height
      $v_m$ of the spike in the left figure.}
  \end{center}
\end{figure}\end{empty}%
\begin{empty}\begin{figure}[htbp]
  \begin{center}
    \mbox{
    \subfigure[$v_m$ versus $\tau$] 
        {\label{vm}
        \includegraphics[width=.4\textwidth]{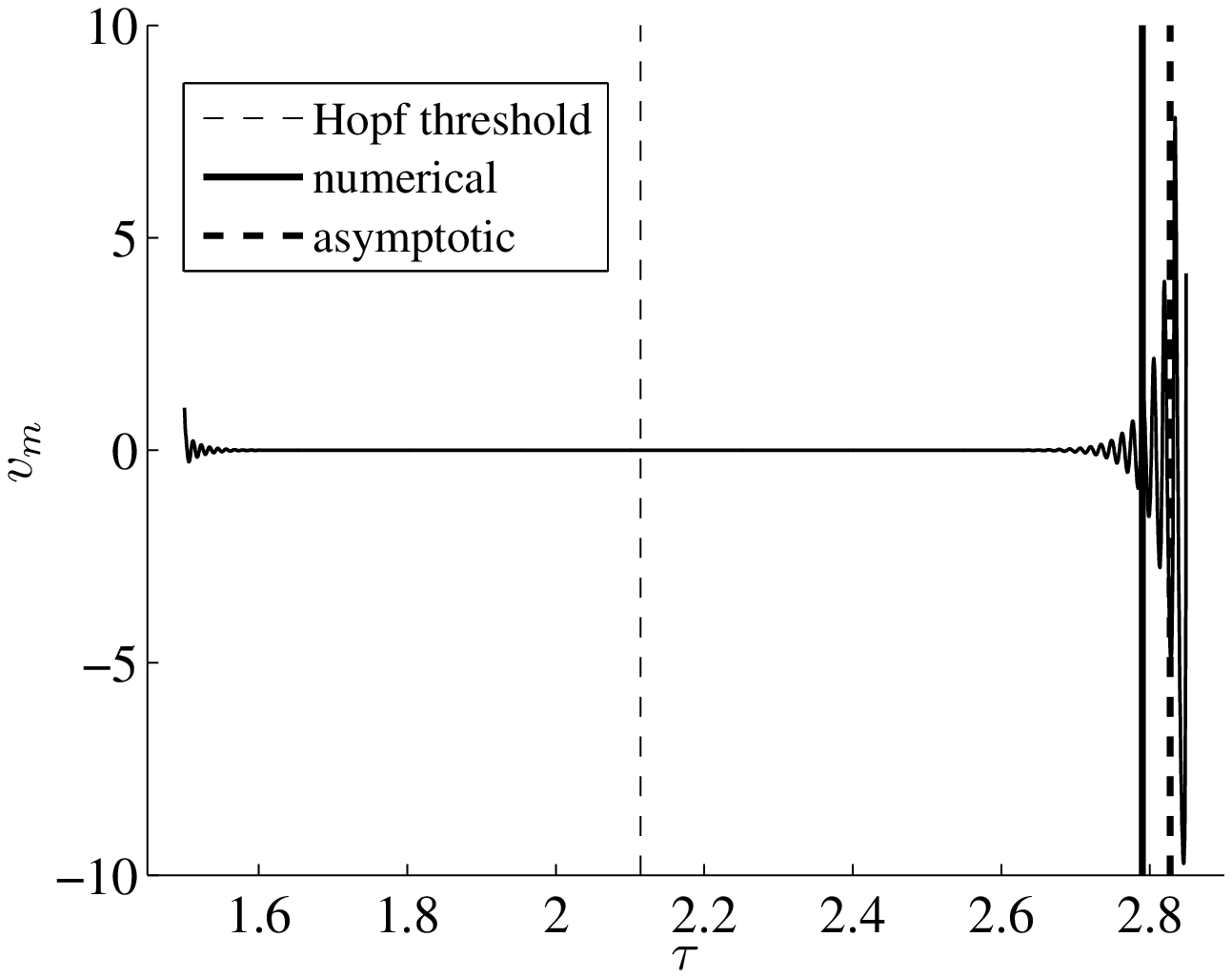}
        }   \hspace{1.5cm}
    \subfigure[numerical versus asymptotic delay] 
        {\label{vmzoom}
        \includegraphics[width=.4\textwidth]{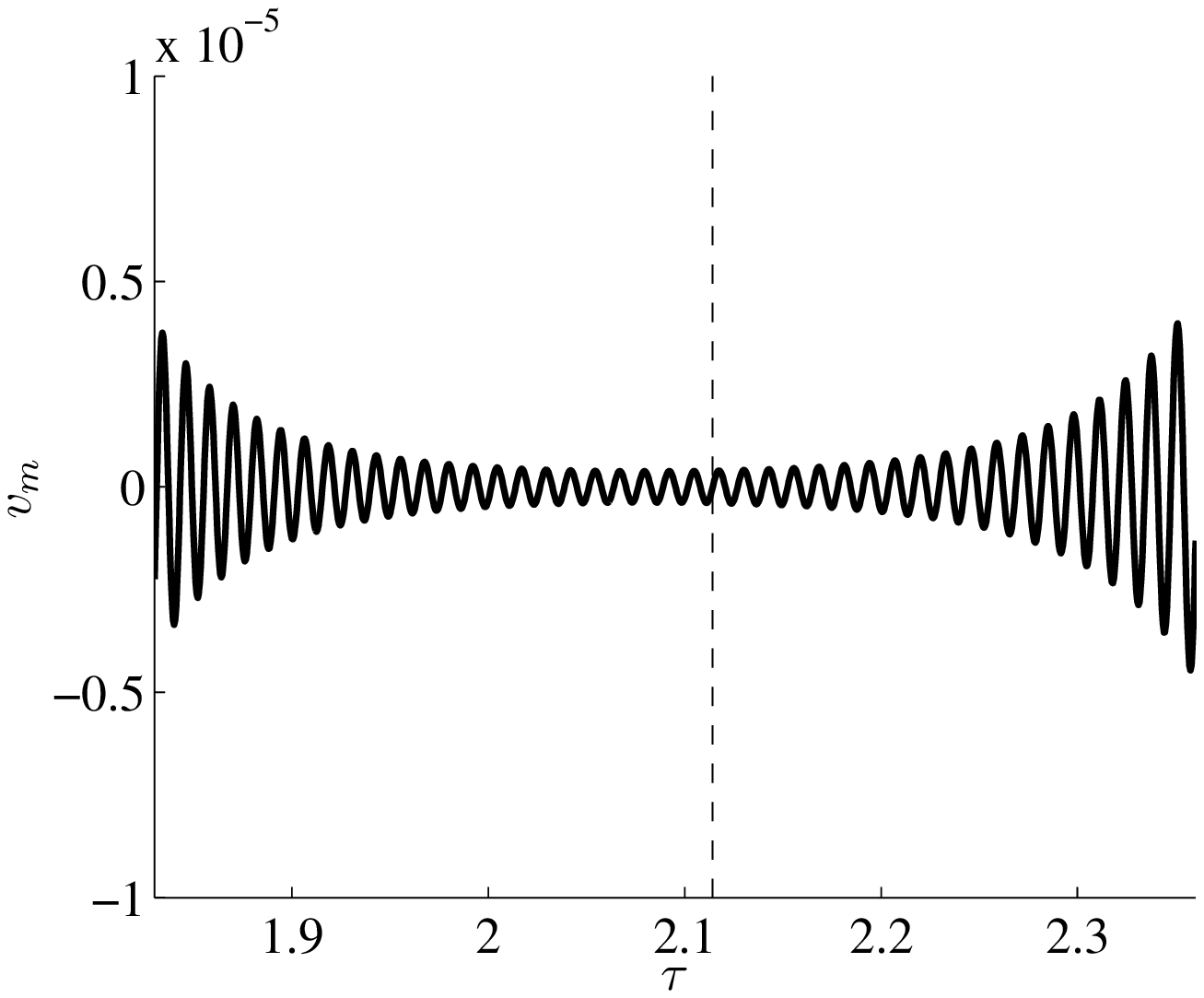}
        }}
    \caption{Delay in the Hopf bifurcation of (\ref{GMinf}). (a) Plot
      of $v_m(\tau)$ as defined in \eqref{vmdef}.  The parameters are
      $\varepsilon = 0.005$ and $\tau = 1.5+\varepsilon t$.  The
      vertical dashed line indicates the Hopf bifurcation value
      $\tau_H \approx 2.114$.  The amplitude first reaches a value of
      one at $\tau^*_m \approx 2.75$ (thick solid line).  The
      asymptotic prediction for $\tau^*$ is $\tau^* \approx 2.828$
      (thick dashed line).  (b) Magnification of (a) on a small
      interval of $\tau$ surrounding $\tau_H$.  The oscillations,
      having decayed when $\tau < \tau_H$, begin growing as $\tau$
      passes $\tau_H$.  Note the scale of the $y$-axis in the right
      figure as compared to that of the left.}
  \end{center}
\end{figure}\end{empty}

In \S \ref{GMfinline}, we consider a quasi-equilibrium one-spike solution of
a GM model centered at $x=x_{0}$ on the domain $|x|<1$. For a spike not
centered at $x=0$, the finite domain induces a slow drift of the spike
toward the origin. Because the drift occurs on an asymptotically slow time
scale while the characteristic time scale of a Hopf bifurcation is $\mathcal{%
O}(1)$, stability analysis may proceed assuming that the spike remains
\textquotedblleft frozen\textquotedblright\ at $x_{0}$. As before, a Hopf
bifurcation threshold $\tau _{H}$ may be derived, but one that is dependent
on the spike location $x_{0}$. That is, $\tau _{H}=\tau _{H}(x_{0};D)$,
where $D$ is the inhibitor diffusivity. We show two typical curves in Figure %
\ref{tauH_vs_x0} for $D=4$ (left)\ and $D=1$ (right). The solution is stable
(unstable) below (above) the $\tau _{H}(x_{0})$ curve, while the arrows
indicate the direction of spike drift. As such, a Hopf bifurcation may be
triggered by dynamics \textit{intrinsic} to the system and not by an
extrinsic tuning of a control parameter.

\begin{empty}\begin{figure}[htbp]
  \begin{center}
    \mbox{
    \subfigure[$\tau_H(x_0)$ for $D = 4$] 
        {\label{tauH_vs_x0_monot}
        \includegraphics[width=.4\textwidth]{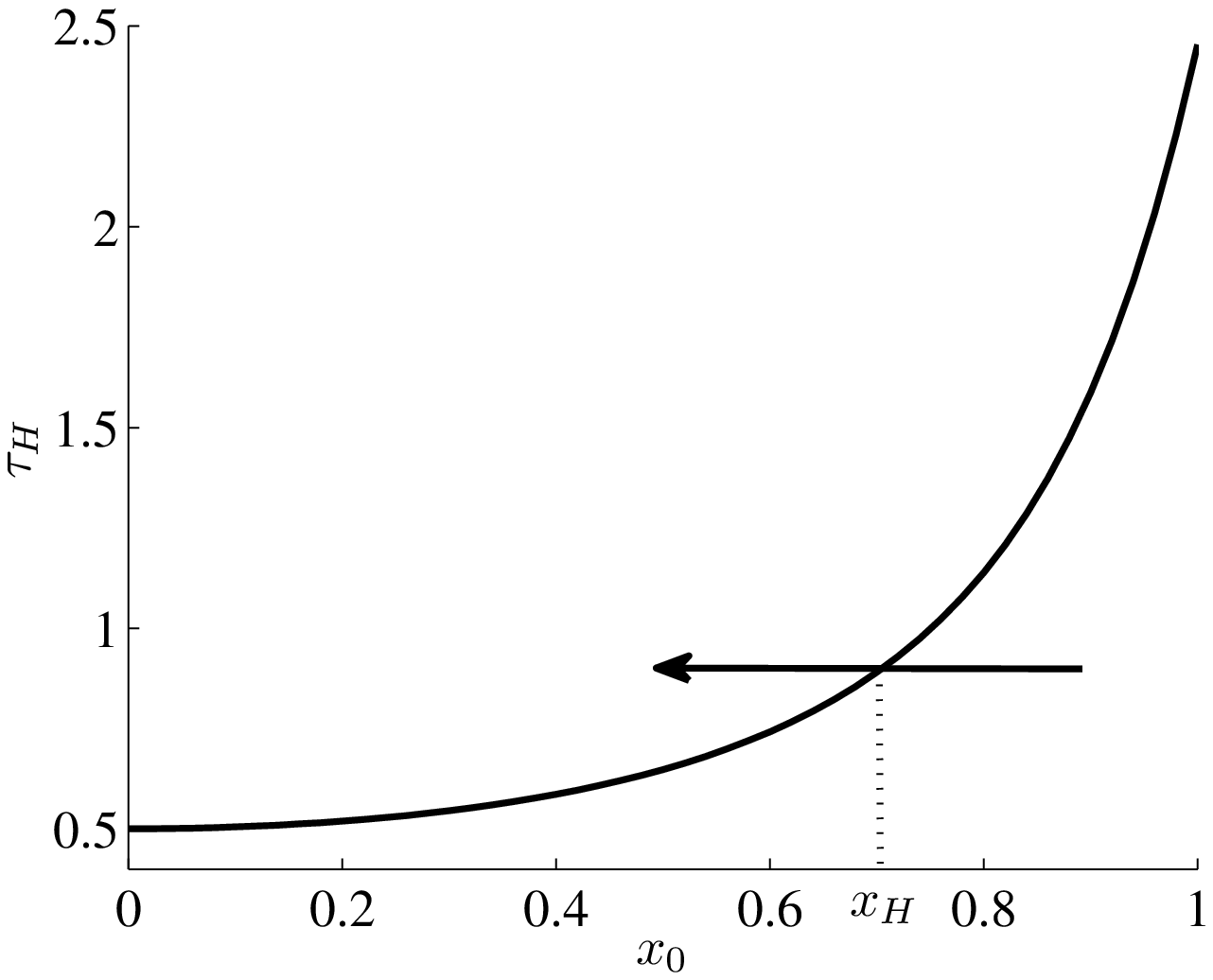}
        }   \hspace{1.5cm}
    \subfigure[$\tau_H(x_0)$ for $D = 1$] 
        {\label{tauH_vs_x0_nonmonot}
        \includegraphics[width=.4\textwidth]{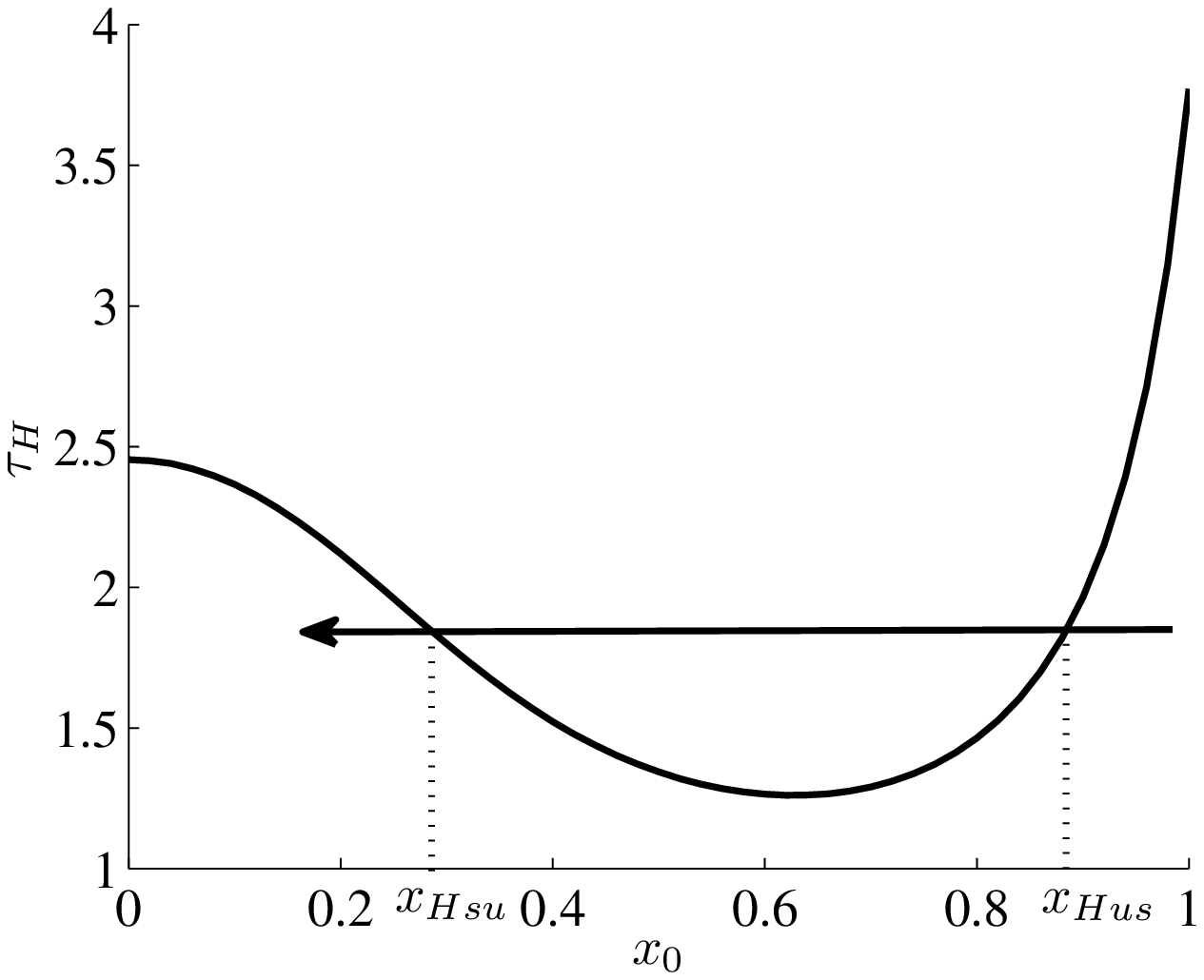}
        }}
    \caption{In both figures, the curve represents the Hopf
      bifurcation threshold $\tau = \tau_H$ plotted against the spike
      location $x_0$ for the GM model. The region below (above) the
      curve is stable (unstable). For a given value of $\tau$, the
      arrows indicate the direction of drift of the spike. Here,
      $(p,q,r,s) = (3, 3, 3, 0)$. In the left figure, with $D = 4$,
      $\tau_H(x_0)$ is monotonic, and once the spike enters the
      unstable zone $x_0 < x_H$, it remains in the unstable zone for
      all time. In the right figure, with $D = 1$, $\tau_H(x_0)$ is
      non-monotonic. For sufficiently small $\tau$, the spike may pass
      one threshold $x_{Hsu}$ from an stable to unstable zone, then
      pass through another threshold $x_{Hus}$ from an unstable to
      stable zone. It then remains in a stable zone for all later times.}
     \label{tauH_vs_x0}
  \end{center}
\end{figure}\end{empty}

For a given value of $\tau $, the scenario in Figure
\ref{tauH_vs_x0_monot} indicates only one threshold crossing as the
spike drifts toward equilibrium. However, the scenario depicted in
Figure \ref{tauH_vs_x0_nonmonot} shows the possibility of two
threshold crossings for sufficiently small $\tau $. In particular, we
find that, by selecting initial conditions to introduce sufficient
delay into the system, the spike may pass \textquotedblleft
safely\textquotedblright\ through the unstable zone without the Hopf
bifurcation ever fully setting in. In doing so, we show that delay has
an important role in determining the dynamics of a system.

In \S \ref{GScomp} we consider a competition instability of a two-spike
equilibrium of a singularly perturbed generalized GS model. Instead
of interior spikes as in the previous examples, two half-spikes are centered
at the boundaries $x = \pm 1$. A typical solution is shown in Figure \ref%
{boundaryspikes}. The solid line depicts two half-spikes in the activator
centered at the two boundaries. Note that the inhibitor component (dashed)
has been scaled by a factor of six to facilitate plotting. The spike
locations remain fixed at the boundaries for all time. In addition to
time-oscillatory Hopf instabilities, a solution containing two or more
spikes may undergo a time-monotonic competition instability leading to the
collapse of one or more spikes. In this example we study the delay in
competition instability as a feed-rate parameter $A$ is decreased through
the stability threshold $A_-$. In Figure \ref{amp_vs_t}, we show a typical
result of such an instability, as the amplitude of the left spike (light
solid) collapses to zero while that of the right (heavy solid) grows.

\begin{empty}\begin{figure}[htbp]
  \begin{center}
    \mbox{
    \subfigure[two boundary spike solution] 
        {\label{boundaryspikes}
        \includegraphics[width=.4\textwidth]{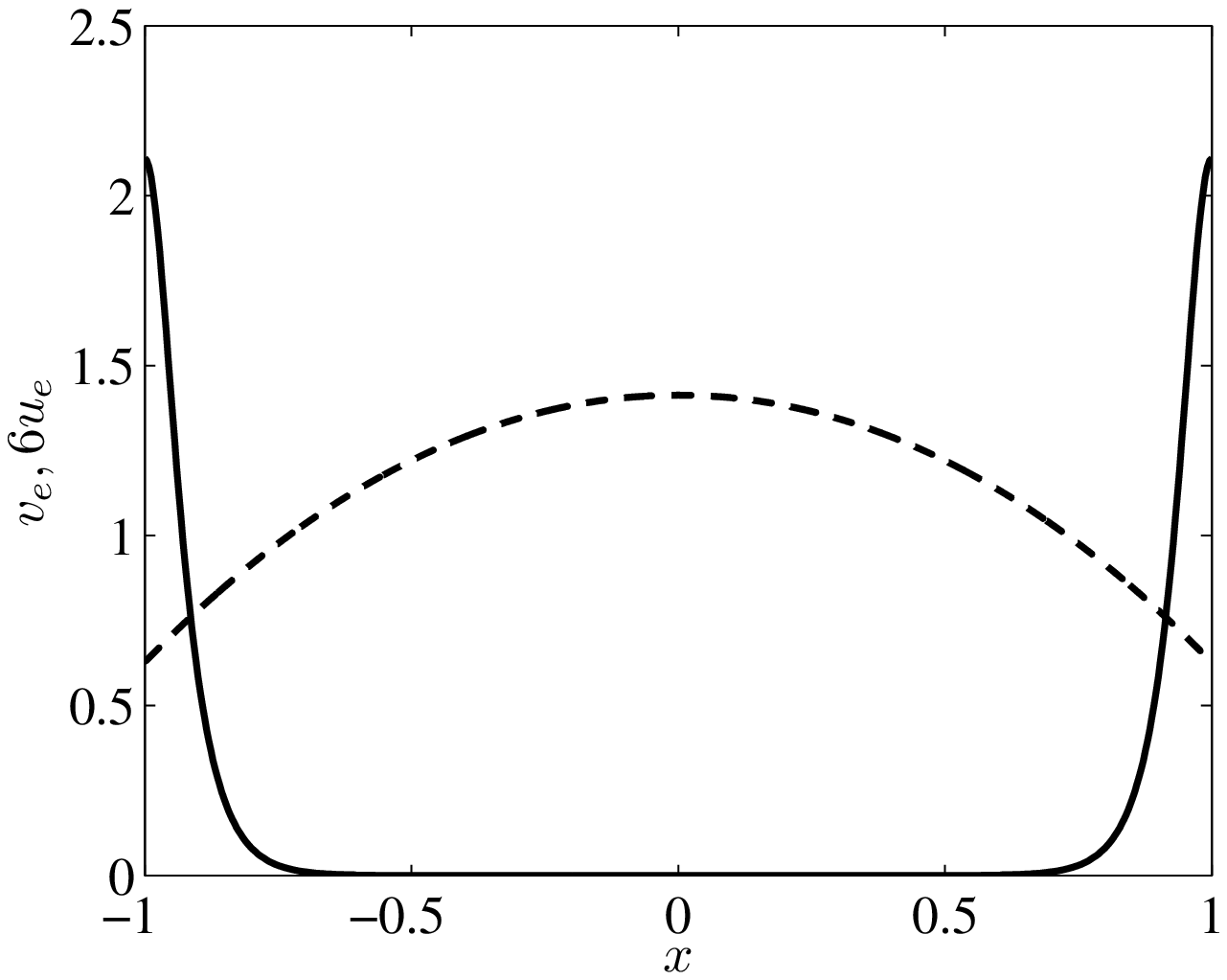}
        }   
    \subfigure[spike amplitudes versus time] 
        {\label{amp_vs_t}
        \includegraphics[width=.4\textwidth]{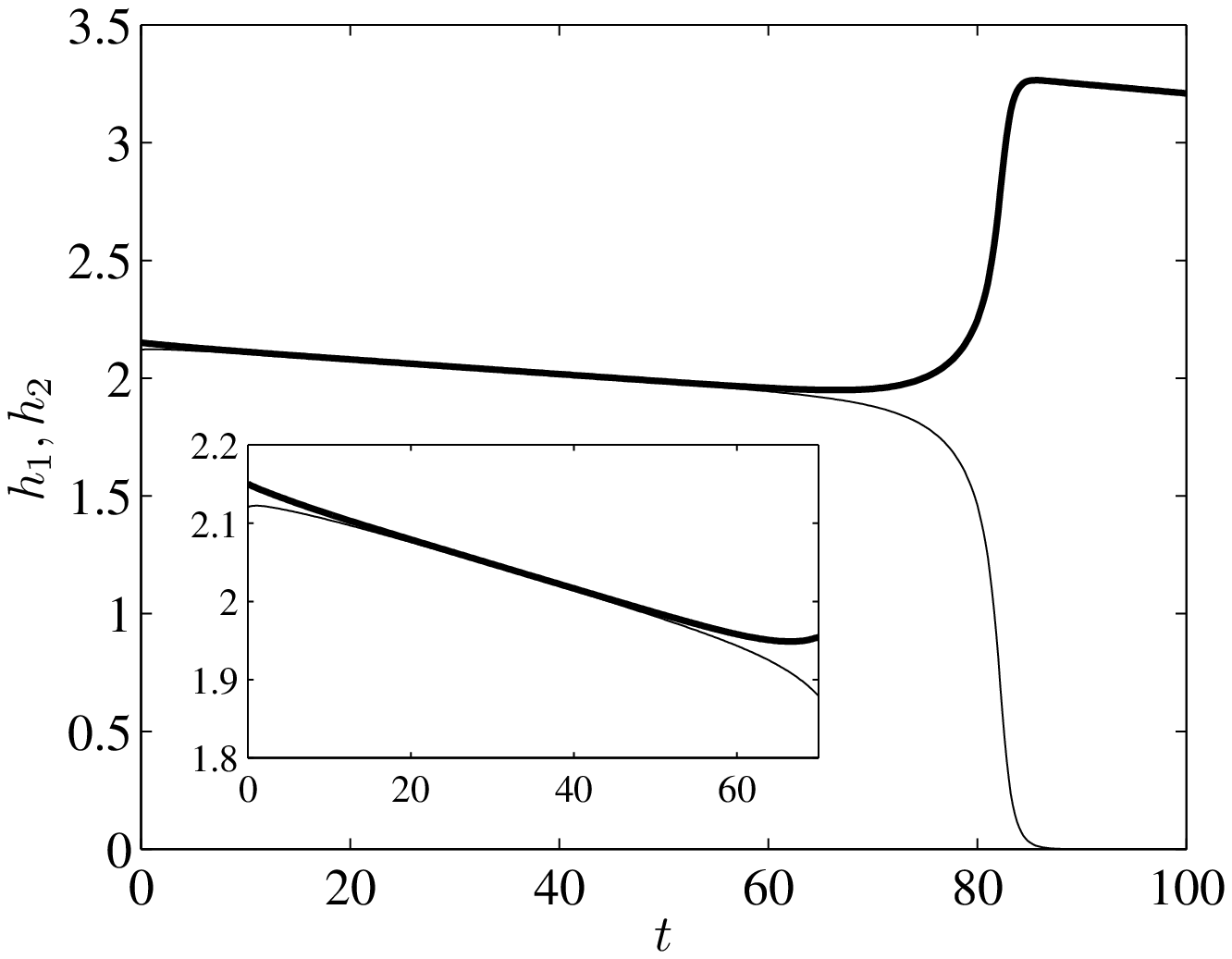}
        }}
    \caption{In the left figure, we show a two boundary spike
      equilibrium solution for $v(x)$ (solid) and $u(x)$ (dashed) in
      \eqref{GS}. The two spikes are of equal height. The $u$
      component has been scaled by a factor of $6$ to facilitate
      plotting. Here, $\varepsilon = 0.05$, $D = 3$ and $A =
      4.1611$. In the right figure, we show the amplitudes of the left
      (light solid) and right (heavy solid) spikes as $A$ is slowly
      decreased past the competition threshold. The inset shows that
      the initial perturbation decreases the amplitude of the left
      spike relative to equilibrium, and increases that of the
      right. With $A$ starting in the stable regime, the amplitudes
      initially grow closer together. As $A$ passes the stability
      threshold, the spikes grow farther apart until the left spike
      amplitude collapses to 0. The results in the right figure are
      for $\varepsilon = 0.004$ and $D = 3$.}
  \end{center}
\end{figure}\end{empty}

A feature of spike solutions in the Gray-Scott model is that there
exists a saddle node in the feed-rate parameter $A$, which we denote
by $A_{m}.$ That is, for $A<A_{m}<A_{-}$, the solution being
considered ceases to exist. We give a typical bifurcation diagram in
Figure \ref{bifdiag} displaying such a saddle node. The horizontal
axis is the bifurcation parameter $A$, while the vertical axis is the
amplitude of the activator boundary spikes. We consider in this
example only the upper solution branch, since the lower branch is known to
be unstable for all $A$. The arrow shows the direction of decrease in
$A$ from a stable regime (heavy solid) to the regime unstable to the
competition mode (light solid). Note that the competition threshold
occurs before the saddle as $A$ decreases. However, as Figure
\ref{bifdiag} suggests, with sufficient delay, the system may reach
the saddle point without the competition instability fully setting
in. We find in this scenario that, while the effect of the saddle is
much weaker in comparison to that of the competition instability,
sufficient delay in the onset of the instability may allow the saddle
effect to dominate. As in the previous example, we thus find that
delay may be critical in determining the eventual fate the system.

\begin{empty}\begin{figure}[htbp]
  \begin{center}
	\includegraphics[width=.4\textwidth]{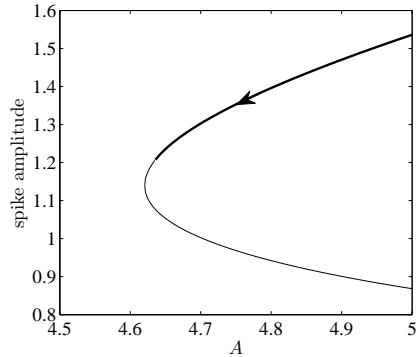}
	\caption{Bifurcation diagram for the two boundary spike
          solution of the GS model when $D = 0.4$. On the upper
          branch, the solid segment indicates stable solutions, while
          the light solid segment indicates solutions unstable to the
          competition mode. The stability transition occurs at $A =
          A_- \approx 4.6351$, while the saddle node occurs at $A =
          A_m \approx 4.6206$. The arrow indicates the evolution of
          the spike amplitude as $A$ is decreased. The lower branch is
          unstable for all values of $A$.}
	\label{bifdiag}
	\end{center}
\end{figure}\end{empty}

In each of the following examples, we focus on three main objectives. We
first seek to demonstrate analytically why a delay in the onset of an
instability occurs when a system is slowly tuned past a stability threshold.
We then show that an explicitly solvable nonlocal eigenvalue problem (NLEP)
allows for an analytic prediction of the magnitude of delay. Finally, we
compare analytic predictions of delay to numerical results obtained from
solving the full PDE systems. The construction of the spike equilibrium and
quasi-equilibrium solutions, as well as the subsequent stability analysis
leading to an explicitly solvable NLEP, follow from similar past problems.
Since our emphasis is on illustrating the delay effect, we include only
enough of the analysis to meet our stated objectives, and relegate the
remaining to the appendix.

\setcounter{equation}{0}

\section{Example 1: Hopf bifurcation of a one-spike solution on the infinite
line} \label{GMinfline}

In the first example, we consider a Hopf bifurcation of a one-spike
equilibrium solution to a particular exponent set of the GM system (\ref%
{gm-intro}) on the infinite real line

\bes \label{GMinf} 
\begin{equation}  \label{GMv}
v_t = \varepsilon^2 v_{xx} - v + \frac{v^3}{u^2} \,, \qquad -\infty < x <
\infty \,, \qquad t > 0 \,, \qquad v \to 0 \quad \mbox{as} \quad |x| \to
\infty \,,
\end{equation}

\begin{equation}
\tau u_{t}=u_{xx}-u+\frac{v^{3}}{\varepsilon }\,,\qquad -\infty <x<\infty
\,,\qquad t>0\,,\qquad u\rightarrow 0\quad \mbox{as}\quad |x|\rightarrow
\infty \,.  \label{GMu}
\end{equation}%
\ees The primary motivation for this choice of exponents is that they
satisfy the key relationship $p=2r-3$ from
\cite{nec2013explicitly}. This relationship allows for an explicit
computation of the large eigenvalue of the NLEP problem associated
with the linearization around the spike equilibrium. Here,
$\varepsilon^2 \ll 1$ is the diffusivity of the activator component
$v$, while the diffusivity of the inhibitor component $u$ is set to
unity without loss of generality. We consider an equilibrium solution
of \eqref{GMinf} for which the activator takes the form of a single
spike of width $\mathcal{O}(\varepsilon)$ centered at $x = 0$ while
the inhibitor varies over an ${\mathcal O}(1)$ spatial scale. The
parameter $\tau$ is taken to be the bifurcation parameter. When $\tau$
is large, the inhibitor responds sluggishly to small activator
deviations from equilibrium, leading to oscillations in the height of
the activator spike. When $\tau$ is below a certain threshold value
$\tau_H$, the response is fast enough such that oscillations decay in
time. When $\tau$ exceeds $\tau_H$, a Hopf bifurcation occurs and
oscillations grow in time. In this section, we analyze the scenario
where $\tau$ is slowly increased past $\tau_H$ starting from $\tau =
\tau_0 < \tau_H$.

\subsection{Analytic calculation of delay}

\label{analyticdelay}

From \cite{nec2013explicitly}, the one-spike equilibrium solution of
\eqref{GMinf} takes the form

\begin{equation}  \label{eqbminf}
v_e \sim U_0w\left(\varepsilon^{-1}x \right) \,, \qquad u_e \sim \frac{U_0}{%
G(0,0)}G(x;0) \,,
\end{equation}

\noI where $w(y)$, $G(x,x_0)$, and $U_0$ are defined by

\begin{equation} \label{eqbminf2}
w(y) = \sqrt{2}\sech y \,; \qquad \intinf w^3 \, dy \equiv b = \pi\sqrt{2}%
\,, \qquad G(x;x_0) = \frac{1}{2}e^{-\lvert x-x_0\rvert} \,, \qquad U_0 = 
\frac{1}{\sqrt{b\,G(0;0)}} \,.
\end{equation}

\noI We plot the solutions for $v$ (solid) and $u$ (dashed) in Figure 
\ref{uvefig} on a domain of length $20$ for $\varepsilon = 0.3$. Note that the
equilibrium solution \eqref{eqbminf} is independent of $\tau$, which only
affects stability.

In Appendix \ref{appA}, we perform a linear stability analysis of the
equilibrium solution \eqref{eqbminf} by perturbing the equilibrium
solution as

\begin{equation}  \label{pert}
v = v_e + e^{\lambda t} \phi \,, \qquad u = u_e + e^{\lambda t} \eta \,;
\qquad \phi, \eta \ll 1 \,,
\end{equation}

\noI where $\lambda$ and $(\phi, \eta)$ are the associated eigenvalue and
eigenfunctions, respectively. From the resulting linearized equation, we
derive a nonlocal eigenvalue problem (NLEP) governing its $\mathcal{O}(1)$
time scale stability to amplitude perturbations. Solving the NLEP
explicitly, we obtain an exact expression for the eigenvalue $\lambda$ in
terms of $\tau$ as

\begin{equation}  \label{NLEPsol}
\lambda(\tau) = 3 - \frac{9}{\sqrt{1+\tau\lambda}} \,.
\end{equation}

\noI The function $\lambda(\tau)$ in \eqref{NLEPsol} may be inverted for $%
\tau$, yielding

\begin{equation}  \label{taulambda}
\tau(\lambda) = \frac{81}{\lambda(3-\lambda)^2} - \frac{1}{\lambda} \equiv
f(\lambda) \,.
\end{equation}

To analyze \eqref{NLEPsol}, we define the function

\begin{equation}  \label{scriptG}
\mathcal{G}(\lambda) \equiv \frac{9}{3-\lambda} \,.
\end{equation}

\noI Then $\lambda$ is a root of the equation
\begin{equation}  
\sqrt{1+\tau\lambda} = \mathcal{G}(\lambda) \,. \label{lambdaeq}
\end{equation}

\noI The function $\mathcal{G}(\lambda)$ is positive (negative) for
$\lambda < 3$ ($\lambda > 3$), and approaches $\pm \infty$ as 
$\lambda\to 3^{\mp}$.  With $\mathcal{G}(0) = 3$, $\mathcal{G}^\prime > 0$ and
$\mathcal{G}^{\prime\prime} > 0$ on $0 <\lambda <3$, we find that
\eqref{lambdaeq} has no positive real roots if $\tau \ll 1$, and two
positive real roots on $0 <\lambda <3$ if $\tau \gg 1$. These two
cases are illustrated schematically in Figure \ref{lambdareal} below.

\begin{empty}\begin{figure}[htbp]
  \begin{center}
	\includegraphics[width=.4\textwidth]{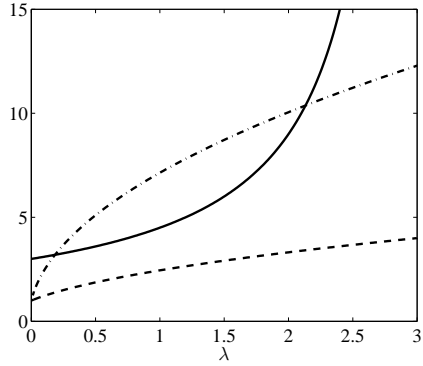}
	\caption{The function $\mathcal{G}(\lambda)$ in
          \eqref{scriptG} is indicated by the solid curve on the
          interval $[0,3]$. The dashed curve depicts the function
          $\sqrt{1+\tau\lambda}$ for $\tau$ sufficiently small so that
          it does not intersect $\mathcal{G}(\lambda)$. The
          dash-dotted curve depicts $\sqrt{1+\tau\lambda}$ for large
          $\tau$. In this case, there are two intersections,
          representing two positive real roots of \eqref{lambdaeq}.}
	\label{lambdareal}
	\end{center}
\end{figure}\end{empty}

The argument principle can be applied to show that the two positive
real roots when $\tau \gg 1$ are the only two roots for $\lambda$ in
the right-half plane (\cite{nec2013explicitly}). Further, it can be
shown that there are no roots in the right-half plane for $\tau$
sufficiently small. Since $\lambda = 0$ is never a solution of
\eqref{lambdaeq} for finite $\tau$, by continuity of the roots of
\eqref{lambdaeq} in $\tau$, there exists a critical value $\tau =
\tau_H$ for which $\lambda = i\lambda_I$ for some positive real
$\lambda_I$. From \eqref{lambdaeq} the unique Hopf bifurcation point is

\begin{equation}  \label{Hopfbifpoint}
\tau_H = \frac{1}{36}\left[2c^2 + 12 + 2c\sqrt{c^2 + 12} \right] > \frac{3}{2%
}, \quad c \equiv \frac{3\sqrt{3}}{2}; \qquad \lambda_I = 3\sqrt{1-\frac{2}{%
3\tau_H}} \,.
\end{equation}

\noI We thus conclude that $\Re(\lambda) < 0$ when $\tau < \tau_H$, and $%
\Re(\lambda) > 0$ when $\tau > \tau_H$.

To understand the phenomenon of delayed Hopf bifurcation as $\tau =
\tau(\sigma t)$, $\sigma \ll 1$, is slowly increased from $\tau =
\tau_0 < \tau_H$ into the unstable regime $\tau > \tau_H$, we must
track the decay of the perturbation in \eqref{pert} during the time
interval that $\tau$ is below $\tau_H$. The longer the system remains
in the stable regime, the more the perturbation decays, and therefore
the more time it requires for the perturbation to grow to its original
amplitude when $\tau > \tau_H$. To analyze this effect, we follow
\cite{baer1989slow} and rewrite the perturbations in \eqref{pert} by
applying the WKB ansatz

\begin{equation}  \label{pertpsi}
v = v_e + e^{\frac{1}{\sigma}\psi(\xi)} \phi \,, \qquad u = u_e + e^{\frac{1%
}{\sigma}\psi(\xi)} \eta \,, \qquad \xi = \sigma t \,, \quad \sigma \ll 1;
\qquad \phi, \eta \ll 1 \,.
\end{equation}

\noI Differentiating \eqref{pertpsi} with respect to $t$, we calculate that

\begin{equation}  \label{pertpsiprime}
v_t = \frac{1}{\sigma} \psi^\prime(\xi) \frac{d\xi}{dt} e^{\frac{1}{\sigma}%
\psi(\xi)}\phi \,, \qquad u_t = \frac{1}{\sigma} \psi^\prime(\xi) \frac{d\xi%
}{dt} e^{\frac{1}{\sigma}\psi(\xi)}\eta \,.
\end{equation}

\noI Noting that $d\xi/dt = \sigma$ in \eqref{pertpsiprime}, and upon
replacing $\tau$ in \eqref{GMinf} by $\tau = \tau(\xi)$ and linearizing, we
find that $\psi^\prime(\xi)$ satisfies the same eigenvalue problem as does
the stationary eigenvalue $\lambda$ in \eqref{NLEPsol}. That is, we obtain
the ordinary differential equation (ODE) for $\psi(\xi)$

\begin{equation}  \label{psipeq}
\psi^\prime(\xi) \equiv \psi_R^\prime(\xi) + i\psi_I^\prime(\xi) = 3 - \frac{%
9}{\sqrt{1 + \tau(\xi)\psi^\prime(\xi)}} \,, \qquad \psi(0) = 0 \,.
\end{equation}

\noI The initial condition for $\psi$ in \eqref{psipeq} is set without
loss of generality by noting that any prefactors in the perturbation
may be absorbed into $\phi$ and $\eta$. In the following, we assume
that $\tau(\xi)$ is a monotonically increasing function of $\xi$ with
$\tau(0) = \tau_0 < \tau_H$.

The correspondence between $\psi^\prime$ with $\lambda$ implies that $%
\psi_R(\sigma t)$ is a decreasing function of time as long as $\tau$ remains
below the threshold $\tau = \tau_H$. This is illustrated in Figure \ref
{psi_p_vs_tau} below, as $\psi_R^\prime$ is negative for all $\tau < \tau_H
\approx 2.114$, where $\tau_H$ is computed from \eqref{Hopfbifpoint}. During
this period, the perturbation decays to an amplitude of order $\mathcal{O}%
(e^{-1/\sigma})$, with $\sigma \ll 1$. The amplitude only begins to grow
once $\tau$ is ramped up past $\tau_H$. The time $t^*>0$ at which the
perturbation grows back to its original amplitude occurs when $\psi_R = 0$.
The longer the system remains in the stable regime, the more $\tau(\sigma t)$
must be ramped up past $\tau_H$ before the perturbation amplitude is
restored and the instability is fully realized. We define the delay to be
the amount by which $\tau(\sigma t^*) \equiv \tau^*$ exceeds $\tau_H$, and
refer to this as the delay effect.

To calculate the value of $\tau^*$ analytically at which $\psi_R = 0$, we
begin by using for $\tau(\xi)$ a linear ramping function

\begin{equation}  \label{tauramp}
\tau(\xi) = \tau_0 + \xi \,, \qquad \xi/t = \sigma \ll 1 \,, \qquad \tau_0 <
\tau_H \,.
\end{equation}

\noI Integrating the relation $\psi^\prime(\xi) = \lambda$ with respect to
slow time $\xi$, we obtain

\begin{equation}  \label{intlambda}
\int_0^{\psi(\tau_1)} \! \psi^\prime \, d\xi = \int_0^{\xi_1} \! \lambda \,
d\xi = \int_{\tau_0}^{\tau_1} \! \lambda \, d\tau \,,
\end{equation}
\noI where $\tau_1 = \tau(\xi_1)$, and where we have used
\eqref{tauramp} to change the variable of integration to $\tau$. Using
\eqref{taulambda} to again change the variable of integration of the
third integral in \eqref{intlambda} from $\tau$ to $\lambda$, we
calculate

\begin{equation}  \label{psitau1}
\psi(\tau_1) = \lbrack\lambda_1f(\lambda_1) - F(\lambda_1)\rbrack -
\lbrack\lambda_0f(\lambda_0) - F(\lambda_0)\rbrack \,,
\end{equation}

\noI where $\lambda_0 = \lambda(\tau_0)$, $f(\lambda)$ is defined in %
\eqref{taulambda}, $\tau_1 = f(\lambda_1)$, and

\begin{equation}  \label{Flambda}
F(\lambda) = \int^\lambda \! f(s) \, ds = 8\log\lambda - 9\log(\lambda-3) - 
\frac{27}{\lambda-3} \,.
\end{equation}

\noI Setting the right-hand side of \eqref{psitau1} to 0 with $F(\lambda)$
defined in \eqref{Flambda} yields an algebraic equation for $\lambda_1 =
\lambda^*$. We then calculate $\tau^* = f(\lambda^*)$ using \eqref{taulambda}%
. Note that $\tau^*$ is independent of $\varepsilon$. That is, the delay in
terms of $\tau$ is independent of the rate at which it is decreased.
However, the duration in time of the delay increases monotonically with $%
1/\varepsilon$, as observed in \cite{baer1989slow}.

Our analysis, confirmed by numerical computations, shows that the farther $%
\tau $ starts below threshold in the stable regime, the farther it must be
increased above threshold for the instability to fully set in. In Figure \ref
{delay}, we illustrate the delay phenomenon for a range of values of $\tau
_{0}$. Denoting $\tau ^{\ast }$ as the value of $\tau $ at which $\psi _{R}$
changes sign from negative to positive, we find that the farther into the
stable regime $\tau _{0}$ is, the farther into the unstable regime $\tau
^{\ast }$ must be for oscillations resulting from the Hopf bifurcation to
grow to the size of the original perturbation. The increasing relationship
between the \textquotedblleft initial buffer\textquotedblright\ $\tau
_{H}-\tau _{0}$ and the distance above threshold before onset $\tau ^{\ast
}-\tau _{H}$ is typical in all of our findings, regardless of the triggering
parameter or mechanism.

\begin{empty}\begin{figure}[htbp]
  \begin{center}
	\includegraphics[width=.4\textwidth]{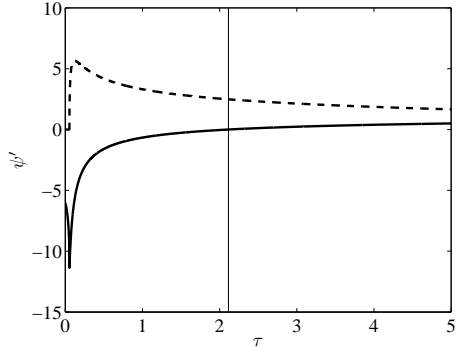}
	\caption{A plot of the real (heavy solid) and imaginary
          (dashed) parts of the solution to the algebraic equation for
          $\psi^\prime$ in \eqref{psipeq}. At $\tau = \tau_H \approx
          2.114$ (solid vertical line), $\Re(\psi^\prime) = 0$, while
          $\Re(\psi^\prime) < 0$ ($\Re(\psi^\prime) > 0$) when $\tau <
          \tau_H$ ($\tau > \tau_H$). }
	\label{psi_p_vs_tau}
	\end{center}
\end{figure}\end{empty}

\subsection{Numerical validation}

In this section, we compare the asymptotic results for delay obtained above
with numerical results computed from the GM model \eqref{GMinf}. We replace $%
\tau $ in \eqref{GMu} with a slowly varying function $\tau =\tau
(\varepsilon t)$ according to \eqref{tauramp}. To solve \eqref{GMinf}
numerically, we used a semi-implicit second order predictor-corrector method
in time and pseudo-spectral Fourier method in space. The following results
did not differ significantly when the number of grid points was doubled
while the time-step was decreased by a factor of four. To approximate the
infinite line, we used a computational domain length of $L=20$. Doubling $L$
did not alter the results significantly.

The initial conditions were taken as a perturbation of the true equilibrium

\begin{equation}  \label{ICinf}
v(x,0) = v_e^*(x)\left[1 + \delta \cos\left(\frac{\pi x}{\varepsilon}%
\right)e^{-\left(\frac{x}{\varepsilon}\right)^2} \right] \,, \qquad u(x,0) =
u_e^*(x) \,,
\end{equation}

\begin{figure}[htbp]
  \begin{center}
    \mbox{
    \subfigure[] 
        {\label{delay}
        \includegraphics[width=.4\textwidth]{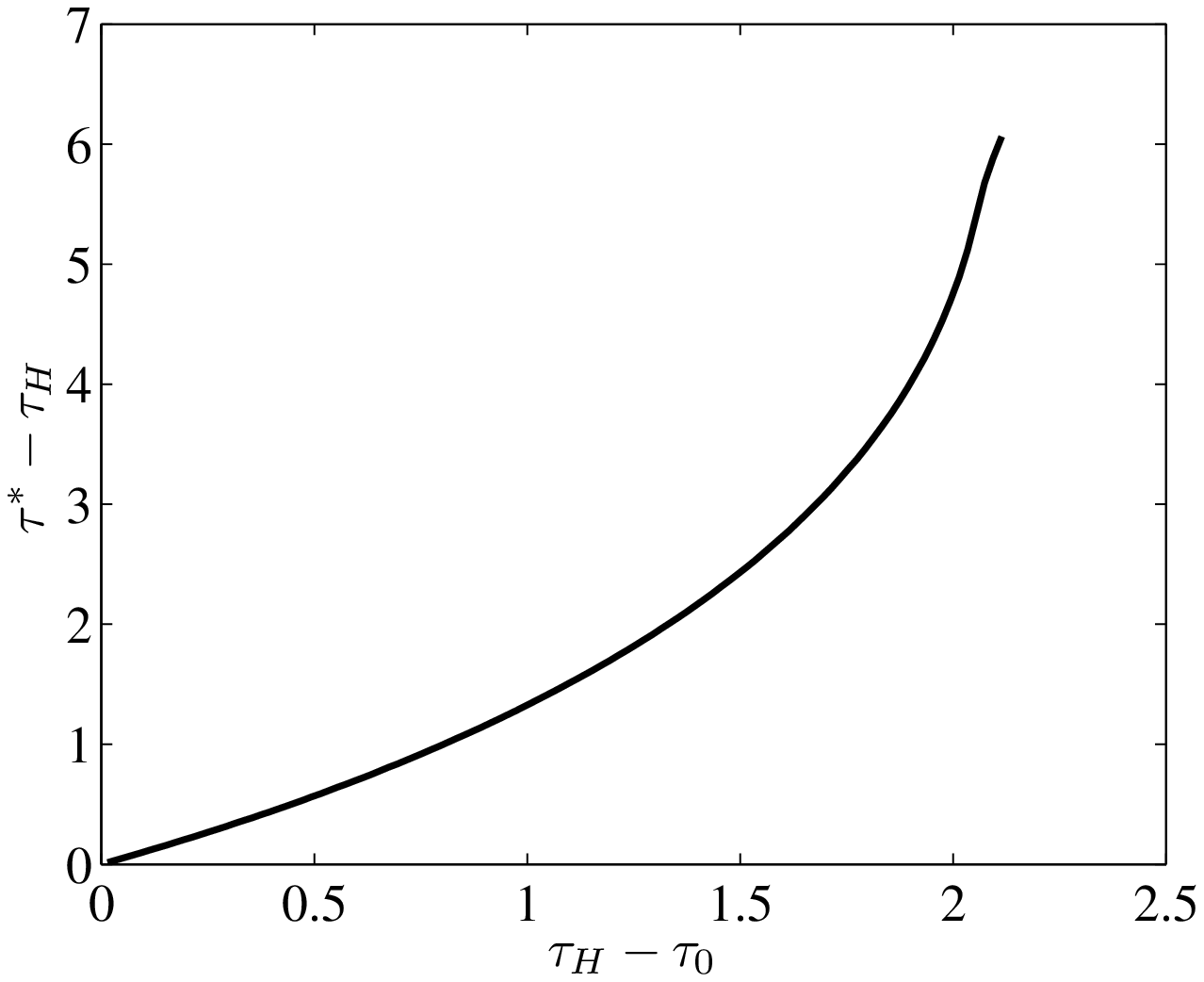}
        }   
    \subfigure[] 
        {\label{delaycomparefig}
        \includegraphics[width=.4\textwidth]{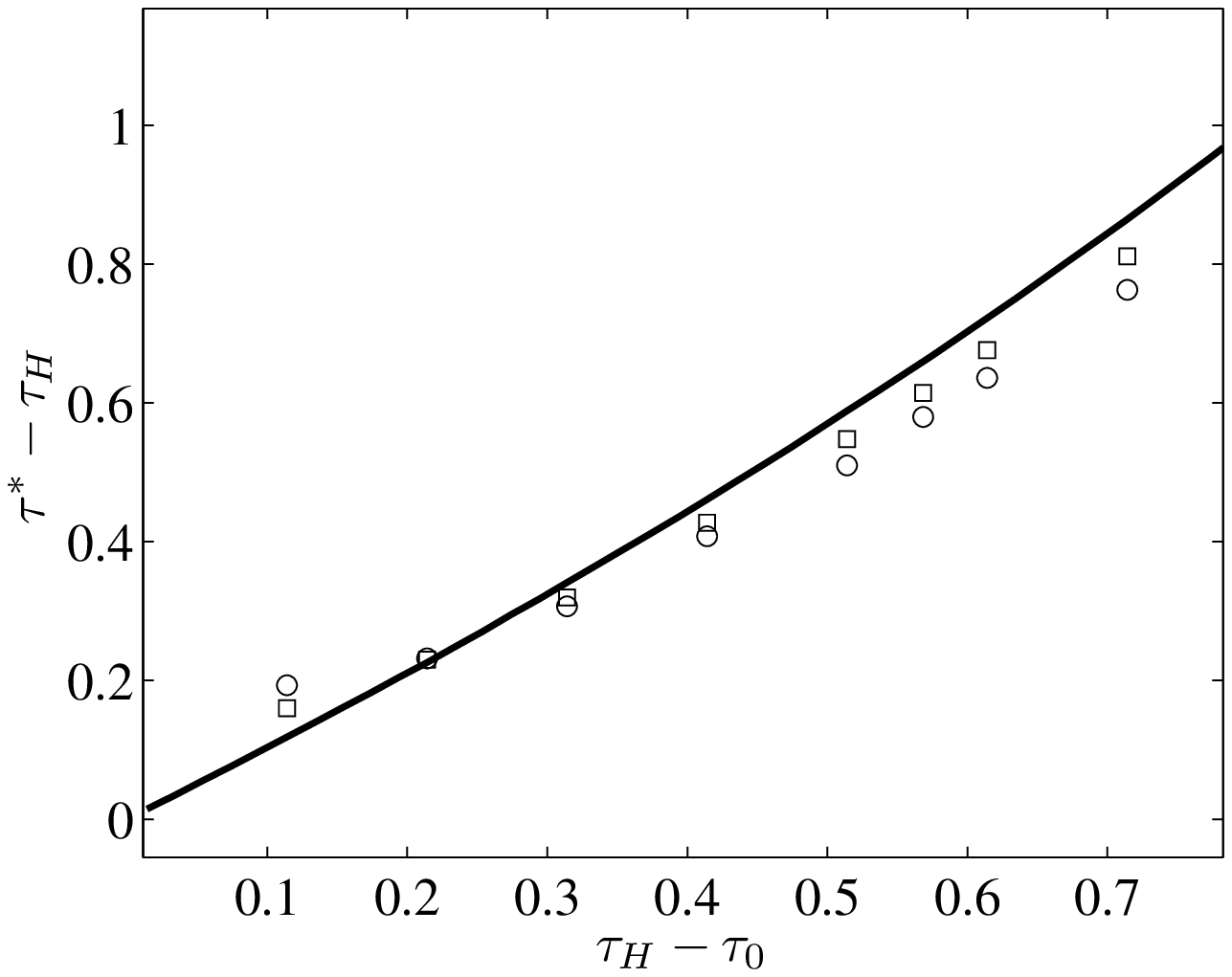}
        }}
\caption{(a) The delay phenomenon obtained by computing the value
$\tau^*$ at which $\Re(\psi)$ changes from negative to positive, for
different values of $\tau_0$. The figure shows that the smaller $\tau_0$
is, the larger $\tau^*$ must be for the Hopf bifurcation to be fully
realized. Here, $\tau_H \approx 2.114$ is the Hopf bifurcation threshold
so that $\tau_H - \tau_0$ is the initial buffer while $\tau^* - \tau_H$
is the distance above threshold.  
(b) Numerical results of delay for $\varepsilon = 0.01$ (circles) and
$\varepsilon = 0.005$ (squares) compared against the asymptotic results
(solid curve) as in (a). The errors for $\varepsilon = 0.005$, as defined in \eqref{error}, for most values of $\tau_0$ are approximately half those for $\varepsilon = 0.01$.}
	\end{center}
\end{figure}

\noI with $\delta$ small. The true equilibrium $(v,u) = (v_e^*, u_e^*)$ was
computed starting from $(v_e, u_e)$ in \eqref{eqbminf} and integrating in
time with fixed $\tau = \tau_0$ until a steady state was reached. In this
way, initial transient oscillations resulting from the error of the leading
order equilibrium solution in \eqref{eqbminf} were removed. To compare
results of numerical computations to the asymptotic results of Figure 
\ref{delay}, we define the oscillation amplitude

\begin{equation}  \label{vmdef}
v_m(\tau(\varepsilon t)) \equiv \frac{v(0,t) - v_e(0)}{v(0,0) - v_e(0)} \,,
\end{equation}

\noI where the denominator in \eqref{vmdef} acts to normalize results over
different values of $\delta$ so that $v_m(\tau_0) = 1$. We found that $%
v_m(\tau)$ behaved rather consistently over a range of values for $\delta$.
According to \eqref{pertpsi}, we define $\tau^*_m$ to be the value of $\tau
> \tau_0$ at which the value of $|v_m(\tau)|$ first exceeds unity. In Figure %
\ref{vm}, we plot a typical case of $v_m(\tau)$ with $\tau_0 = 1.5 < \tau_H$
and $\varepsilon = 0.005$. The vertical dashed line indicates the critical
Hopf bifurcation value $\tau_H$. We found in this instance that $\tau^*_m
\approx 2.75$, while the asymptotic result gives $\tau^* \approx 2.828$.
These two values are indicated by the thick solid and thick dashed lines in Figure %
\ref{vm}, respectively. Defining the percentage error as

\BE \label{error}
\mbox{error} \equiv \frac{(\tau^*-\tau_H) - (\tau^*_m-\tau_H)}{\tau^* - \tau_H} \,,
\EE

\noI we calculate an error of approximately $5.26\%$. Repeating the same run
with double the value of $\varepsilon$ yielded an error of approximately $%
10.86\%$. In most cases, we found the error to approximately double as $%
\varepsilon$ was doubled.

It can be seen in Figure \ref{vm} that the oscillations only become
observable well after $\tau$ has increased past the Hopf bifurcation value $%
\tau_H$. However, with sufficient enlargement as shown in Figure \ref{vmzoom}%
, we find that oscillations decay up until $\tau$ has increased to $\tau_H$,
and then begin to grow thereafter. Since $\tau$ remains in the stable regime
for an extended time, the oscillation amplitude decays to order $%
1\times10^{-6}$ at its smallest value when $\tau = \tau_H$, thereby delaying
the time it takes for it to grow back to its original value.

Repeating the above procedure for various $\tau_0$, we obtain the results
presented in Figure \ref{delaycomparefig}. We observe excellent agreement
between the asymptotic and numerical results over the range of $\tau_0$ for
which we were able to obtain data. Numerical results for larger values of $%
\tau_H - \tau_0$ were generally difficult to obtain, especially for small
values of $\varepsilon$. The reason is that the smaller $\tau_0$ and $%
\varepsilon$ are, the more time the system spends in the stable regime and
so the more time over which the perturbation decays. Once the oscillation
amplitude decays to below machine precision, we observe no ensuing
instabilities even when $\tau$ was increased far past $\tau^*$. In effect,
the system loses the memory of its history accounted for in the asymptotic
analysis, which then would no longer apply.

In this section, we considered a bifurcation triggered by an \textit{%
extrinsic} tuning of the control parameter $\tau$. In contrast, the next
section will consider the triggering of a Hopf bifurcation by dynamics 
\textit{intrinsic} to the system. On a finite domain, we find the
possibility of a non-monotonicity in the Hopf bifurcation threshold, a
feature not present in the example just considered. By carefully setting
initial conditions to induce sufficient delay, we find that this feature
allows a spike to pass safely through a Hopf-unstable zone into a stable
zone with no subsequent instabilities.

\setcounter{equation}{0}

\section{Example 2: Hopf bifurcation of a one-spike solution on a finite
domain}

\label{GMfinline}

In this section, we consider the general GM system on a finite
one-dimensional domain

\bes \label{GMfin} 
\begin{equation}  \label{GMvfin}
v_t = \varepsilon^2 v_{xx} - v + \frac{v^p}{u^q} \,, \qquad -1 < x < 1 \,,
\qquad v_x(\pm 1, t) = 0 \,, \qquad t > 0 \,,
\end{equation}
\begin{equation}  \label{GMufin}
\tau u_t = Du_{xx} - u + \frac{1}{\varepsilon}\frac{v^r}{u^s} \,, \qquad -1<
x < 1 \,, \qquad u_x(\pm 1, t) = 0 \,, \qquad t > 0 \,,
\end{equation}
\ees

\noI where the exponents, $p, q, r, s \geq 0,$ satisfy the relation $%
qr/(p-1) - s - 1 > 0$. To obtain an explicitly solvable NLEP as in Section %
\ref{GMinfline}, we require the additional relation

\begin{equation}  \label{prexplic}
p = 2r-3 \,, \quad r > 2 \,.
\end{equation}

\noI In the previous section, a Hopf bifurcation was triggered by an
extrinsic tuning of the parameter $\tau$. In contrast, the Hopf bifurcation
that we consider in this section is intrinsically triggered by slow spike
dynamics. That is, an initially stable quasi-equilibrium profile centered at 
$x = x_0 >0$ undergoes a slow $\mathcal{O}(\varepsilon^2)$ drift towards its
equilibrium location of $x_0 = 0$ and triggers a Hopf bifurcation before
reaching equilibrium. At the Hopf bifurcation, the associated eigenvalue is
of $\mathcal{O}(1)$ and imaginary. We emphasize that all parameters in %
\eqref{GMfin} remain constant, with only the intrinsic motion of the spike
able to trigger a bifurcation.

Two scenarios are possible. The first is illustrated schematically in Figure %
\ref{tauH_vs_x0_monot} for $(p,q,r,s) = (3, 3, 3, 0)$ and $D = 4$. The black
curve represents the Hopf bifurcation threshold $\tau = \tau_H$ plotted
against the spike location $x_0$. The quasi-equilibrium solution is stable
(unstable) when $\tau$ is below (above) the threshold vale $\tau_H$.
Alternatively, for a given value of $\tau$, the quasi-equilibrium solution
is stable (unstable) when $x_0 > x_H(\tau)$ ($x_0 < x_H(\tau)$). Starting at 
$x_0(0) > x_H$, Figure \ref{tauH_vs_x0_monot} illustrates schematically the
intrinsic triggering of a Hopf bifurcation due to the direction of drift,
indicated by the arrow. As in the case of \S \ref{GMinfline}, oscillations
are expected to decay while $x_0 < x_H$, beginning to grow only when the
spike enters the unstable zone. The amplitude of oscillations when $x_0 =
x_H $ must then be smaller than that of the original perturbation at $x_0 =
x_0(0)$. The delay refers to how far the spike must travel into the unstable
zone before the oscillation amplitude is restored to that of the original
perturbation and the Hopf bifurcation is considered to be fully realized.

For the same exponent set, Figure \ref{tauH_vs_x0_nonmonot} shows an example
of the second scenario where the function $\tau_H(x_0)$ is non-monotonic
when $D = 1$. For a given $\tau$ sufficiently small, there exists two
Hopf-stability thresholds. The first, $x_{Hsu}$, occurs as the spike drifts
from a stable to unstable zone. The second, $x_{Hus}$, occurs as the spike
re-enters a stable region from an unstable region. If the predicted delay is
sufficiently large, the spike may pass ``safely'' through the unstable zone
without the Hopf bifurcation ever being fully realized. Both of these
scenarios are demonstrated numerically in the following section.

In Appendix \ref{appB}, we construct a quasi-equilibrium one-spike solution to 
\eqref{GMfin} and derive an ODE describing the slow drift of the spike
profile. Assuming that the spike location remains frozen with respect to an $%
\mathcal{O}(1)$ time scale, we perform a linear stability analysis to
calculate the Hopf bifurcation threshold $\tau_H(x_0)$, examples of which
are shown in Figure \ref{tauH_vs_x0}. By similar arguments to \S \ref
{analyticdelay}, we obtain a coupled system for the spike location and the
time-dependent eigenvalue $\psi(\varepsilon^2t)$, from which we compute the
asymptotic prediction of delay. As before, we present only the results of
this analysis, and refer the reader to Appendix \ref{appB} for more details.

\subsection{Asymptotic prediction of delay}

\label{GSanalyticdelay}

The one-spike quasi-equilibrium solution to \eqref{GMfin}, with spike
centered at $x = x_0$, is given by

\begin{equation}  \label{uvqe}
v_{qe} = U_0^{q/(p-1)}w(\varepsilon^{-1}(x-x_0)) \,, \qquad u_{qe} = \frac{%
U_0}{G_{00}}G(x;x_0) \,.
\end{equation}

\noI Here, $w(y)$ is the solution of the equation

\begin{equation} \label{weqp}
w^{\prime\prime} - w + w^p = 0 \,, \qquad -\infty < y < \infty \,, \qquad
w(0) > 0 \,, \qquad w^\prime(0) = 0 \,, \qquad w \to 0 \quad \mbox{as} \quad
|y| \to \infty \,,
\end{equation}

\noI given by \cite{doelman2001large}

\begin{equation}  \label{wsolp}
w(y) = \left\lbrace \frac{p+1}{2} \sech^2 \left(\frac{p-1}{2}y \right)
\right\rbrace^{1/(p-1)} \,; \qquad b_r \equiv \intinf w^r \, dy \,.
\end{equation}

\noI In \eqref{uvqe}, $G(x;x_0)$ is given by

\begin{equation} \label{G}
G(x;x_0) = G_{00}\left\{ 
\begin{array}{lr}
\frac{\cosh\left(\theta_0(1+x) \right)}{\cosh\left(\theta_0(1+x_0) \right)}
\,, & x<x_0 \,, \\ 
\frac{\cosh\left(\theta_0(1-x) \right)}{\cosh\left(\theta_0(1-x_0) \right)}
\,, & x>x_0 \,,%
\end{array}
\right.
\end{equation}

\noI while $G_{00}$ and $U_0$ are given by

\begin{equation}  \label{G00}
G_{00} = \frac{1}{\sqrt{D}\left\lbrack \tanh\left(\theta_0(1+x_0) \right) +
\tanh\left(\theta_0(1-x_0) \right) \right\rbrack} \,; \qquad \theta_0 \equiv 
\frac{1}{\sqrt{D}} \,,
\end{equation}

\noI and

\begin{equation}  \label{U0fin}
U_0 = \frac{1}{(b_rG_{00})^{1/M}} \,; \qquad M \equiv \frac{qr}{p-1} - s - 1
\,,
\end{equation}

\noI respectively, where $b_r$ is defined in \eqref{wsolp}.

\begin{figure}[htbp]
  \begin{center}
	\includegraphics[width=.4\textwidth]{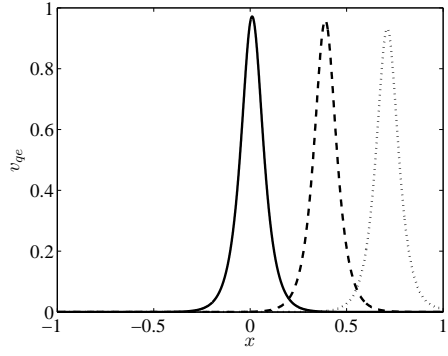}
	\caption{Plots of $v_{qe}$ at various times during its evolution according to \eqref{x0ODE}. The spike increases in height as it drifts toward $x_0 = 0$. The parameters are $(p,q,r,s) = (3,3,3,0)$, $\varepsilon = 0.05$, $D = 4$, and $\tau = 0.01$. By Figure \ref{tauH_vs_x0_monot}, this value of $\tau$ is well below threshold for all $x_0$, and so no oscillations in spike amplitude are present. The times depicted are $t = 0$ (dotted), $t = 400$ (dashed), and $t = 2800$ (solid).}
	\label{vqe}
	\end{center}
\end{figure}

When $x_0 \neq 0$, the spike profile drifts on a slow time scale according
to the equation

\begin{equation}  \label{x0ODE}
\frac{dx_0}{d\sigma} = -\frac{q}{(p-1)\sqrt{D}}\left\lbrack\tanh\left(%
\theta_0(1+x_0)\right) - \tanh\left(\theta_0(1-x_0) \right) \right\rbrack
\equiv F(x_0) \,; \qquad \sigma \equiv \varepsilon^2 t \,,
\end{equation}
\noI where $\theta_0$ is defined as in \eqref{G00}. Note that $F(x_0) < 0$ ($F(x_0) > 0$) when $x_0 > 0$ ($x_0 <0$) with $%
F(0) = 0$ so that the dynamics of the spike are always monotonic toward the
equilibrium point $x = 0$. The corresponding evolution of the spike
amplitude can be obtained from \eqref{uvqe}, \eqref{G00}, and \eqref{U0fin}.
In Figure \ref{vqe}, we show the spike at three different times during its
evolution, beginning at $x_0(0) = 0.7055$. As time increase, the spike
drifts toward the origin while keeping a constant profile, changing only in
height. The parameters are $(p,q,r,s) = (3,3,3,0)$, $D = 4$, and $\tau =
0.01 $. By Figure \ref{tauH_vs_x0_monot}, this value of $\tau$ is well below
threshold for all $0<x_0<1$, and so no oscillations in spike amplitude are
present.

To find the Hopf bifurcation threshold, we perturb the quasi-equilibrium
solution \eqref{uvqe} by

\begin{equation}  \label{pertfin}
v = v_{qe} + e^{\lambda t}\phi \,, \qquad u = u_{qe} + e^{\lambda t} \eta
\,; \qquad \phi,\eta \ll 1 \,.
\end{equation}

\noI Analysis of the resulting linearized equation with $p$ satisfying %
\eqref{prexplic} leads to an explicitly solvable NLEP, from which we obtain
the equation for the eigenvalue $\lambda$

\begin{equation}  \label{lambdafin}
\lambda = \beta - \frac{r}{2}\chi(\lambda, x_0) \,, \qquad \beta \equiv r^2
- 2r > 0 \,,
\end{equation}

\noI where $\chi(\lambda, x_0)$ is given by

\begin{equation*}
\chi = rq \frac{G_{\lambda00}}{G_{00}}\frac{1}{1 + sG_{\lambda00}I_r} \,;
\qquad I_r \equiv U_0^{Rr-s-1} \intinf w^r \, dy \,, \qquad R \equiv \frac{q%
}{p-1} \,.
\end{equation*}

\noI Here, $G_{00}$ is given by \eqref{G00}, while $G_{\lambda 00}$ is
defined as

\begin{equation*}
G_{\lambda00} = \frac{1}{\sqrt{D(1+\tau\lambda)}\left\lbrack
\tanh\left(\theta_\lambda(1+x_0) \right) + \tanh\left(\theta_\lambda(1-x_0)
\right) \right\rbrack} \,; \qquad \theta_\lambda \equiv \theta_0\sqrt{%
1+\tau\lambda} \,,
\end{equation*}

\noI with $\theta_0$ defined in \eqref{G00}. By setting $\lambda =
i\lambda_I$, we may solve the real and imaginary parts of \eqref{lambdafin}
for $\lambda_I \in \mathbb{R}$ and the Hopf bifurcation threshold $\tau_H$
as functions of $x_0$. The relation $\tau_H(x_0)$ for two different values
of $D$ is shown in Figure \ref{tauH_vs_x0}.

To account for the slow dynamics and the dependence of $\lambda$ on $x_0$,
we proceed as in \S \ref{analyticdelay} and replace \eqref{pertfin} with the
WKB ansatz

\begin{equation}  \label{pertfinWKB}
v = v_{qe} + e^{\frac{1}{\varepsilon^2}\psi(\sigma)}\phi \,, \qquad u =
u_{qe} + e^{\frac{1}{\varepsilon^2}\psi(\sigma)} \eta \,, \qquad \sigma
\equiv \varepsilon^2 t \,.
\end{equation}

\noI Substituting \eqref{pertfinWKB} into \eqref{GMfin} and linearizing to
identify the equivalence $\psi^\prime = \lambda$, we obtain for $%
\psi(\sigma) $

\begin{equation}  \label{psiofsigma}
\psi(\sigma) = \int_0^{\sigma} \! \lambda \, d\sigma =
\int_{x_0(0)}^{x_0(\sigma)} \lambda(x_0) \frac{1}{F(x_0)} \, dx_0 \,.
\end{equation}

\noI In \eqref{psiofsigma}, we have taken $\psi(0) = 0$ without loss of
generality, and used \eqref{x0ODE} to change the variable of integration
from $\sigma$ to $x_0$. The delay phenomenon may be understood in the same
manner as in \S \ref{GMinfline}. By setting $x_0(0) > x_H$ in the
Hopf-stable regime so that $\Re(\lambda) < 0$, $\psi(\sigma)$ will be
negative and decreasing until $x_0(\sigma)$ reaches $x_H$. During this time,
the oscillations decay to an $\mathcal{O}(e^{-1/\varepsilon^2})$ amplitude.
The spike will then enter the unstable regime, at which time $\psi(\sigma)$
will begin to increase towards $0$. Assuming the scenario depicted in Figure %
\ref{tauH_vs_x0_monot}, $\psi(\sigma)$ will then reach 0 for some $\sigma =
\sigma^*$ for which $x_0(\sigma^*) = x_0^* < x_H$. We define this as the
time when the Hopf bifurcation is fully realized. That is,

\begin{equation}  \label{x0star}
\int_{x_0(0)}^{x_0(\sigma^*)} \lambda(x_0) \frac{1}{F(x_0)} \, dx_0 = 0.
\end{equation}

\begin{figure}[htbp]
  \begin{center}
	\includegraphics[width=.4\textwidth]{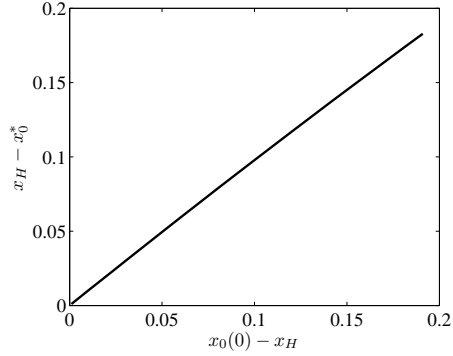}
	\caption{A plot of delay, $x_H - x_0^*$, versus the ``initial buffer,'' $x_0(0) - x_H$, where $x_0(0)$ is the initial location of the spike, $x_H$ is the Hopf bifurcation value, and $x_0^*$ is the spike location at which the oscillation amplitude recovers to the size of the original perturbation. The exponents ($(p,q,r,s) = (3,3,3,0)$) and value of $D$ ($D = 4)$ correspond to the monotonic $\tau_H(x_0)$ depicted in Figure \ref{tauH_vs_x0_monot}, while $\tau$ is set at $0.891$.}
	\label{delayvsx0}
	\end{center}
\end{figure}

\noI Along with \eqref{lambdafin}, \eqref{x0star} constitutes a set of
algebraic equations for $x_0^*$ as a function of $x_0(0)$. As in \S \ref%
{analyticdelay}, the delay in terms of $x_0$ is independent of $\varepsilon$%
. For $(p,q,r,s) = (3,3,3,0),$ we show in Figure \ref{delayvsx0} the
relation between the delay $x_H - x_0^*$ and $x_0(0) - x_H$, the ``initial
buffer,'' or how far into the stable zone the spike is located at $t = 0$.
The increasing function indicates that the larger the initial buffer, the
larger the delay. Qualitatively, the more time the spike remains in the
stable zone, the more its oscillation amplitude decays, and so the more time
it must spend in the unstable zone for the oscillations to recover to their
original amplitude.

For the scenario depicted in Figure \ref{tauH_vs_x0_nonmonot}, initial
conditions may be chosen to induce sufficient delay so that $\psi$ will not
increase past $0$ before it passes through the unstable zone. In this case,
the spike can pass safely through the unstable zone without the Hopf
bifurcation ever being fully realized. In the following section, we present
numerical examples of both scenarios. Due to the sensitive nature of the
numerical computations, we compare the numerical results to asymptotic
results only for the case where $\tau_H(x_0)$ is monotonic. Numerical
results for the non-monotonic case serve only to illustrate the qualitative
aspect of the theory.

\subsection{Numerical validation}

We illustrate the theory by numerically solving \eqref{GMfin} for two
exponent sets $(p,q,r,s) = (3, 2, 3, 0)$ and $(p,q,r,s) = (3, 3, 3, 0)$. The
time integration was performed using the \texttt{MATLAB} \texttt{pdepe()}
routine. The initial conditions were taken as a perturbation of a ``true
quasi-equilibrium'' state $(v(x,0), u(x,0))$ = $(v_{eq}^*(x), u_{eq}^*(x))$,
similar to that of \eqref{ICinf}. To obtain $(v_{eq}^*(x), u_{eq}^*(x))$, we
started from initial conditions $(v_{eq}, u_{eq})$, the asymptotic result
given in \eqref{uvqe}, and integrated in time to allow for transient effects
to decay. The spike location in $(v_{eq}, u_{eq})$ was set so that, after
the initial integration, $(v_{eq}^*(x), u_{eq}^*(x))$ had the desired spike
location. All values for the initial spike locations stated below are
reflected in $(v_{eq}^*(x), u_{eq}^*(x))$. We first present results for the
scenario in Figure \ref{tauH_vs_x0_monot}, where $\tau_H(x_0)$ is monotonic.

The results below for $\varepsilon = 0.007$ were obtained with $2000$ grid
points, while those for $\varepsilon = 0.005$ were obtained with $3000$ grid
points. Unlike the static problem of Section \eqref{GMinfline}, we found
that this problem displayed sensitivity to the number of grid points used.
In particular, we found that decreasing mesh size tended to trigger the Hopf
bifurcation earlier than expected. We conjecture this may be due to rounding
errors associated with a large number of grid points. Further, while the
asymptotic results become more accurate as $\varepsilon$ is decreased, we
found that small $\varepsilon$ caused spike oscillations to decay so much
that the grid was unable to resolve the oscillations as the spike moved from
one grid location to the next. To compensate for small $\varepsilon$, we set
initial spike locations close to threshold so that oscillations remained of
sufficient amplitude when the spike reached threshold.

\begin{figure}[htbp]
  \begin{center}
    \mbox{
    \subfigure[$x_0$ versus $t$] 
        {\label{x0dyn}
        \includegraphics[width=.4\textwidth]{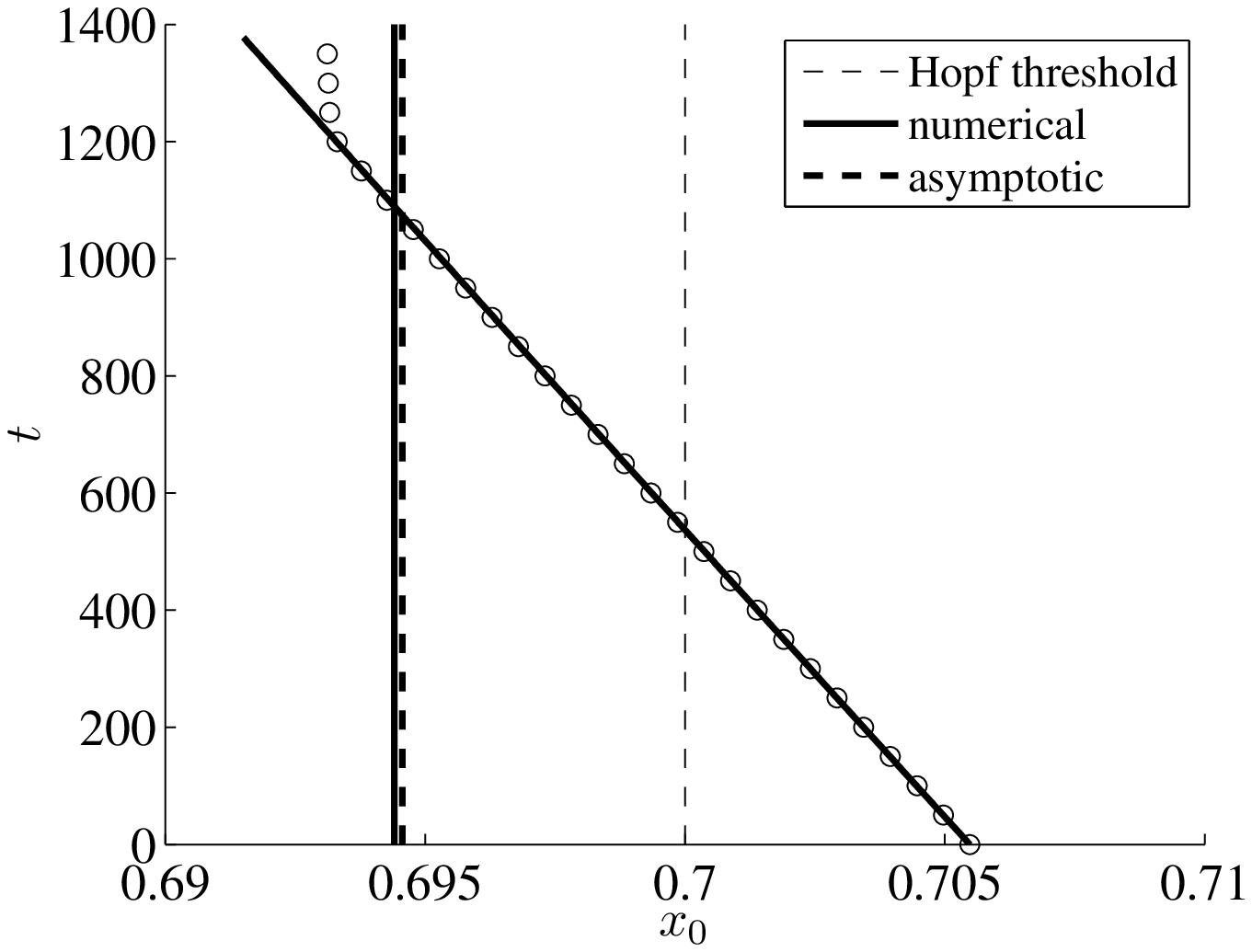}
        }   
    \subfigure[amplitude of oscillation versus $t$] 
        {\label{oscamp}
        \includegraphics[width=.4\textwidth]{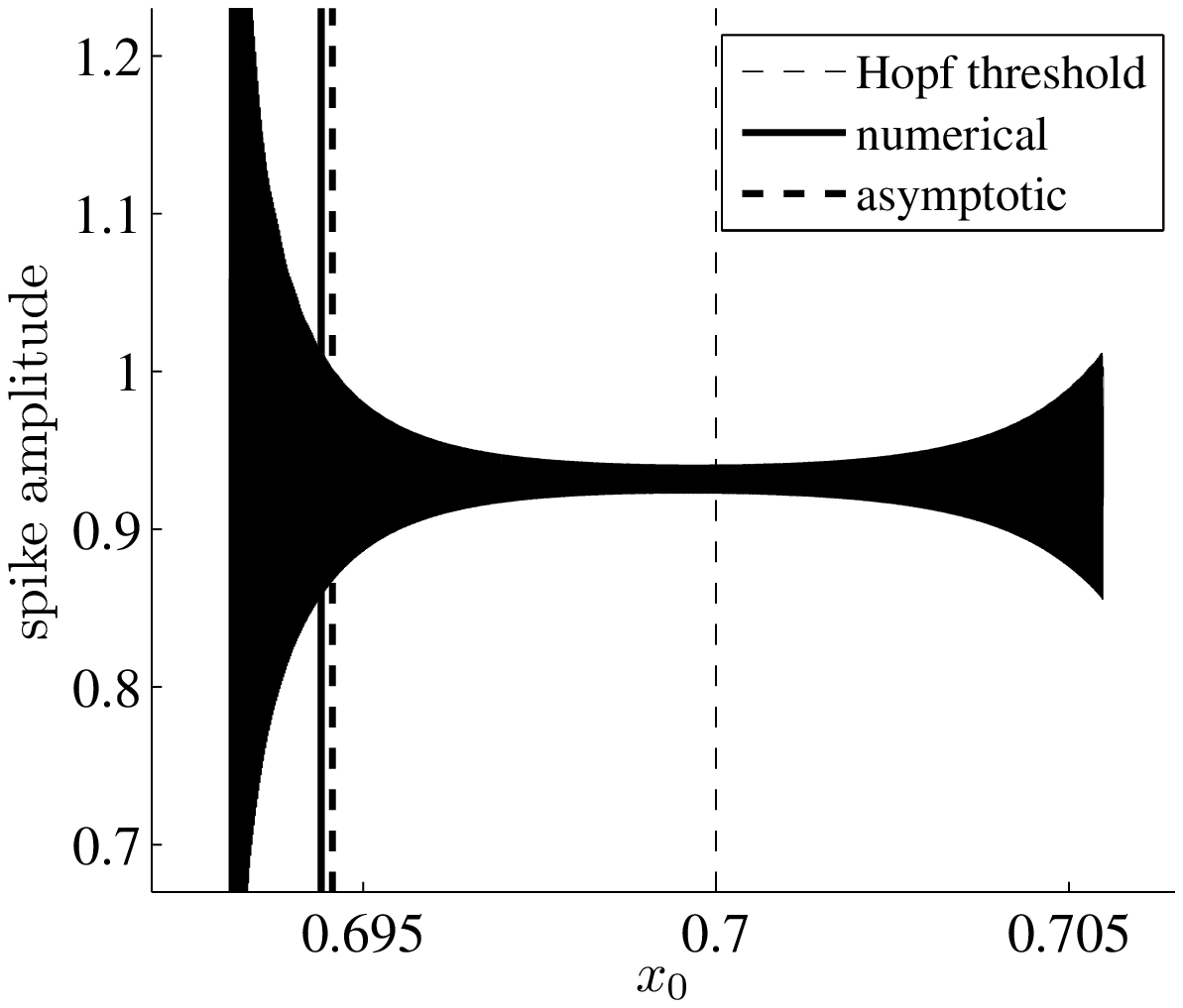}
        }}
    \caption{In the left figure, the curve represents the asymptotic prediction for the spike location given by \eqref{x0ODE}, while the circles were obtained by numerically solving the PDE system \eqref{GMfin}. The deviation beginning near $t = 1200$ is due to oscillations growing beyond the asymptotic regime. Note that the deviation occurs well after the Hopf bifurcation (vertical dashed line). In the right figure, we show the corresponding oscillations in spike amplitude. The initial decay in the amplitude reflects the initial stability of the solution. Near $x_0 = 0.7$ (vertical dashed line), a Hopf bifurcation occurs, at which point the spike oscillations begin to grow. When $x_0 \approx 0.6944$ (heavy solid line), the oscillations grow to their original size. The asymptotic prediction is $x_0^* \approx 0.6946$ (heavy dashed line). Here, $(p, q, r, s) = (3, 3, 3,0)$, $\varepsilon = 0.005$, $D = 4$, and $\tau = 0.891$.} 
     \label{dyn}
  \end{center}
\end{figure}

A typical numerical result is shown in Figure \ref{dyn}. In Figure \ref%
{x0dyn} we compare the asymptotic result for spike location \eqref{x0ODE}
(black curve) to that found by numerically solving the PDE system %
\eqref{GMfin} (circles) with $(p, q, r, s) = (3, 3, 3,0)$, $\varepsilon =
0.005$, $D = 4$, and $\tau = 0.891$. Beginning at $x_0(0) = 0.7055$, the
spike drifts toward $x_0 = 0$ on an $\mathcal{O}(\varepsilon^2)$ time scale.
We observe excellent agreement until $x_0 \approx 0.6932$, at which point
the oscillations grow beyond the asymptotic regime. Note that the asymptotic
prediction for the spike location remains valid well after the Hopf
bifurcation takes place (vertical dashed line).

For the parameters of the simulation, a Hopf bifurcation occurs at
approximately $x_H = 0.7$. Figure \ref{oscamp} shows that the amplitude of
oscillations decays from the original size of the perturbation when $x_0 >
x_H$, reaching a minimum at $x_0 \approx x_H$. Once $x_0$ crosses into the
unstable regime $x_0 < x_H$, the amplitude begins to grow. However, the
Hopf-bifurcation is not fully realized until $x_0 \approx 0.6944 < x_H$ (heavy solid line), when the oscillations returns to their original amplitude.
By solving \eqref{x0star} along with \eqref{lambdafin}, we find that $x_0^*\approx 0.6946$ (heavy dashed line), indicating good agreement between
asymptotic and numerical results. The oscillations occur on an
asymptotically shorter time scale compared to that over which they drift,
and are thus not visible in Figure \ref{oscamp}.

\begin{figure}[htbp]
  \begin{center}
	\includegraphics[width=.4\textwidth]{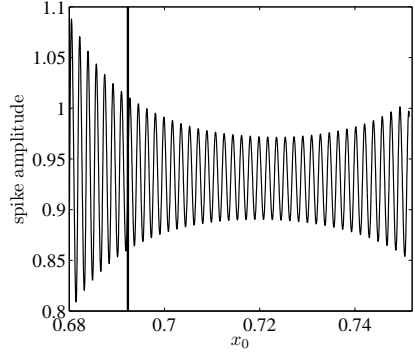}
	\caption{The same parameters as in Figure \ref{dyn} except with $\varepsilon = 0.05$. Due to larger $\varepsilon$, the Hopf bifurcation is triggered at $x_0 \approx 0.72$ before the spike reaches the theoretical threshold of $x_H \approx 0.7$. The Hopf bifurcation is then fully realized at $x_0 \approx 0.6923 < 0.72$ (heavy vertical line) when the oscillations return to their original amplitude. The asymptotic result of $x_0^* = 0.6492$ is not shown.}
	\label{oscamp_largeeps}
	\end{center}
\end{figure}

In Figure \ref{oscamp_largeeps}, we show a case with the same parameters
except with $\varepsilon = 0.05$. The time scale of the drift is much faster
in this case so that individual oscillations are visible. Further, the
starting point may be set farther in the stable regime ($x_0(0) = 0.75$)
without danger of the oscillation amplitude becoming too small at a later time. However, the
Hopf-bifurcation threshold is not as sharp due to larger $\varepsilon$,
causing oscillations to begin growing at $x_0 \approx 0.72$ instead of at $%
x_0 \approx 0.7$ as in Figure \ref{oscamp} for smaller $\varepsilon$. As
such, the predicted value of $x_0^* \approx 0.6492$ is rather far from the
numerical value of $0.6923$ (heavy solid). The delay in bifurcation is still evident, as
the spike must move well past the (numerical) bifurcation point before the
bifurcation is fully realized. This illustration shows the difficulty in
balancing the small $\varepsilon$ required for asymptotic accuracy and the
larger $\varepsilon$ required for numerical workability.

\begin{figure}[htbp]
  \begin{center}
    \mbox{
    \subfigure[$(p, q, r, s) = (3, 3, 3,0)$] 
        {\label{epscompilationQ3}
        \includegraphics[width=.4\textwidth]{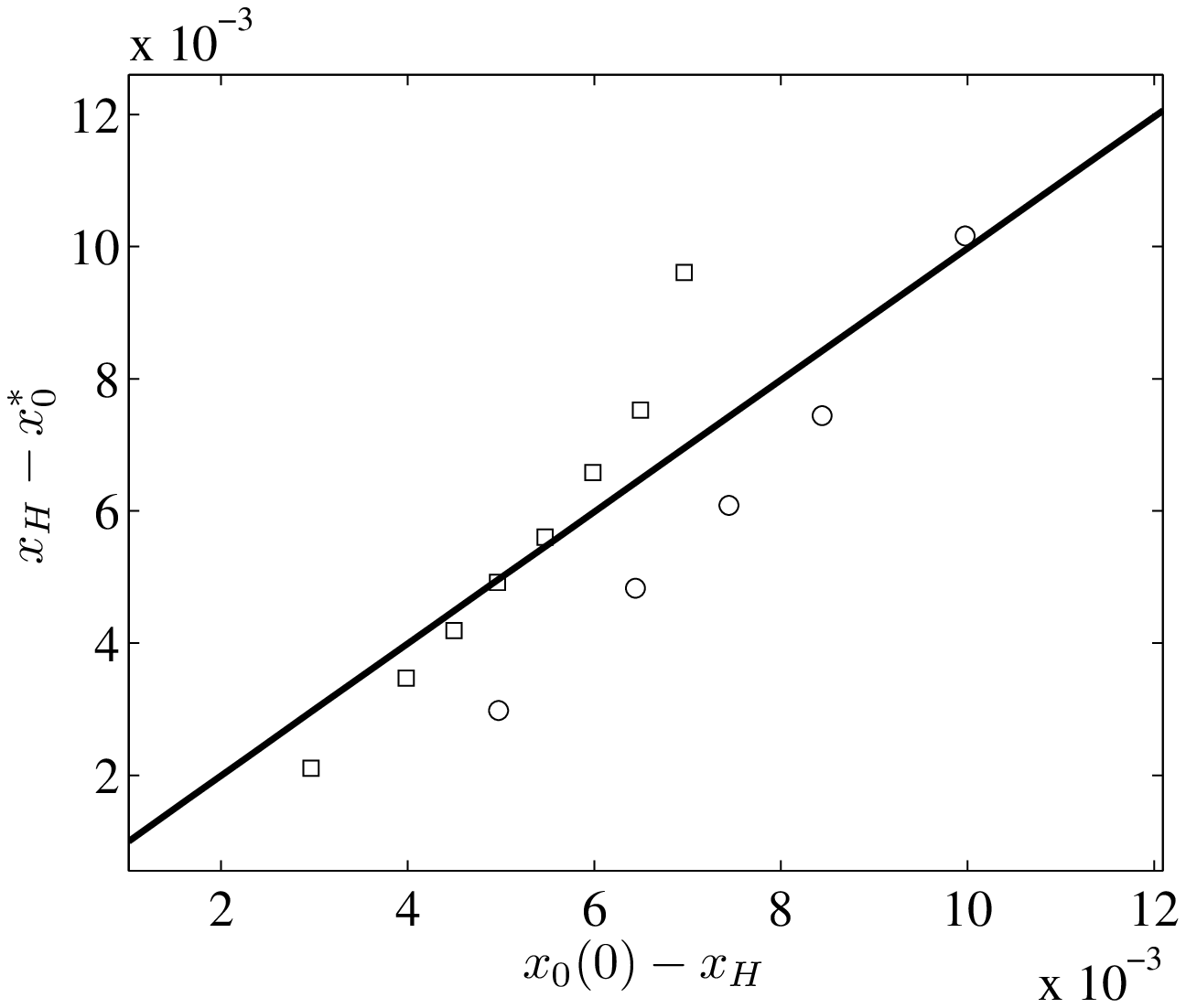}
        }   
    \subfigure[$(p,q,r,s) = (3, 2, 3, 0)$] 
        {\label{epscompilationQ2}
        \includegraphics[width=.4\textwidth]{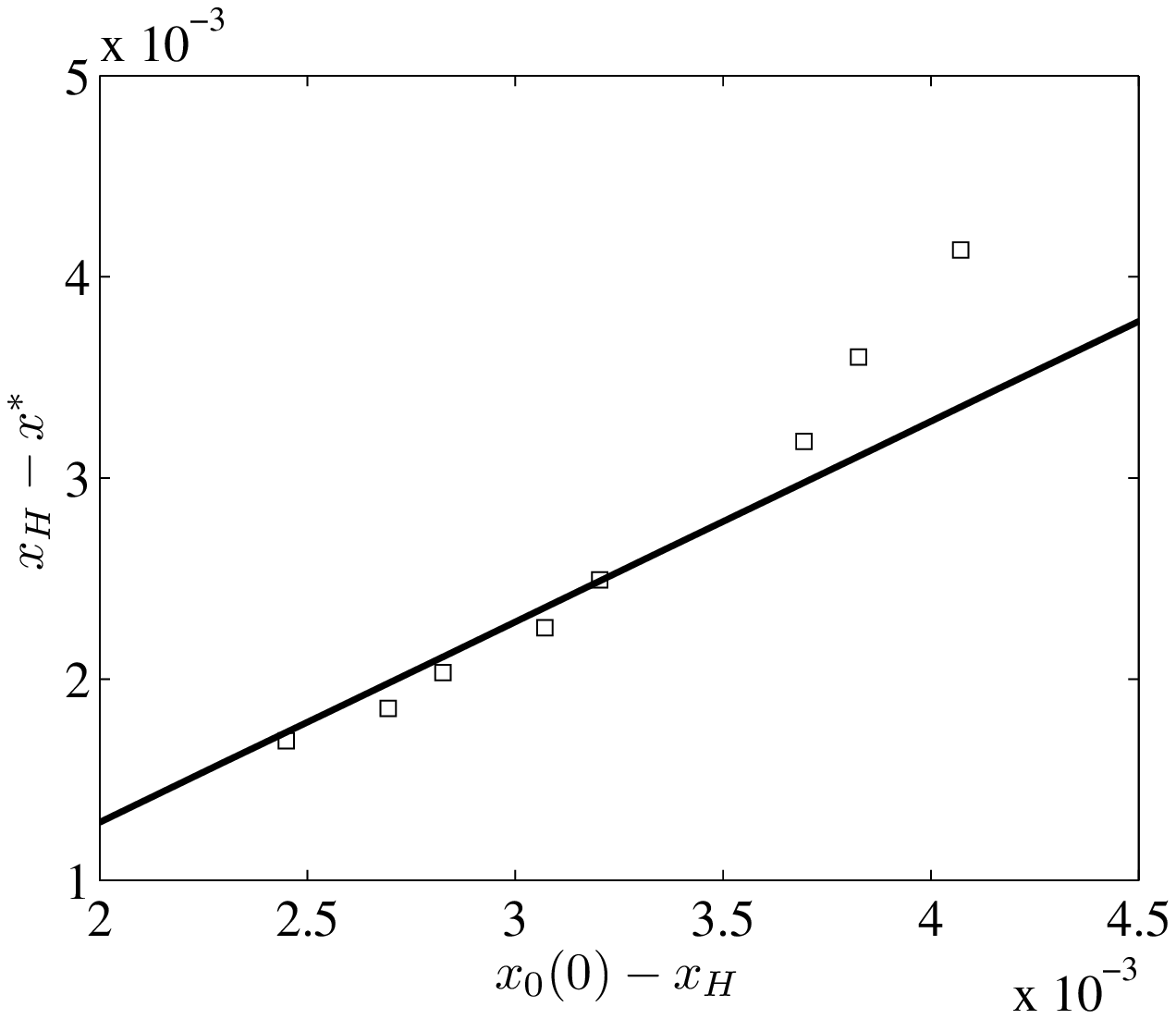}
        }}
    \caption{In the left figure, we compare the asymptotic prediction (solid line) for the delay with the numerical results. The circles (squares) indicate results for $\varepsilon = 0.007$ ($\varepsilon = 0.005$). The Hopf bifurcation occurs when $x_0 = x_H \approx 0.7$. The spike is stable (unstable) when $x_0 > x_H$ ($x_0 < x_H$). Oscillations return to their initial amplitude when $x_0 = x_0^* < x_H$. While results for both values of $\varepsilon$ follow the trend of the asymptotic result, the results for $\varepsilon = 0.005$ show better agreement. Here, $(p, q, r, s) = (3, 3, 3,0)$, $D = 4$, and $\tau = 0.891$. In the right figure, we show similar results for $(p,q,r,s) = (3, 2, 3, 0)$, $\varepsilon = 0.005$, $D = 4$, and $\tau = 0.5$.} 
  \end{center}
\end{figure}

In Figure \ref{epscompilationQ3}, we compile results for $\varepsilon =
0.007 $ (circles) and $\varepsilon = 0.005$ (squares) for various starting
locations $x_0(0)$. The curve represents the asymptotic result show in
Figure \ref{delayvsx0}. We observe good agreement between asymptotic and
numerical results, with the results for $\varepsilon = 0.005$ appearing to
yield closer agreement. In Figure \ref{epscompilationQ2}, we show similar
results for $(p,q,r,s) = (3, 2, 3, 0)$ and $\varepsilon = 0.005$. Because
the character of oscillations at the beginning appeared slightly different
from that of Figure \ref{oscamp}, we defined the numerical result for $x_0^*$
in a slightly different manner. However, the delay effect, illustrated by
the increasing relation between $x_H - x^*$ and $x_0(0) - x_H$, is still
evident and agreeable with asymptotic results.

Finally, we give an example of a scenario where $\tau_H(x_0)$ is
non-monotonic, as in Figure \ref{tauH_vs_x0_nonmonot}. Qualitatively, the
theory suggests that the larger $x_0(0) - x_H$ is, the farther into the
unstable zone the spike can penetrate before the Hopf bifurcation is fully
realized. Figure \ref{tauH_vs_x0_nonmonot} shows that, for appropriate $\tau$
and $x_0(0)$ sufficiently large, it is possible for $\psi$ never to reach 0
in the unstable zone. In such a case, no solution for $x_0^*$ of %
\eqref{x0star} would exist. That is, if the spike starts far enough into the
stable zone to the right of $x_{Hsu}$, it may pass safely through the
unstable zone $x_0 \in (x_{Hus}, x_{Hsu})$ without the Hopf bifurcation ever
being fully realized.

\begin{figure}[htbp]
  \begin{center}
	\includegraphics[width=.4\textwidth]{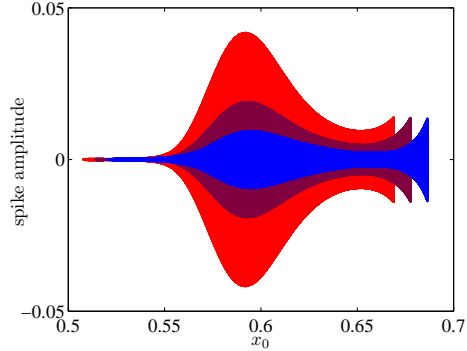}
	\caption{Plots of spike amplitude versus $x_0$ with $(p,q,r,s) = (3, 3, 3, 0)$, $\varepsilon = 0.005$, $D = 1$, and $\tau = 1.245$. The three colors differ only in the starting location $x_0(0)$. In the red plot, starting closest to the bifurcation threshold, oscillations grow in the unstable regime to beyond their original value. In this case, the Hopf bifurcation is fully realized before the spike passes through the unstable zone. The purple plot shows that starting farther into the stable zone reduces the maximum oscillation amplitude attained in the unstable zone. Starting even farther back, the blue plot shows that the spike may pass safely through the unstable zone without the oscillation amplitude ever reaching its original value.}
	\label{safepassage}
	\end{center}
\end{figure}

The theory is illustrated in Figure \ref{safepassage} for $(p,q,r,s) = (3,
3, 3, 0)$, $\varepsilon = 0.005$, $D = 1$, and $\tau = 1.245$. The three
colors differ only in the starting location $x_0(0)$. In the red plot,
starting closest to the bifurcation threshold, oscillations initially decay
while the spike is in the stable regime. Upon crossing $x_{Hsu}$ into the
unstable regime, the oscillations grow to beyond their original value. In
this case, the Hopf bifurcation has been fully realized before the spike has
passed through the unstable zone. Upon crossing $x_{Hus}$ into the stable
regime, the oscillations then decay. The purple plot shows that starting
farther into the stable zone reduces the maximum oscillation amplitude
attained in the unstable zone. However, the amplitude still exceeds its
original value while in the unstable zone. The blue plot shows that starting
sufficiently far in the stable regime allows the spike to pass safely
through the unstable zone without the Hopf bifurcation being fully realized.
This behavior may be explained by noting in Figure \ref{safepassage} that
the farther into the stable regime the spike is initially set, the more the
oscillation amplitude has decayed by the time the Hopf bifurcation is
triggered, thus requiring more time in the unstable zone to recover to its
original value. We have shown in this scenario that the phenomenon of delay
makes it possible to pass safely through an unstable regime into a stable
zone.

In the next section, we consider the delay of a monotonic competition
instability of a two boundary spike equilibrium solution in a generalized
Gray-Scott model. Unlike the GM model, the Gray-Scott model exhibits a
saddle node structure associated with weak dynamics just beyond the saddle.
Analogous to the second scenario just considered, by introducing sufficient
delay into the system through careful choice of initial conditions, we find
that the weak saddle node dynamics may dominate the dynamics of the more
dominant competition instability.

\setcounter{equation}{0}

\section{Example 3: Competition instability of a two boundary spike solution}

\label{GScomp}

For this example, we consider a two boundary spike solution of the
generalized Gray-Scott (GS) model

\bes \label{GS} 
\begin{equation}  \label{GSv}
v_t = \varepsilon^2 v_{xx} - v + Auv^3 \,, \qquad -1 < x < 1 \,, \qquad
v_x(\pm 1, t) = 0 \,, \qquad t > 0 \,,
\end{equation}
\begin{equation}  \label{GSu}
\tau u_t = Du_{xx} + (1-u) + \frac{1}{\varepsilon}uv^3 \,, \qquad -1< x < 1
\,, \qquad u_x(\pm 1, t) = 0 \,, \qquad t > 0 \,.
\end{equation}
\ees

\noI As in the previous examples, the diffusivity $\varepsilon^2$ of the
activator component ($v$) is asymptotically small compared to the
diffusivity $D$ of the inhibitor component ($u$). The $uv^3$ nonlinearity
replaces the usual $uv^2$ term, and leads to an explicitly solvable NLEP. In
this rescaled form of the GS model, the parameter $A$ is referred to as the
feed-rate parameter, as it is a measure of how strongly the inhibitor is fed
into the system from an external reservoir. In the context of solutions
characterized by spikes in the activator component, if the feed-rate $A$ is
too small, the process that fuels the activator spikes becomes insufficient,
and one or more spikes collapse monotonically in time. In Figure \ref%
{boundaryspikes}, we show a two boundary spike equilibrium solution of %
\eqref{GS} for $v(x)$ (solid) and $u(x)$ (dashed, and scaled by a factor of $%
6$ to facilitate plotting). The two spikes are of equal amplitude, are
stable to slow drift instabilities, and remain centered at $x = \pm 1$ for
all time. Figure \ref{amp_vs_t} plots their amplitudes as $A$ is decreased
past a stability threshold at which the feed-rate becomes insufficient to
support two spikes. Note that the collapse of the left spike (light solid)
is monotonic in time.

This type of instability, referred to as a competition instability due the
local conservation of spike amplitudes at onset, occurs when a single
eigenvalue crosses into the right-half plane through the origin. This is in
contrast to the Hopf-bifurcations studied in the previous sections, where
two complex conjugate eigenvalues crossed through the imaginary axis,
leading to an oscillatory instability. As $A$ is decreased sufficiently past
the competition threshold $A_-$, the solution encounters a saddle node
bifurcation at $A = A_m < A_-$, past which point the two boundary spike
solution ceases to exist. An example of the saddle node structure is shown
in Figure \ref{bifdiag}. On the upper branch, the heavy solid segment
indicates stable solutions. The light solid segment indicates solutions
unstable to the competition mode. The stability transition occurs at $A =
A_- \approx 4.6351$, while the saddle node occurs at $A = A_m \approx 4.6206$%
. The arrow indicates the evolution of the spike amplitude as $A$ is
decreased. The lower branch is always unstable, and will not be considered.

As in the previous two sections, because $A$ starts in the stable regime $A
> A_-$, a delay is expected to occur such that the competition instability
is fully realized only when $A$ has been decreased sufficiently past $A_-$
to $A = A^* < A_-$. This gives rise to the two scenarios, $A^* > A_m$ and $%
A^* < A_m$. In the first scenario, the instability fully sets in before the
system reaches the saddle node so that the solution has been driven
relatively far from equilibrium by the instability. In the second scenario,
the instability does not fully set in, leaving the solution still very close
to equilibrium when it reaches the saddle node. These two scenarios differ
markedly in their response to amplitude perturbations slightly past the
saddle node. We illustrate both of these scenarios numerically in later
sections. We note that, since no solution exists below $A = A_m$, the
statement $A^* < A_m$ only serves to state that the instability is not
expected to set in before the system reaches the saddle node. No
quantitative predictions of delay can be made in this case.

In what follows, we take $A$ to be the bifurcation parameter, and study the
delay that occurs as it is slowly decreased through the competition
threshold. The parameters $D$ and $\tau$ remain constant. In the analysis, $\tau$ is set to $0$ while in the
numerical computations of \S \ref{GSnumval}, $\tau$ is taken to be a value
much smaller than one. We begin by first stating the two boundary spike
solution and deriving values for $A_m$, $A_-$, and the expected delay $A_- -
A_*$. As in the previous sections, we present only key steps of the
analysis. Full derivations may be found in Appendix \ref{appC}.

\subsection{Two boundary spike equilibrium and prediction of delay}

\label{GSpred}

For constant $A$, the two boundary spike equilibrium solution of \eqref{GS}
is

\begin{equation}  \label{GSeqbm}
v_e \sim \frac{1}{\sqrt{AU_-}} w\left(\varepsilon^{-1}(x+1) \right) + \frac{1%
}{\sqrt{AU_-}} w\left(\varepsilon^{-1}(x-1) \right) \,, \qquad u_e \sim 1 - 
\frac{b}{A^{3/2}U_-^{1/2}}G(x) \,,
\end{equation}

\noI where $w(y)$ and $G(x)$ are given by

\begin{equation*}
w(y) = \sqrt{2}\sech y \,; \qquad \intinf w^3 \, dy \equiv b = \pi\sqrt{2}%
\,, \qquad G(x) = \frac{(\theta_0/2)\cosh(\theta_0 x)}{\sinh \theta_0} \,.
\end{equation*}

\noI In \eqref{GSeqbm}, $0 < U_- < 1/3$ is the smaller solution of the
equation

\begin{equation}  \label{H}
H(U) \equiv \sqrt{U}(1-U) = \frac{b}{A^{3/2}}G(0) \,.
\end{equation}

\noI The upper branch in Figure \ref{bifdiag} is a plot of the spike
amplitude $\sqrt{2/(AU_-)}$ as a function of $A$, while the bottom is a plot
of $\sqrt{2/(AU_+)}$, where $1/3 < U_+ < 1$ is the larger solution of %
\eqref{H}. To compute the value of $A$ at the saddle point, we note that $%
H(U)$ in \eqref{H} has a global maximum at $U=1/3$ where $H(1/3) = 2/(3\sqrt{%
3})$. For a solution to \eqref{H} to exist, $A$ must satisfy $A > A_m$,
where $A_m$ is the value at the saddle given by

\begin{equation}  \label{Am}
A_m = \left\lbrack \frac{3\sqrt{3}\,b\,G(0)}{2}\right\rbrack^{2/3} \,.
\end{equation}

\noI Here, $G(x)$ is defined in \eqref{GSeqbm}.

To determine the stability of \eqref{GSeqbm} for constant $A$, we perturb
the equilibrium by

\begin{equation}  \label{GSpert}
v = v_e + e^{\lambda t}\phi \,, \qquad u = u_e + e^{\lambda t} \eta \,;
\qquad \phi,\eta \ll 1 \,.
\end{equation}

\noI With $\tau = 0$, two modes of instability are possible
corresponding to odd and even eigenfunction $\phi$. The odd
competition mode satisfies $\phi(x) = -\phi(-x)$ and $\eta(x) = -\eta(-x)$.  As described above, the
competition instability leads to the growth of one spike at the
expense of the collapse of the other. The even mode, referred to as
the synchronous mode, satisfies $\phi\,^\prime(0) = \eta\,^\prime(0) =
0$ with $\phi(x) = \phi(-x)$ and $\eta(x) = \eta(-x)$. The synchronous
mode leads to the simultaneous collapse of both spikes. In Appendix
\ref{appC}, we show that the lower branch is always unstable to both
modes of instability, while the upper branch is always stable to the
synchronous mode. We now obtain the condition for which the upper
branch is stable to the competition mode.

For $\tau = 0$, we obtain from the explicitly solvable NLEP

\begin{equation}  \label{GSlambda}
\lambda = 3 - \frac{9}{2\left\lbrack 1 + \frac{U^{3/2}}{H(U)}\coth^2\theta_0
\right\rbrack} \,.
\end{equation}

\noI The condition $\lambda = 0$ yields that, at the competition instability
threshold,

\begin{equation}  \label{Ueminus}
U_- = 1 - \frac{1}{1 + C} \, \equiv \, U_{e-} \,; \qquad C \equiv \frac{1}{%
2\coth^2\theta_0} \,,
\end{equation}

\noI with $\lambda < 0$ ($\lambda > 0$) when $U_- < U_{e-}$ ($U_- > U_{e-}$%
). We note that, with $C < 1/2$ for all $\theta_0 > 0$, we have that $0 <
U_{e-} < 1/3$, corresponding to a solution on the upper branch of Figure \ref%
{bifdiag}. The lower branch is thus always unstable to the competition mode. As $D \to 0$, $\theta_0 = 1/\sqrt{D} \to \infty$ so that $U_{e-}
\to 1/3$. Thus, on an infinitely long domain, the entire upper branch is
always stable to both modes of instability. The stability to the competition
mode on an infinite domain may be interpreted as the lack of a ``crowding
out'' effect between the spikes. That is, the larger the domain size (or
similarly, the smaller the value of $D$), the weaker is the interaction
between the spikes, and the greater the number of spikes that may co-exist.
For this reason, the competition instability is sometimes referred to as an
``overcrowding'' instability. With $A_m$ defined in \eqref{Am}, we have from %
\eqref{H} that the value of $A$ at the competition threshold is given by

\begin{equation}  \label{Aminus}
A_- = A_m \left\lbrack \frac{3\sqrt{3}H(U_{e-})}{2} \right\rbrack^{-2/3} \,,
\end{equation}

\noI with $U_{e-}$ given in \eqref{Ueminus}. As the bifurcation diagram in
Figure \ref{bifdiag} suggests, $\lambda < 0$ ($\lambda > 0$) when $A > A_-$ (%
$A < A_-$).

We note that, had we considered the case of two interior spikes for $v_e$
and $u_e$, the spectrum of the linearized equation for $\phi$ and $\eta$
would also contain small eigenvalues of $\mathcal{O}(\varepsilon^2)$. The
largest of these eigenvalues is associated with a slow drift instability,
with corresponding eigenfunctions $\phi$ and $\eta$ being locally odd about
the center of the spikes. It can be shown that the drift instability
threshold occurs at a larger value of $A$ than does the competition
threshold. As $A$ decreases past $A_-$, it must then first trigger the drift
instability. By considering spikes located at the two boundaries where we
impose pure Neumann conditions, drift instabilities are eliminated. Doing so
made the numerical validations significantly less difficult.

To calculate the delay that results from slowly decreasing $A$ past $A_-$
according to,

\begin{equation}  \label{Aramp}
A = A_0 - \xi \,; \qquad A_0 > A_- \,, \qquad \xi = \varepsilon t \,,
\end{equation}

\noI we replace \eqref{GSpert} by the WKB ansatz

\begin{equation*}
v = v_e + e^{\frac{1}{\varepsilon} \psi(\xi)}\phi \,, \qquad u = u_e + e^{%
\frac{1}{\varepsilon} \psi(\xi)} \eta \,; \qquad \phi,\eta \ll 1 \,.
\end{equation*}

\noI As in the previous two examples, we draw the equivalence $\psi^\prime =
\lambda$, from which we obtain

\begin{equation}  \label{psiA}
\psi(A) = \int_0^\xi \! \lambda \, d\xi = -\int_{A_0}^{A} \! \lambda(A) \,
dA \,,
\end{equation}

\noI where we set $\psi(0) = 0$ and have used \eqref{Aramp} to change the
variable of integration from $\xi$ to $A$. In \eqref{psiA}, $\lambda(A)$ may
be obtained by explicitly solving \eqref{H} for $U(A)$, and using $U(A)$ in %
\eqref{GSlambda}. Since $\lambda < 0$ when $A > A_-$, $\psi$ will be
negative and decreasing until $A$ is decreased to $A_-$. At $A = A_-$, $\psi$
will begin to increase, reaching $0$ only when $A = A^* < A_-$. We define $%
A^*$ as the value of $A$ at which the competition instability has fully set
in.

\begin{figure}[htbp]
  \begin{center}
    \mbox{
    \subfigure[$A_- - A^*$ versus $A_0 - A_-$] 
        {\label{GSdelayD3}
        \includegraphics[width=.4\textwidth]{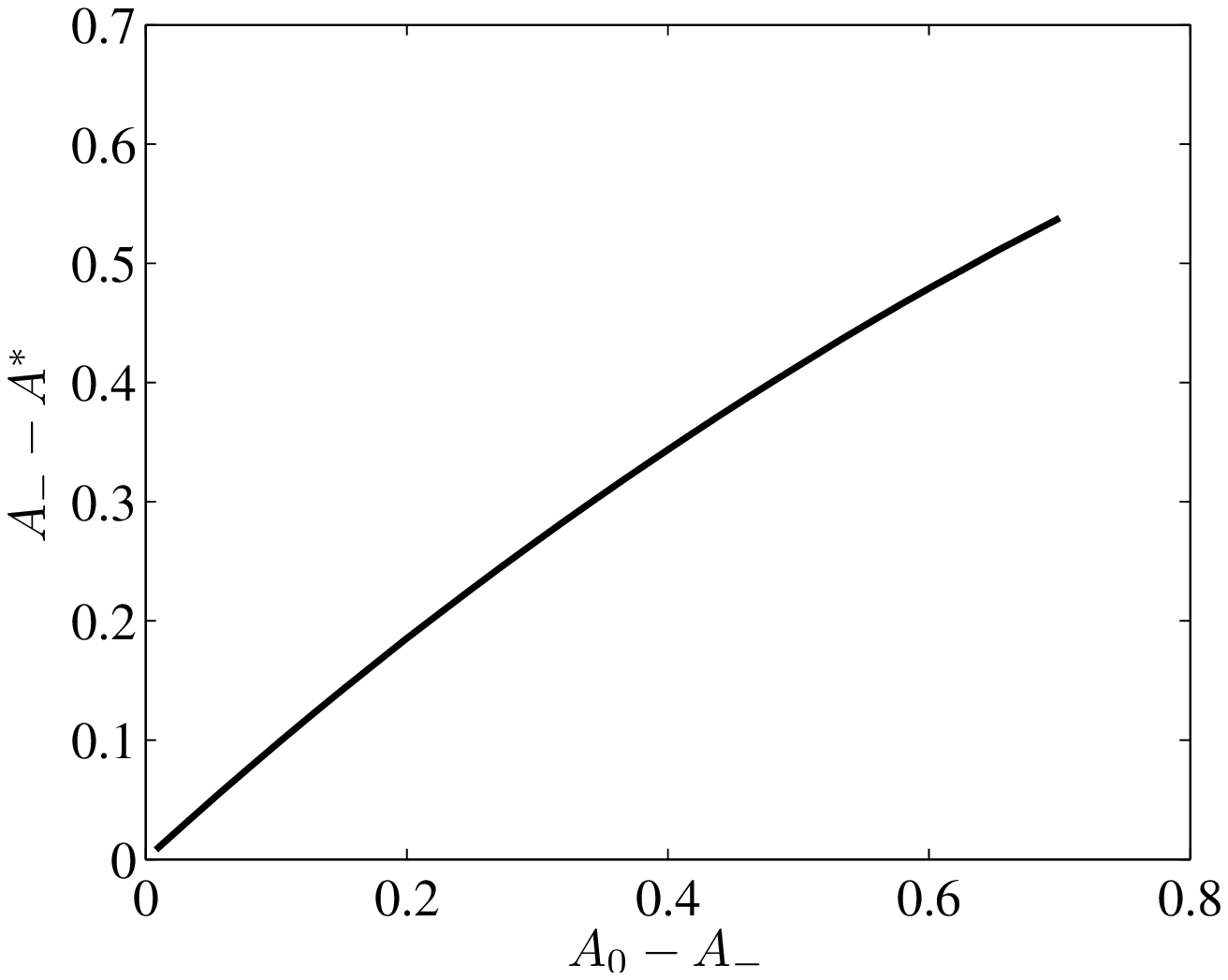}
        }   
    \subfigure[$A_0^m$ versus $D$] 
        {\label{A0vsD}
        \includegraphics[width=.4\textwidth]{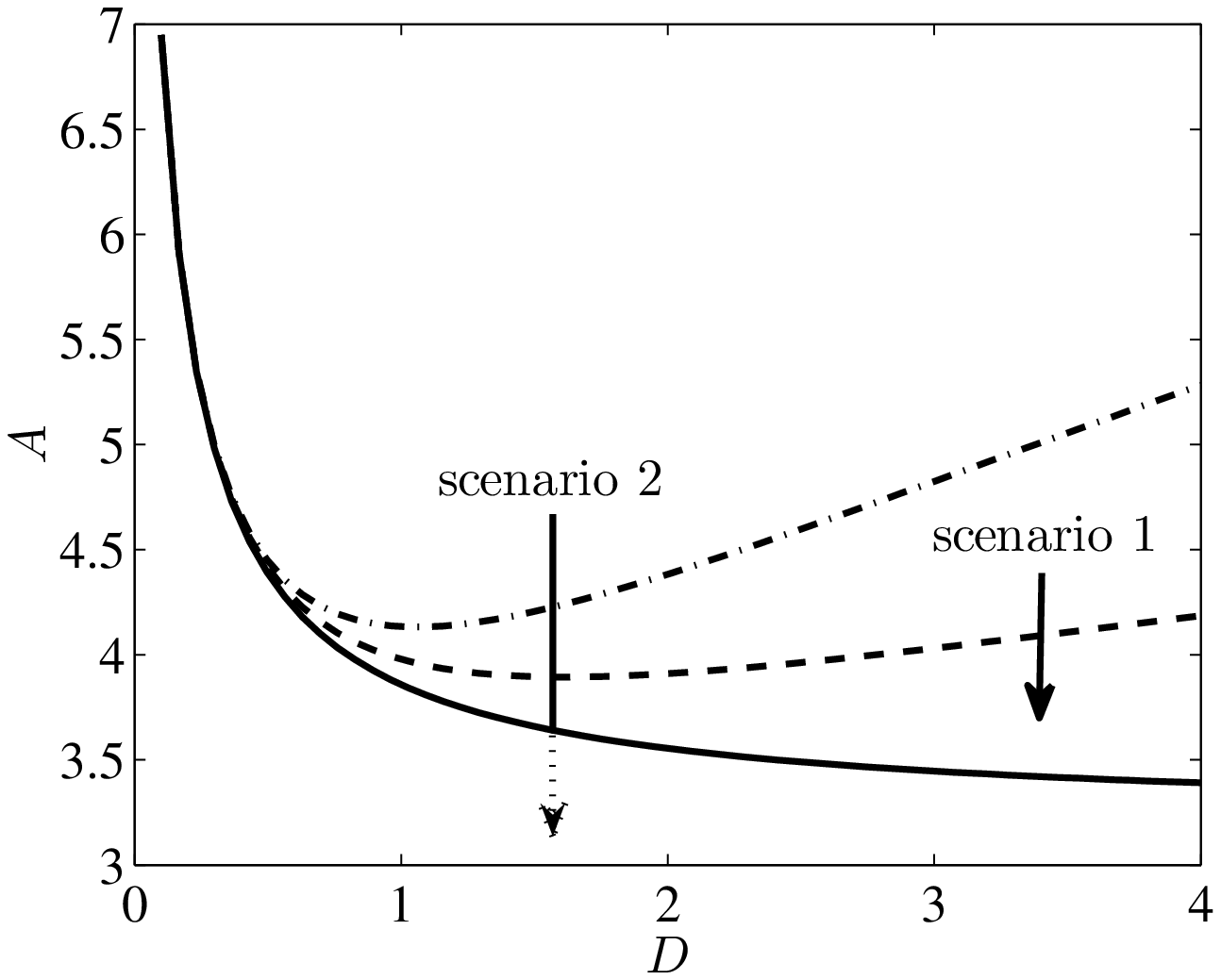}
        }}
    \caption{In the left figure, we show the expected delay $A_- - A^*$ as a function of the initial buffer $A_0 - A_-$ for $D = 3$ and $\tau = 0$. The increasing function indicates that the larger the initial buffer, the larger the expected delay. In the right figure we plot, as functions of $D$, $A_m$ (solid), $A_-$ (dashed), and $A_0^m$ (dash-dotted), where $A_0^m$ is the starting value of $A$ for which $A^* = A_m$. If $A_0 < A_0^m$ (scenario 1), the delay sets in before $A_m$. If $A_0 > A_0^m$ (scenario 2), the delay does not set in by the time $A$ has decreased to $A_m$.} 
  \end{center}
\end{figure}

Setting $\psi(A^*) = 0$ in \eqref{psiA} and solving the resulting algebraic
equation for $A^* < A_-$, we obtain a relation between the delay $A_- - A^*$
and the ``initial buffer'' $A_0 - A_-$. An example of a typical relationship
is shown in Figure \ref{GSdelayD3} for $D = 3$. Note that, as in \S \ref%
{analyticdelay} and \S \ref{GSanalyticdelay}, the delay in terms of $A$ are
independent of the rate at which it is decreased. The increasing function
shows that, the larger the initial buffer, the larger the expected delay.
The values of $A_0$ in Figure \ref{GSdelayD3} are such that $A^* > A_m$ so
that the instability sets in before the system reaches the saddle node. In
Figure \ref{A0vsD} we plot, for various $D$, $A_m$ (solid), $A_-$ (dashed),
and the starting value of $A_0 = A_0^m$ (dash-dotted) such that $A^* = A_m$.
For $A_0 < A_0^m$, the systems starts sufficiently close to threshold such
that the delay is expected to be small and the instability sets in before $A$
reaches its saddle value $A_m$. This is illustrated schematically as
scenario 1 in Figure \ref{A0vsD}, where the arrow ending above the $A_m$
curve indicates that the instability sets in before $A_m$. When $A_0 > A_0^m$%
, the delay increases to the point where the instability does not fully set
in by the time $A = A_m$. This is illustrated as scenario 2 in Figure \ref%
{A0vsD}. Here, the arrow extends below $A_m$, with the dotted segment
indicating the delay that may have occurred in the absence of a saddle. In
the next section, we show that the asymptotic prediction in Figure \ref%
{GSdelayD3} agrees with results obtained by numerically solving \eqref{GS}.
We also highlight the differences between scenarios 1 and 2.

\subsection{Numerical validation}

\label{GSnumval}

In this section, we illustrate the theory of \S \ref{GSpred} by numerically
solving the PDE system \eqref{GS} with $A$ taken to be the slowly decreasing
function of time given in \eqref{Aramp}. The parameter $\tau$ was taken to
be a small positive number much less than one. The time integration was
performed using the \texttt{MATLAB} \texttt{pdepe()} routine. The initial
conditions were taken as a perturbation of a true equilibrium state $%
(v(x,0), u(x,0))$ = $(v_e^*(x), u_e^*(x))$,

\begin{equation}  \label{GSnumpert}
v(x,0) = v_e^*(x)\left(1 - \delta\sech(\varepsilon^{-1}(x+1)) + \delta\sech%
(\varepsilon^{-1}(x-1))\right) \,, \qquad u(x,0) = u_e^*(x) \,; \qquad 0 <
\delta \ll 1 \,.
\end{equation}

\noI The equilibrium state $(v_e^*(x), u_e^*(x))$ was computed by
integrating \eqref{GS} to equilibrium starting from \eqref{GSeqbm}. The
perturbation in \eqref{GSnumpert} decreases the amplitude of the spike
centered at $x = -1$, and increases by an equal amount that of the spike
centered at $x = 1$. We begin with an example of scenario 1 with $A_0 <
A_0^m $.

\begin{figure}[htbp]
  \begin{center}
    \mbox{
    \subfigure[spike amplitudes versus $A$] 
        {\label{amp_vs_A}
        \includegraphics[width=.4\textwidth]{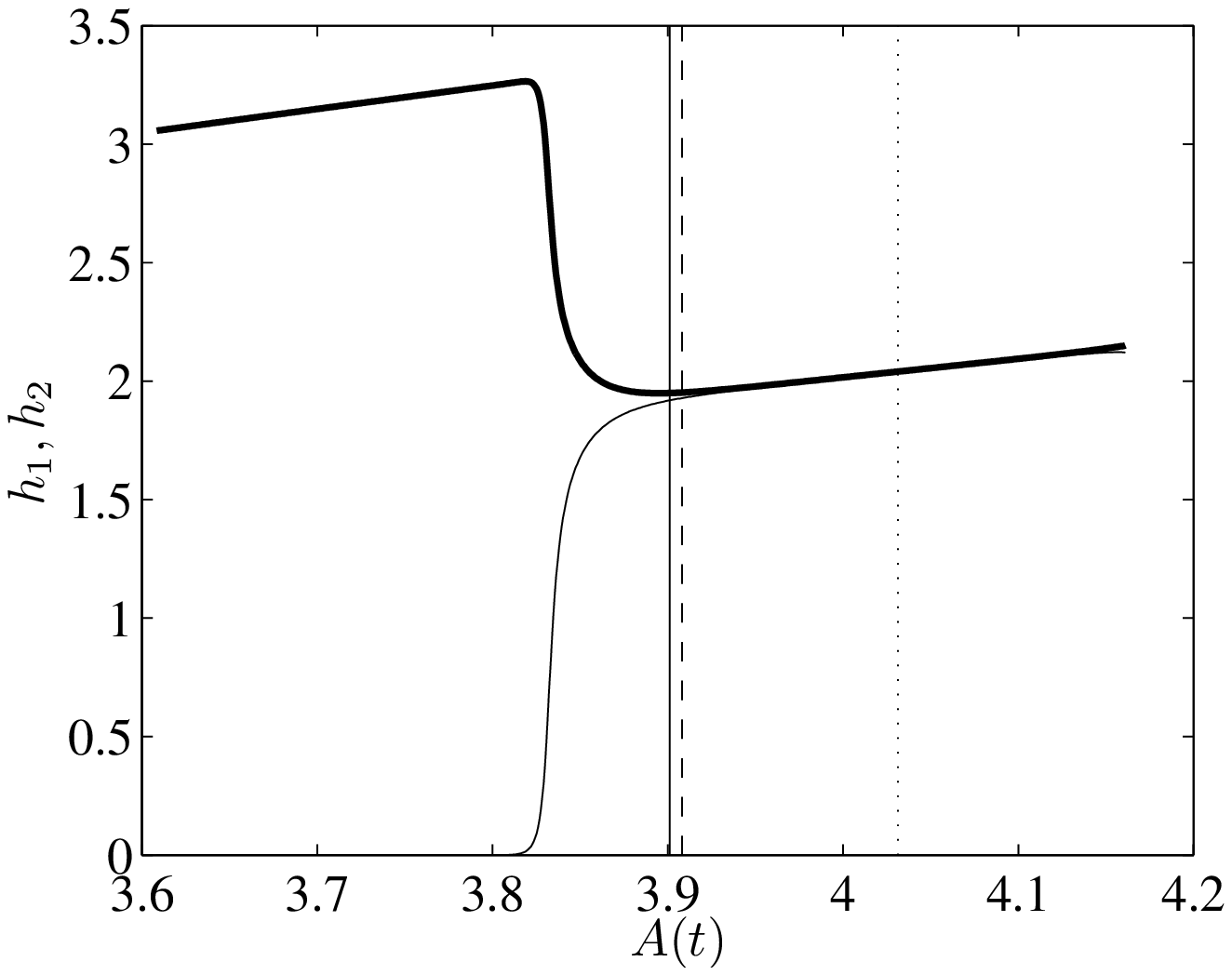}
        }   
    \subfigure[amplitude difference versus $A$] 
        {\label{amp_diff}
        \includegraphics[width=.4\textwidth]{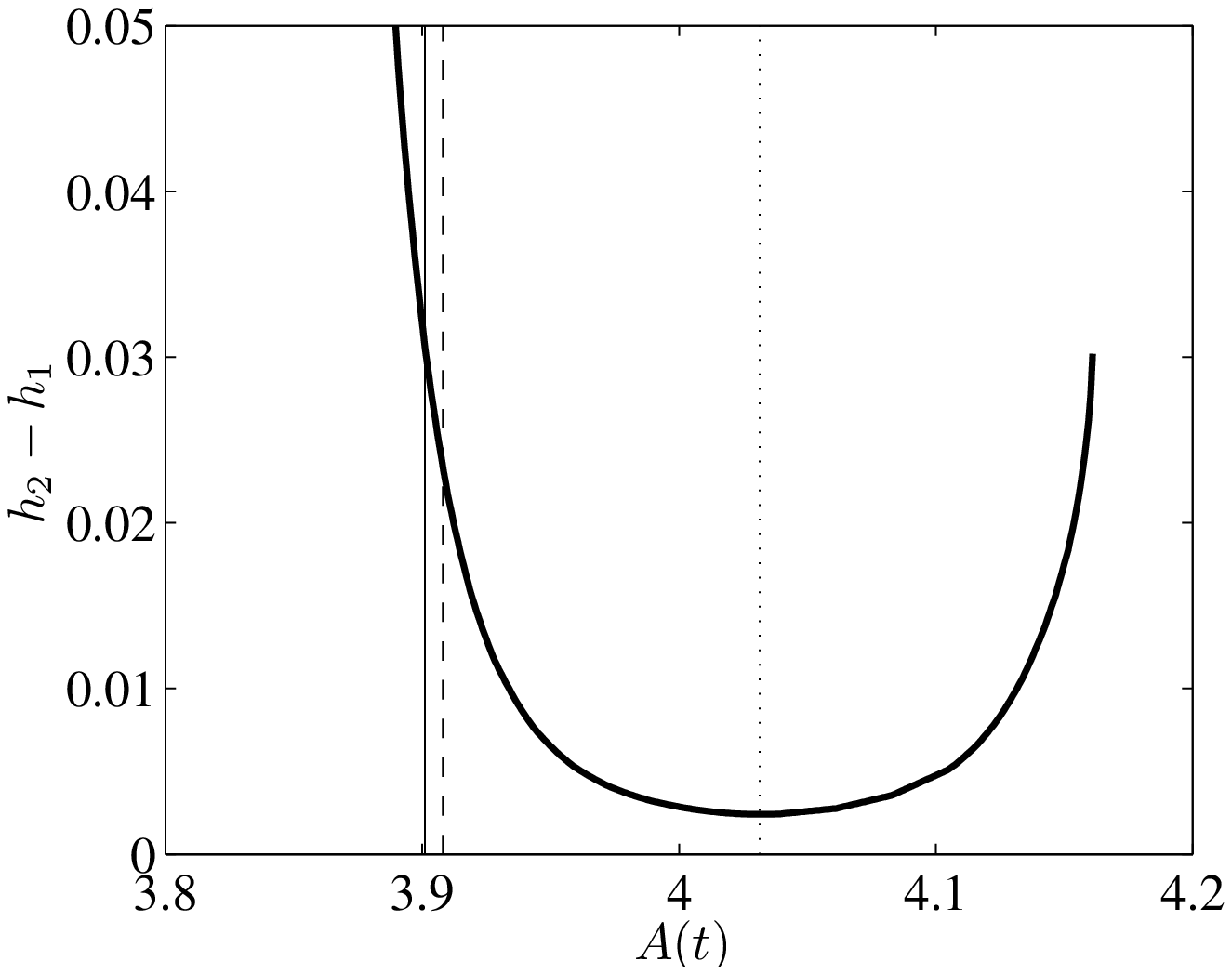}
        }}
    \caption{In the left figure, we plot the amplitude of the left (light solid) and right (heavy solid) spikes as a function of $A$, with $A$ being decreased according to \eqref{Aramp}. Increasing time is to the left. The stability threshold $A_-$ is indicated by the vertical dotted line, the asymptotic prediction of $A^*$ by the vertical dashed line, and the numerical value of $A^*$ by the vertical solid line. The amplitude difference is plotted in the right figure, with the vertical lines corresponding to those in the left figure. Starting at $A_0 > A_-$, the amplitudes grow closer together until the stability threshold $A = A_- \approx 4.03$ is reached. Here, $A_-$ is computed using \eqref{Aminus}. For $A < A_-$, the solution enters the unstable regime, causing the amplitudes to diverge. The difference in amplitudes does not reach their original value until well after $A = A_-$, indicating delay.} 
  \end{center}
\end{figure}

In Figure \ref{amp_vs_A}, we show the same typical result with $\varepsilon
= 0.004$ and $D = 3$ as in Figure \ref{amp_vs_t} except with $A(t)$ plotted
on the horizontal axis. Note that, since $A$ is a decreasing function of
time, the direction of time increase is to the left. As $A$ decreases, both
amplitudes decrease as indicated by Figure \ref{bifdiag}. The stability
threshold $A^-$ is indicated by the vertical dotted line, the asymptotic
prediction of $A^*$ by the vertical dashed line, and the numerical value of $%
A^*$ by the vertical solid line. We observe good agreement between the
asymptotic prediction and numerical value of $A^*$. As predicted, the
amplitudes do not appear to diverge until $A \approx A^*$, well after the
instability has been triggered. This illustrates the delay in competition
instability. The instability then leads to the eventual collapse of the left
spike along with the growth in amplitude of the right spike. The amplitude
of the remaining spike continues to decrease with the continued decrease of $%
A$.

In Figure \ref{amp_diff}, we illustrate the phenomenon more clearly by
plotting the difference in amplitudes as a function of $A$. The vertical
lines correspond to those in Figure \ref{amp_vs_A}. When $A > A_-$, the
system is stable, causing the initial perturbation to decay and the
amplitudes to grow closer together. When $A = A_-$, the instability is
triggered and the amplitudes begin to diverge. However, since the amplitudes
grew closer together on the interval $A_0 \geq A > A_-$, $A$ must be
decreased well beyond $A_-$ for the amplitude difference to grow back to its
initial size at $t = 0$. For $D = 3$, we find from Figure \ref{A0vsD} that $%
A_- \approx 4.03$, which matches almost exactly the location of the minimum
in Figure \ref{amp_diff}, indicating again excellent agreement between
between asymptotic and numerical results.

We repeat the computations with $D = 3$ and find the delay for various
values of the initial buffer. The results are compiled in Figure \ref%
{GSdelay_compareD3}, where we compare the results to asymptotic result of
Figure \ref{GSdelayD3} for $\varepsilon = 0.008$ (circles) and $\varepsilon
= 0.004$ (squares). We observe excellent agreement, with the numerical
results for $\varepsilon = 0.004$ matching the asymptotic result (solid
curve) more closely for small $A_0$. The deviation of the squares from the
curve for larger $A_0$ is likely due to the small $\mathcal{O}%
(e^{-1/\varepsilon})$ amplitude difference being obscured by numerical
errors.

\begin{figure}[htbp]
  \begin{center}
	\includegraphics[width=.4\textwidth]{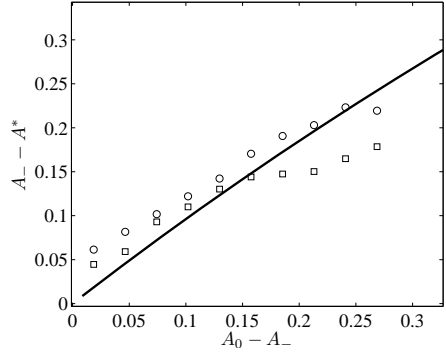}
	\caption{Comparison of numerical and asymptotic (solid) results for $D = 3$ and $\varepsilon = 0.008$ (circles) and $\varepsilon = 0.004$ (squares). The results for $\varepsilon = 0.004$ match the asymptotic result (solid curve) more closely for small $A_0$. The deviation of the squares from the curve for larger $A_0$ is likely due to the small $\mathcal{O}(e^{-1/\varepsilon})$ amplitude difference being obscured by numerical errors.}
	\label{GSdelay_compareD3}
	\end{center}
\end{figure}

\begin{figure}[htbp]
  \begin{center}
    \mbox{
    \subfigure[spike amplitude versus $A$] 
        {\label{saddledyn_A}
        \includegraphics[width=.4\textwidth]{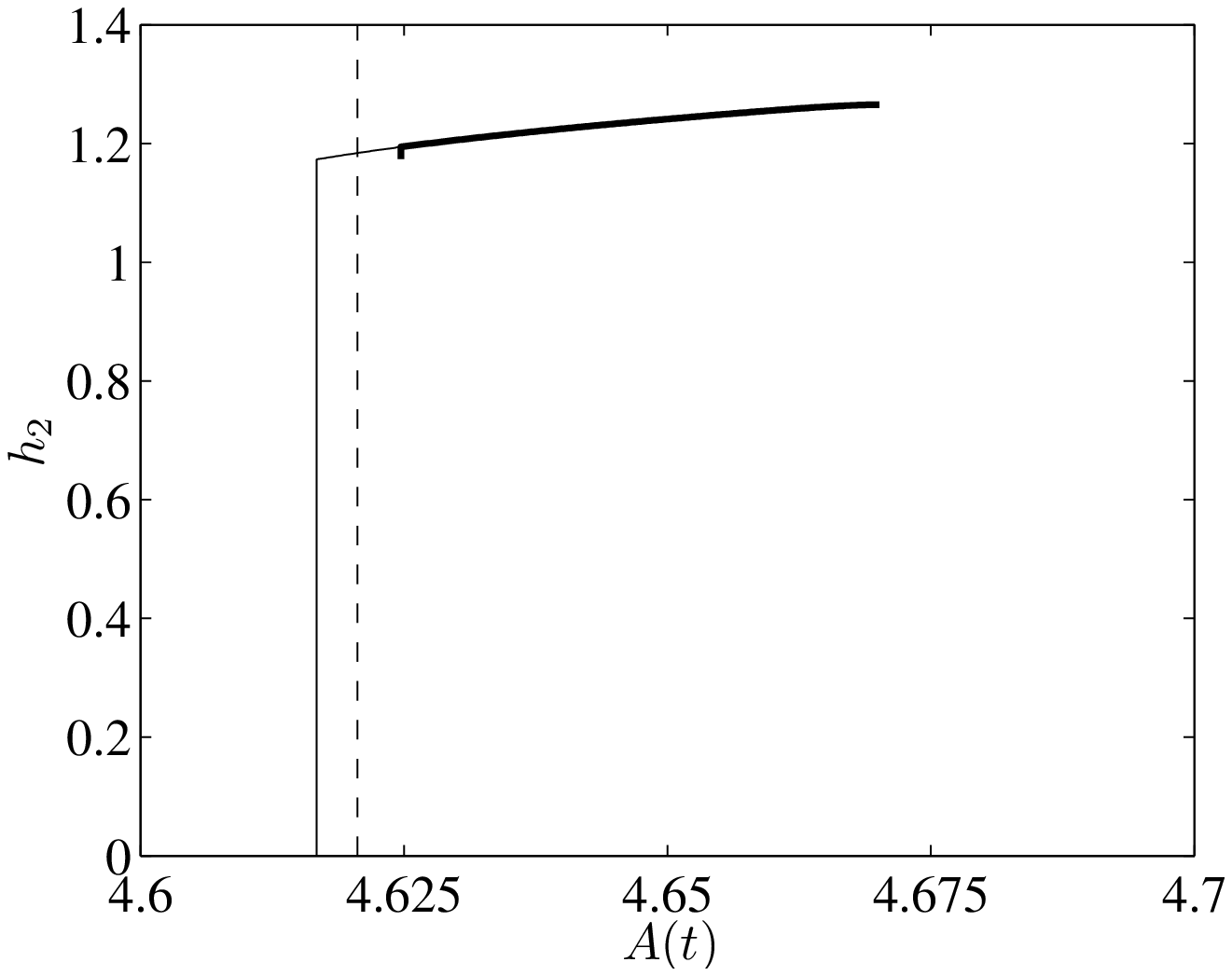}
        }   
    \subfigure[spike amplitude versus $t$] 
        {\label{saddledyn_t}
        \includegraphics[width=.4\textwidth]{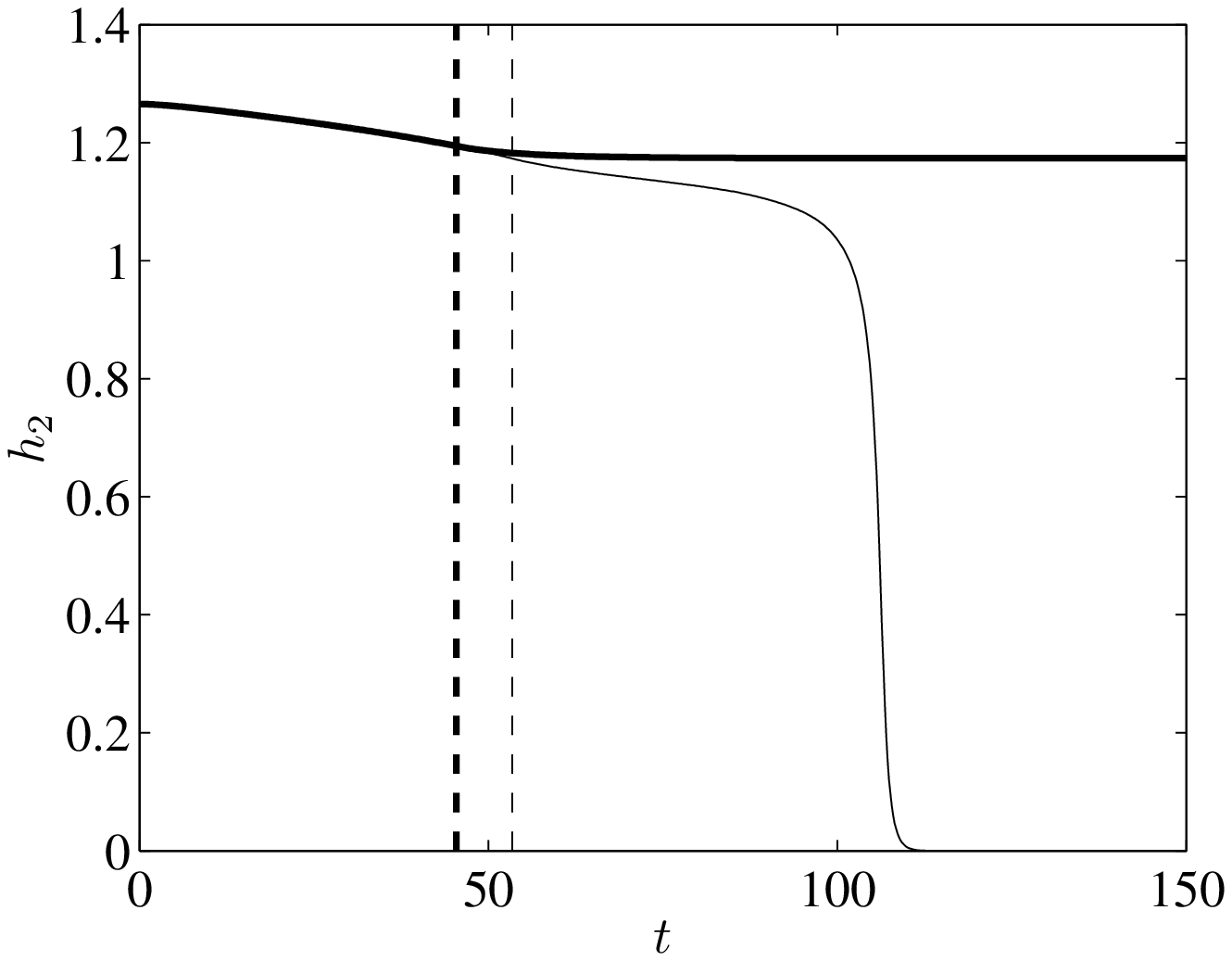}
        }}
    \caption{Evolution of the amplitude of a one boundary spike quasi-equilibrium solution as $A$ is slowly decreased. In the left figure, we plot the amplitude of the spike as a function of $A$. The saddle value $A_m \approx 4.6206$ is indicated by the vertical dashed line. In the case that $A$ stops decreasing at $A = A_m + 0.004$ (heavy solid curve), the spike settles to a constant non-zero value. If $A$ is decreased past $A_m$ to $A = A_m - 0.004$, the spike collapses. The corresponding evolution as a function of time is shown in the right figure. The heavy (light) dashed line indicates the time that $A$ reaches its terminal value of $A_m + 0.004$ ($A_m - 0.004$). The starting value of $A$ in both instances is $A_0 = 4.6701$. Here, $\varepsilon = 0.001$, and $D = 0.4$.} 
  \end{center}
\end{figure}

To illustrate the second scenario where $A^* < A_m$, we first confirm
numerically the location of the saddle. To do so, we solve \eqref{GS} on the
domain $0 < x < 1$ with pure Neumann boundary conditions for one boundary
spike centered at $x = 1$. In this way, we eliminate the possibility of the
odd competition instability and isolate the effects of the saddle node. In
Figure \ref{saddledyn_A}, we show the evolution of the spike amplitude as $A$
is decreased starting from a true one boundary spike equilibrium, analogous
to $(v_e^*(x), u_e^*(x))$, with $D = 0.4$ and no initial perturbations. The
heavy solid curve shows the case where the decrease of $A$ is stopped at $A
= A_m + 0.004$, slightly before it reaches its value at the saddle. The
value for $A_m \approx 4.6206$ may be computed from \eqref{Am} and is
indicated by the vertical dashed line. As shown in Figure \ref{saddledyn_t},
the spike amplitude settles to a constant non-zero value after the time that
the decrease of $A$ has ceased (heavy dashed line). The light solid curves
in Figures \ref{saddledyn_A} and \ref{saddledyn_t} show the case where $A$
is decreased slightly past the saddle to $A = A_m - 0.004$. Contrary to the
first case, the spike collapses after the decrease of $A$ has ceased (light
dashed line in Figure \ref{saddledyn_t}). We thus conclude that the true
location of the saddle is close to that predicted by the asymptotic result %
\eqref{Am}, and that in a two-spike equilibrium, the dynamics beyond the
saddle induce the simultaneous collapse of both spikes. We emphasize that
the simultaneous collapse is due to the effect of the saddle, not the
synchronous instability described in \S \ref{GSpred}.

We now contrast the two scenarios $A_0 < A_0^m$ and $A_0 > A_0^m$. Recall
that if $A_0 < A_0^m$, the instability is expected to set in before $A$
reaches the saddle, while if $A_0 > A_0^m$, the delay is sufficiently large
so that the instability does not fully set in when $A$ reaches $A_m$. When
two spikes are present, two competing effects take place slightly beyond the
saddle node. The less dominant effect is that just described, which leads to
the simultaneous collapse of both spikes. The more dominant is the residual
effect of the competition instability, which leads to the collapse of one
spike and the growth of the other.

The relative dominance may be attributed to the zero eigenvalue of the
synchronous mode exactly at the saddle node. Recall that the lower solution
branch is always unstable to the even synchronous mode while the upper
branch is always stable to the even synchronous mode. Where they meet, the
eigenvalue of the even mode must be zero. While no spike solutions exist
beyond the saddle, the dynamics associated with an even perturbation will be
slow due to the nearby presence of the zero eigenvalue. Similarly, the
dynamics associated with an odd perturbation will be relatively fast due to
the nearby presence of the positive eigenvalue of the competition mode.

\begin{figure}[htbp]
  \begin{center}
    \mbox{
    \subfigure[scenario 1: $A_0 = 4.6451 < A_0^m$] 
        {\label{AltA0m}
        \includegraphics[width=.4\textwidth]{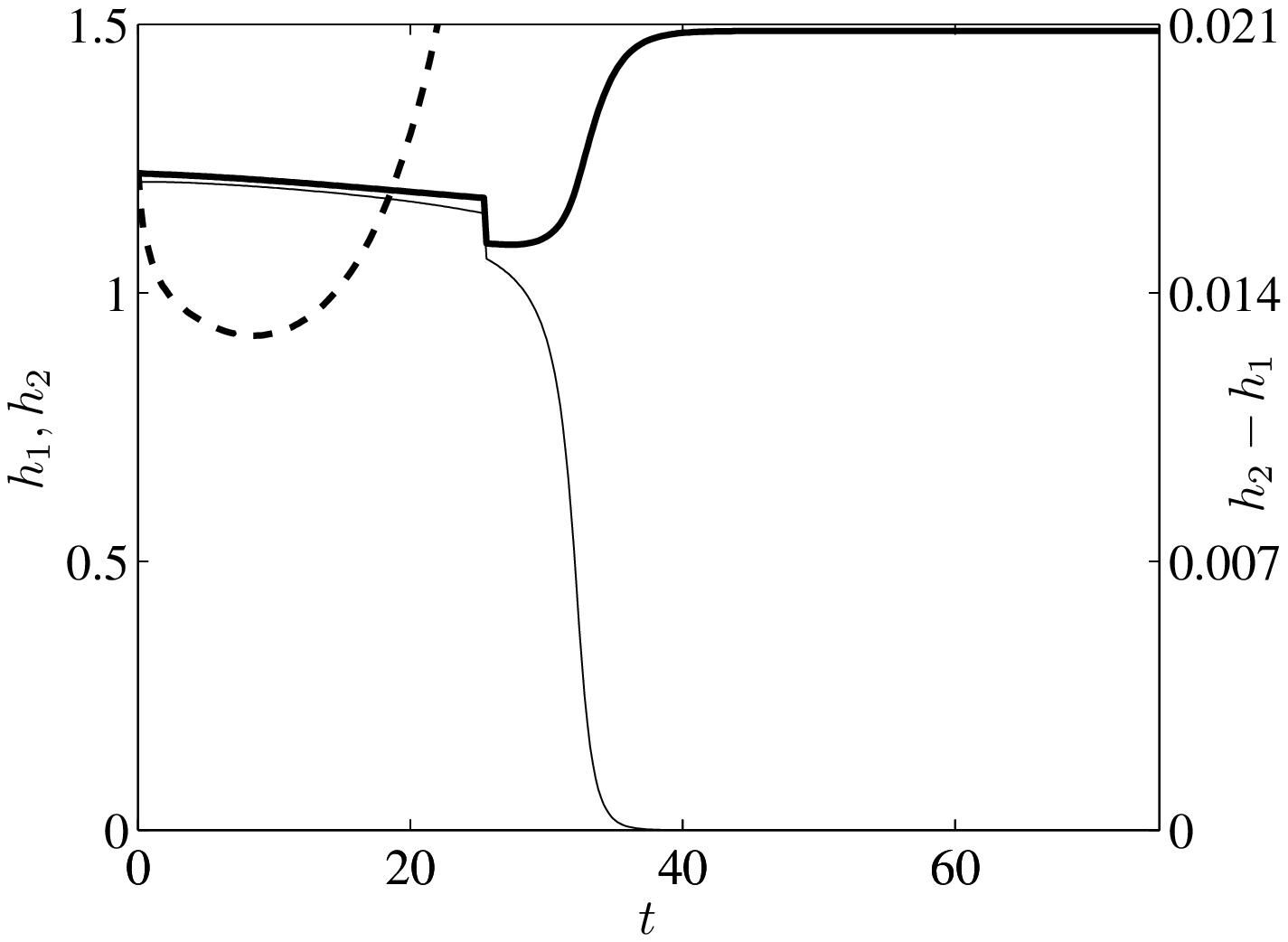}
        }   
    \subfigure[scenario 2: $A_0 = 4.6701 > A_0^m$] 
        {\label{AgtA0m}
        \includegraphics[width=.4\textwidth]{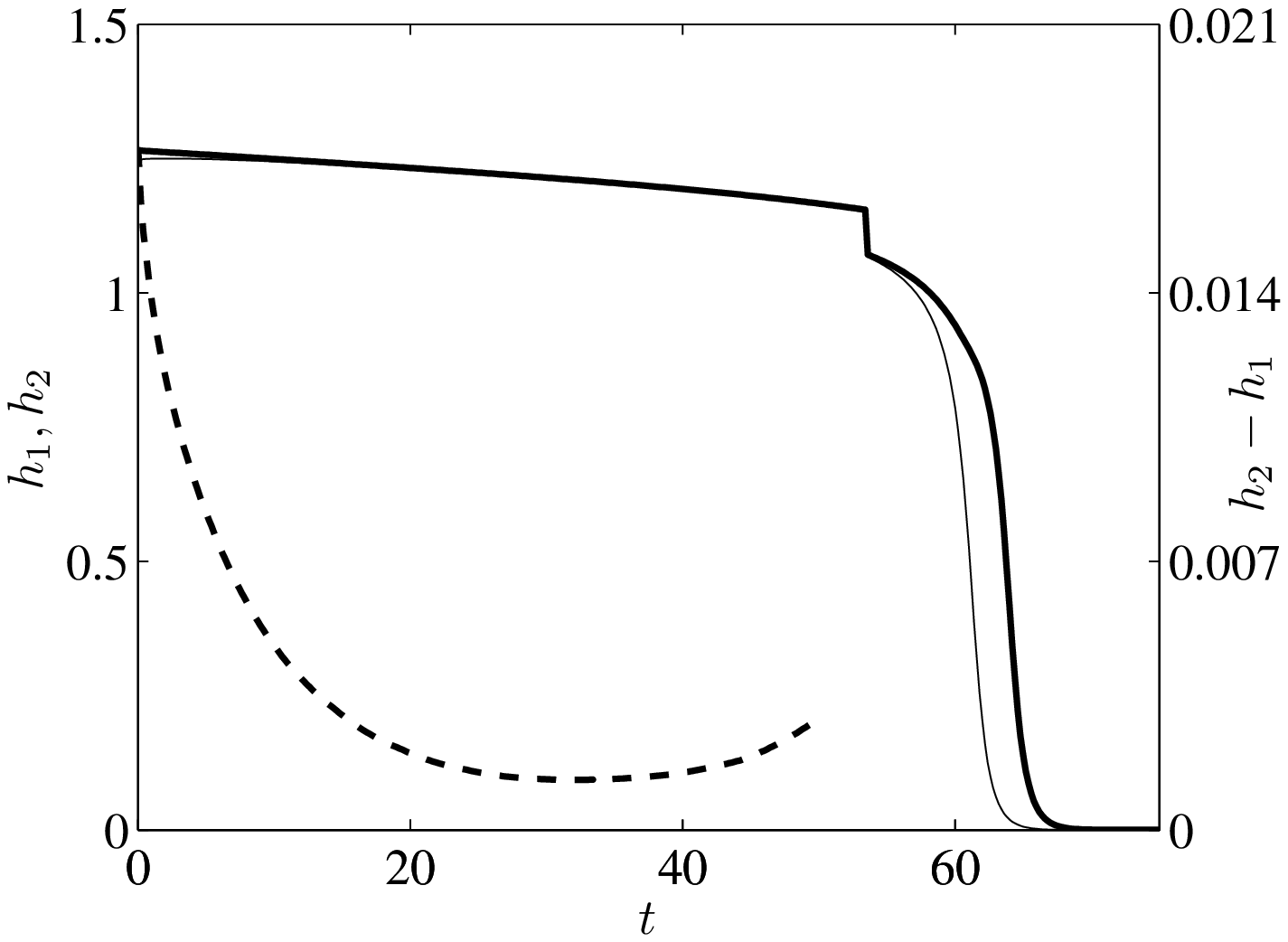}
        }}
    \caption{Plots of the amplitude of the left (light solid) and right (heavy solid) spikes versus time as $A$ is decreased from $A_0$ to $A_m - 0.004$. For $D = 0.4$, we find from Figure \ref{A0vsD} that $A_- = 4.6351$, $A_m = 4.6206$, and $A_0^m = 4.6534$. The near-vertical kinks at $t \approx 25.4$ in the left figure and $t \approx 53.4$ in the right are the result of the addition of an even perturbation when $A$ reaches $A_m - 0.004$.  The dashed curve, plotted against the right vertical axis, is a plot of the difference in spike amplitudes. In the left figure where $A_0 < A_0^m$, the competition mode sets in early and dominates the dynamics near the saddle. As a result, only one spike collapses. In the right figure, the competition mode has not set in by the time the even perturbation is added. The resulting dynamics are near that of a simultaneous collapse of both spikes. The presence of the competition mode causes the right spike to collapse slightly more slowly than the left.} 
  \end{center}
\end{figure}

We therefore expect for a two spike solution that when $A$ is decreased to
below $A_m$, perturbing the solution with both an odd and even perturbation
would result in dynamics mirroring that of the dominant competition
instability. One spike would collapse while the other would survive. This is
depicted as scenario 1 in Figure \ref{AltA0m}, where $A_0 < A_0^m$. On the
left vertical axis, we plot the amplitude of the left (light solid) and
right (heavy solid) spikes as $A$ is decreased, stopping at $A = A_m - 0.004$%
, slightly off the saddle. On the right vertical axis, we plot the amplitude
difference (dashed). The horizontal axis is time. The simultaneous decrease
in both amplitudes at $t \approx 25.4$ is a result of an even perturbation
added when $A$ reaches its terminal value of $A = A_m - 0.004$. As expected,
because the competition instability sets in before $A$ reaches $A_m$, the
dynamics of the competition mode dominate beyond the saddle and only one
spike collapses.

If the size of the odd perturbation were to be sufficiently small relative
to that of the even, the slower growth of the even mode would be compensated
for by its larger initial size. We would then expect the resulting dynamics
to reflect that of the one spike solution, with both spikes collapsing
almost simultaneously. This is depicted as scenario 2 in Figure \ref{AgtA0m}%
, where $A_0 > A_0^m$. As shown by the dashed curve, the instability has not
fully set in by the time the $A$ reaches $A_m - 0.004$ and the even
perturbation is added. As a result, the even mode added at $t \approx 53.4$
dominates, and both spikes collapse. The presence of the competition mode
causes the right spike to collapse slightly more slowly than the left. This
scenario illustrates that the dynamics of a comparatively weak mode may
prevail over that of a dominant mode due solely to the phenomenon of delay.

\section{Discussion}

We have presented three examples of delayed bifurcations for spike solutions
of reaction-diffusion systems. In the first example with a single stationary
spike, we considered the case where a model parameter $\tau$ was
extrinsically tuned slowly past a Hopf bifurcation threshold. In the second
example with a slowly drifting single spike, we studied the case where all
model parameters were held constant and a Hopf bifurcation with $\mathcal{O}%
(1)$ time scale oscillations was triggered by \textit{intrinsic} $\mathcal{O}%
(\varepsilon^2)$ drift dynamics. A feature of this example not present in
the first was that of a non-monotonic Hopf bifurcation threshold curve.
Introducing sufficient delay into the system by careful selection of initial
conditions, we found that the non-monotonicity allowed the spike to drift
safely through a Hopf-unstable zone without the Hopf bifurcation fully
setting in. In the third example with two stationary boundary spikes, we
considered the delay of a competition instability as a feed rate parameter $%
A $ was tuned slowly past a stability threshold $A_-$. In addition to the
competition threshold, there existed a saddle node bifurcation at $A = A_m$
past which no two-spike solutions exist. The presence of two critical values
of $A$ led to two competing effects near the saddle node. We found that the
delay played a critical role in determining which effect prevailed. In
particular, we showed that a delay in the onset of the competition
instability allowed the effect of the saddle node to dominate despite being
comparatively weak.

In all three examples, linear stability analysis of the equilibrium or
quasi-equilibrium solutions led to an explicitly solvable NLEP. By obtaining
an explicit expression for the eigenvalue, we were able to formulate an
algebraic problem for how far above a stability threshold the system must be
in order for the instability to be fully realized. This delay in terms of
the parameter was independent of the rate at which the system crossed the
stability threshold. For all three examples, we solved the full PDE system
numerically and observed excellent agreement with asymptotic predictions for
the magnitude of delay. A key numerical challenge involved obtaining results
not obscured by numerical errors when the system started far below
threshold. For such computations, more digits of precision may be beneficial.

An interesting open problem in regards to Example 2 would be to understand
the oscillations that occur well after the Hopf bifurcation has set in. For
example, a weakly nonlinear theory may be developed to determine whether the
bifurcation is subcritical or supercritical. In the case shown in Figure \ref%
{oscamp}, we find that, well after the onset of the Hopf bifurcation, the
oscillations exhibit a repeating pattern of series of five successively
growing peaks, with peaks in each subsequent series slightly larger than the
corresponding peaks in the previous series. This is shown in Figure \ref%
{nonlinosc}.

\begin{figure}[htbp]
  \begin{center}
	\includegraphics[width=.4\textwidth]{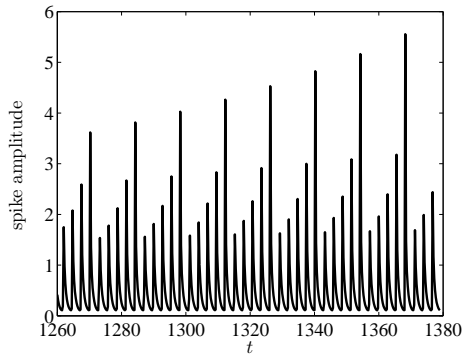}
	\caption{Nonlinear oscillations well after the intrinsically triggered Hopf bifurcations of Example 2 have set in. The parameters are the same as those of Figure \ref{oscamp}. The oscillations exhibit a pattern of series of five successively growing peaks, with peaks in each subsequent series slightly larger than the corresponding peaks in the previous series. The oscillations also appear to be relaxational.}
	\label{nonlinosc}
	\end{center}
\end{figure}

Another interesting problem would be to quantify the effect of a periodic
forcing function on the delay of a Hopf bifurcation. For the ODE system
considered in \cite{baer1989slow}, it was found that a small amplitude
sinusoidal forcing function with frequency equal to that of the Hopf
frequency reduced the magnitude of the delay. For the example shown in
Figure \ref{oscamp}, we added to the right-hand side of \eqref{GMvfin} a
small forcing function $f(x,t)$ of the form

\begin{equation*}
f(x,t) = 0.001 \sin(\omega_H t) \, w(\varepsilon^{-1}(x-x_0(t))) \,,
\end{equation*}

\noI with $w(y)$ given in \eqref{wsolp}. Here, $\omega _{H}$ is the resonant
Hopf frequency, and $x_{0}(t)$ is the center of the spike at time $t$. We
observed in this case a beat phenomenon in the amplitude oscillations, with
the low frequency envelope decaying in the Hopf stable region. Unlike that observed in \cite{baer1989slow}, the forcing resulted in only a very slight decrease in the magnitude of delay. However, as part of a more
detailed study of how delay varies with changes in forcing amplitude and
frequency, the above result may help identify methods for more accurately
determining bifurcation thresholds in experimental systems.

A related issue is the effect of noise on dynamics and
bifurcations. In the context of ODE's, a number of works elucidate the
role that stochastic noise can play in pushing the system through
tipping points; see for ex.  \cite{van1994noise, muratov2008noise,
  kuehn2011mathematical, gentz} and the references therein. Some
recent papers also explore how the noise changes the dynamics in the
context of stochastic PDE's \cite{muratov2007noise, kuehn2012warning,
  hairer2012triviality}. However much work remains to be
done in this direction. In particular the effect of noise on the
stability of spikes in RD systems remains largely unexplored.

\appendix

\setcounter{equation}{0}
\section{Construction and stability of a one-spike equilibrium of the GM model on the infinite line} \label{appA}

Here, we construct a one-spike solution of \eqref{GMinf} and derive an
explicitly solvable nonlocal eigenvalue problem (NLEP) governing its
stability to $\mathcal{O}(1)$ eigenvalues. Solving the NLEP, we derive
\eqref{NLEPsol} of \S \ref{GMinfline}. In the inner region of the
spike centered at $x = 0$, we transform to a stretched variable $y =
x/\varepsilon$ and let

\BE \label{innUV}
u \sim U(y) \,, \qquad v \sim V(y) \,.
\EE

\noI The equilibrium problem on $-\infty<y<\infty$ is
\BE \label{innprob}
 V^{\prime\prime} - V + \frac{V^3}{U^2} = 0 \,, \qquad 
 \frac{1}{\varepsilon^2}U^{\prime\prime} - U + \frac{V^3}{\varepsilon} = 0 \,,
\EE
with $V\to 0$ and $U$ bounded as $|y|\to\infty$. From \eqref{innprob} for
$U$, we have to leading order that $U = U_0$ is a constant and
\BE \label{V}
	V = U_0 w(y) \,,
\EE
\noI where $w(y)$ is the homoclinic solution of
\BE \label{weq}
 w^{\prime\prime} - w + w^3 = 0, \qquad -\infty < y < \infty \,, 
\qquad w(0) > 0 \,, \qquad w^\prime(0) = 0 \,, \qquad w \to 0 \quad 
\mbox{as} \quad |y| \to \infty \,.
\EE

\noI Equation \eqref{weq} may be solved explicitly, with the solution
given in \eqref{eqbminf2}.

In the outer region where $|x| = \mathcal{O}(1)$, the term
$\varepsilon^{-1}v^3$ in \eqref{GMu} is exponentially small. As
$\varepsilon \to 0$, its mass becomes concentrated in an
$\mathcal{O}(\varepsilon)$ width region around $x = 0$ with height
$\mathcal{O}(\varepsilon^{-1})$ at $x = 0$. In the sense of
distributions, using \eqref{innUV} and \eqref{V}, we calculate
$\varepsilon^{-1} v^3 \to \left(\frac{1}{\varepsilon} \varepsilon \intinf 
U_0^3 w^3 \, dy\right) \delta(x) = b \, U_0^3\delta(x)$, 
where $b$ is defined in \eqref{eqbminf2}, and $\delta(x)$ is the
Dirac-delta function centered at $x = 0$. Substituting this expression
into \eqref{GMu}, we find that the outer solution for $u$ satisfies
\BE \label{uoutereq} 
  u_{0xx} - u_0 = b\,U_0^3\delta(x) \,, \qquad
-\infty < x < \infty \,, \qquad u_0 \to 0 \quad \mbox{as} \quad |x|
\to \infty \,, 
\EE 
\noI with the matching condition $u_0(0) = U_0$.
The solution to \eqref{uoutereq} is written in terms of a Green's
function $G(x;x_0)$ as $u_0(x) = b\,U_0^3G(x;0)$, where where $G(x;0)$
satisfies
\BE \label{Ginfeq}
G_{xx} - G = -\delta(x) \,, \qquad -\infty < x < \infty \,, \qquad G \to 0 
\quad \mbox{as} \quad |x| \to \infty \,.
\EE

\noI The solution to \eqref{Ginfeq} is $G(x;0) = {e^{-\lvert
    x\rvert}/2}$. Applying the matching condition $u_0(0)=U_0$, we
calculate $U_0 ={1/\left(\sqrt{b\,G(0;0)}\right)}$. In this way, we
obtain the results \eqref{eqbminf} and \eqref{eqbminf2} of \S
\ref{GMinfline}.

To derive the transcendental equation for the eigenvalue in
\eqref{NLEPsol}, we linearize \eqref{GMinf} by perturbing the
equilibrium solution as in \eqref{pert}. The linearized equation is
then
\bes \label{GMlin}
\BE \label{GMlinv}
\lambda \phi = \varepsilon^2\phi_{xx} - \phi + \frac{3v_e^2}{u_e^2}
\phi - \frac{2v_e^3}{u_e^3} \eta \,, \qquad -\infty < x < \infty \,,
\qquad \phi \to 0 \quad \mbox{as} \quad |x| \to \infty \,,
\EE

\BE \label{GMlinu} 
\tau \lambda \eta = \eta_{xx} - \eta +
\frac{3v_e^2}{\varepsilon}\phi \,, \qquad -\infty < x < \infty \,,
\qquad \eta \to 0 \quad \mbox{as} \quad |x| \to \infty \,.  
\EE \ees

\noI Here, $v_e$ and $u_e$ are given in \eqref{eqbminf}. Since the
coefficients $v_e^2/u_e^2$ and $v_e^3/u_e^3$ are localized near $x =
0$, we seek solutions to \eqref{GMlinv} where $\phi$ is localized near
$x = 0$ and $\eta$ varies over the same scale as does $u_e(x)$. With
$\phi = \Phi(y)$ and $\eta(x) \sim \eta(0)$ as $x \to 0$, we obtain
the following equation for $\Phi(y)$,

\BE \label{Phieq}
L_0 \Phi - 2w^3\eta(0) = \lambda \Phi \,, \qquad -\infty < y < \infty \,, 
\qquad \Phi \to 0 \quad \mbox{as} \quad |y| \to \infty \,,
\EE
\noI where the linear operator $L_0$ is defined as
\BE \label{L0}
L_0\psi \equiv \psi^{\prime\prime} - \psi + 3w^2\psi \,.
\EE

To determine $\eta(0)$ in \eqref{Phieq}, we solve \eqref{GMlinu} for
$\eta(x)$. Since the term $v_e^2\phi$ is localized near $x = 0$, we
have in the sense of distributions that $\varepsilon^{-1}v_e^2\phi
\sim \left[\intinf U_0^2 w^2 \Phi(y) \, dy\right] \delta(x)$, where we
have used \eqref{eqbminf} for $v_e$ in \eqref{GMlinu}.  The resulting
equation for $\eta(x)$ is then 
\BE \label{etaeq} 
\eta_{xx} - (1+\tau\lambda)\eta = -3U_0^2 \left[\intinf w^2\Phi \,dy
  \right]\delta(x) \,, \qquad -\infty < x < \infty \,, \qquad \eta \to
0 \quad \mbox{as} \quad |x| \to \infty \,.  
\EE

\noI We write the solution to \eqref{etaeq} in terms of the Green's
function $G_\lambda(x;0)$ as

\BE \label{etasol}
\eta(x) \sim 3U_0^2 \left[\intinf w^2\Phi \, dy \right]G_\lambda(x;0) \,,
\EE

\noI where $G_\lambda(x;0)$ satisfies

\BE \label{Glambdaeq} 
G_{\lambda xx} - (1+\tau \lambda)G_{\lambda} =
-\delta(x) \,, \qquad -\infty < x < \infty \,, \qquad G_\lambda \to 0
\quad \mbox{as} \quad |x| \to \infty \,.  
\EE

\noI The solution of \eqref{Glambdaeq} is

\BE \label{Glambdasol}
G_{\lambda}(x;0) = \frac{1}{2\sqrt{1+\tau\lambda}}e^{-\theta_\lambda \lvert x \rvert} \,,
\qquad  \theta_\lambda \equiv \sqrt{1+\lambda\tau} \,.
\EE

\noI Using \eqref{etasol} to compute $\eta(0)$, and using
$U_0 ={1/\left(\sqrt{b\,G(0;0)}\right)}$, we obtain the nonlocal eigenvalue 
problem (NLEP) 
\BE \label{NLEP}
L_0 \Phi - \chi w^3 \frac{\intinf w^2 \Phi \, dy}{\intinf w^3 \, dy} = 
\lambda\Phi \,,
\qquad
\chi \equiv 6\frac{G_\lambda(0,0)}{G(0;0)} \,,
\EE

\noI with $L_0\psi$ is defined in \eqref{L0}. From $G(0;0)={1/2}$ and 
\eqref{Glambdasol}, we calculate $\chi$ in \eqref{NLEP} as

\BE \label{chi2}
\chi = \frac{6}{\sqrt{1+\tau\lambda}} \,.
\EE

From \cite{nec2013explicitly}, the specific choice of powers of the GM
model in \eqref{GMinf} allows the NLEP \eqref{NLEP} to be solved
explicitly. We begin by noting that, in addition to the zero
eigenvalue with associated eigenfunction $w^\prime(y)$ that changes
sign once on $-\infty < y < \infty$, $L_0 \psi = \nu \psi$ has a
unique positive eigenvalue $\nu_0 = 3$ with eigenfunction $\psi_0 =
w^2$ of constant sign. To show this, we first multiply \eqref{weq} by
$w^\prime$ and integrate to compute that $(w^\prime)^2 = w^2 -
{w^4/2}$. We then calculate

\BE \label{L0wsq}
L_0 w^2 = 2(w^\prime)^2 + 2ww^{\prime\prime} - w^2 + 3w^4 \,.
\EE

\noI Then, by using \eqref{weq} for $w^{\prime\prime}$ and the expression above
for $w^\prime$, we find from \eqref{L0wsq} that indeed

\BE \label{L0eve}
L_0 w^2 = 3w^2 \,.
\EE

Next, we multiply \eqref{NLEP} by $w^2$ and integrate over the real
line to obtain

\BE \label{NLEPint} 
\intinf w^2 L_0 \Phi dy = \chi \frac{\intinf w^5
  dy \intinf w^2 \Phi dy }{\intinf w^3 dy} + \lambda \intinf w^2 \Phi
dy \,.  \EE

\noI With $\Phi(y)$, $\Phi^\prime(y)$, $w(y)$, and $w^\prime(y)$ all
decaying exponentially to zero at infinity, Green's second identity
yields $\intinf \Phi L_0w^2 dy = \intinf w^2L_0\Phi dy$. With this
identity, together with \eqref{L0eve}, we obtain for the left-hand
side of \eqref{NLEPint} that
$\intinf w^2L_0\Phi dy = 3\intinf \Phi w^2 dy$. With this expression,
the NLEP \eqref{NLEPint} then becomes

\BE \label{NLEPfactor}
\intinf \Phi w^2 dy \left\lbrack 3 - \chi 
\frac{\intinf w^5 dy}{\intinf w^3 dy} - \lambda \right\rbrack = 0\,.
\EE

\noI Calculating $\intinf w^5dy / \intinf w^3 dy = 3/2$, we conclude
that any eigenvalue of \eqref{NLEP} for which the eigenfunction
satisfies $\intinf \Phi w^2 dy \neq 0$ must satisfy the expression
given in \eqref{NLEPsol} of \S \ref{GMinfline}, where we use
\eqref{chi2} for $\chi$ in \eqref{NLEPfactor}.

\setcounter{equation}{0}

\section{One-spike quasi-equilibrium and slow dynamics of the GM model on a finite domain} \label{appB}

Here, we construct the one-spike quasi-equilibrium solution of
\eqref{GMfin} and derive the ODE \eqref{x0ODE} describing its slow
dynamics. For the inner solution of a one-spike quasi-equilibrium
solution centered at $x = x_0$, we let
\BE \label{uvinnerfin}
u \sim U_0(y) + \varepsilon U_1(y) + \cdots \,, \qquad v \sim V_0(y) + 
\varepsilon V_1(y) + \cdots \,, \qquad y = \frac{x-x_0(\sigma)}{\varepsilon} 
\,; \qquad \sigma \equiv \varepsilon^\alpha t \,,
\EE

\noI to obtain

\bes \label{uvinneqfin}
\BE \label{uvinneqfinv}
V^{\prime\prime} - V + \frac{V^p}{U^q} = 0 \,, \qquad V \to 0 \quad \mbox{as} 
\quad |y| \to \infty \,,
\EE
\BE \label{uvinneqfinu}
\frac{1}{\varepsilon^2}DU^{\prime\prime} - U + 
\frac{1}{\varepsilon}\frac{V^r}{U^s} = 0 \,.
\EE
\ees

\noI The limiting conditions for \eqref{uvinneqfinu} come from
matching conditions with the outer solution. From \eqref{uvinneqfinu},
we have that $U \sim U_0$ is a constant to leading order so that $V_0$
satisfies

\BE \label{V0eq}
V_0^{\prime\prime} - V_0 + \frac{V_0^p}{U_0^q} = 0 \,, \qquad V \to 0 \quad 
\mbox{as} \quad |y| \to \infty \,.
\EE

\noI The solution of \eqref{V0eq} can be written

\BE \label{V0}
	V_0 \sim U_0^{\frac{q}{p-1}}w(y) \,,
\EE

\noI where $w(y)$ is the solution of the equation in \eqref{weqp} with
solution given in \eqref{wsolp} of \S \ref{GMfinline}.

To compute the outer solution for $u = u_0(x)$ in \eqref{GMufin}, we
proceed as in Appendix \ref{appA} and represent the $v^r/u^s$ term as
a weighted Dirac-delta function centered at $x = x_0$. We then have

\BE \label{uouteq}
Du_{0xx} - u_0 = -U_0^{\frac{qr}{p-1} -s} b_r \, \delta(x - x_0) \,,
\EE

\noI where $b_r$ is defined in \eqref{wsolp}. The solution of
\eqref{uouteq} may be written in terms of a Green's function
$G(x;x_0)$ as

\BE \label{uoutsol}
	u_0(x) = U_0^{\frac{qr}{p-1} -s} b_r G(x;x_0) \,,
\EE

\noI where $G(x;x_0)$ satisfies

\BE \label{Gfin}
	DG_{xx} - G = -\delta(x-x_0) \,, \qquad G_x(\pm1,x_0) = 0 \,.
\EE

\noI The solution of \eqref{Gfin} is given by \eqref{G} of \S
\ref{GMfinline}. The constant $G_{00}$ in \eqref{G00} is found by
imposing the jump condition $DG_x(x_0^+;x_0) - DG_x(x_0^-;x_0) =
-1$. Finally, by imposing the matching condition $u(x_0) = U_0$ in
\eqref{uoutsol}, we arrive at \eqref{U0fin} of \S
\ref{GMfinline}. With \eqref{V0}, \eqref{wsolp}, \eqref{uoutsol}, and
\eqref{G}, the one-spike quasi-equilibrium is then given by
\eqref{uvqe}.

To derive \eqref{x0ODE} for the drift of the spike center, we consider
the next order in $\varepsilon$ of \eqref{uvinneqfin} with
\eqref{uvinnerfin}. We calculate that $dV_0/dt =
-\varepsilon^{\alpha-1}V_0^\prime x_0^\prime$, while $dU_0/dt =
\mathcal{O}(\varepsilon^2)$. To match orders, we must take $\alpha =
2$ so that $\sigma \equiv \varepsilon^2 t$. We then have at the next
order

\bes \label{nextorder}
	\BE	\label{nextorderv}
 L V_1 \equiv V_1^{\prime\prime} - V_1 + p\frac{V_0^{p-1}}{U_0^q}V_1 = 
 -V_0^\prime x_0^\prime  + q\frac{V_0^p}{U_0^{q+1}} \,,
	\EE
	\BE	\label{nextorderu}
		DU_1^{\prime\prime} = -\frac{1}{U_0^s}V_0^r \,.
	\EE
\ees

\noI By differentiating \eqref{V0eq} with respect to $y$, we find that
$L V_0^\prime = 0$, or equivalently, $Lw^\prime = 0$. The right-hand
side of \eqref{nextorderv} must then satisfy the solvability condition

\BE \label{solvability}
 \intinf w^\prime \left\lbrack -x_0^\prime U_0^R w^\prime + qU_0^Q w^p U_1 
\right\rbrack \, dy = 0 \,; \qquad R \equiv \frac{q}{p-1} \,, \quad 
Q \equiv \frac{pq}{p-1} - q - 1 \,.
\EE

\noI With $R - Q = 1$ and $w^\prime w^p = (p+1)^{-1}dw^{p+1}/dy$, 
we have from \eqref{solvability}

\BE \label{solvability2}
	x_0^\prime U_0 \intinf \left(w^\prime\right)^2 \, dy = \frac{q}{p+1} 
\intinf \left(w^{p+1} \right)^\prime U_1 \, dy \,.
\EE

\noI Integrating by parts once on the right-hand side of
\eqref{solvability2} and using that $w \to 0$ as $|y| \to \infty$, we
obtain

\BE \label{solvability3}
 x_0^\prime U_0 \intinf \left(w^\prime\right)^2  \, dy = 
-\frac{q}{p+1} \intinf w^{p+1} U_1^\prime \, dy \,.
\EE

\noI Integrating by parts again on the right-hand side of
\eqref{solvability3} and letting $v(y) \equiv \int_0^y w^{p+1} \, ds$,
we calculate

\BE \label{solvability4}
 x_0^\prime U_0 \intinf \left(w^\prime\right)^2  \, dy = 
-\frac{q}{p+1} \left\lbrack U_1^\prime v \bigg\rvert_{-\infty}^\infty 
- \intinf v U_1^{\prime\prime} \, dy \right \rbrack\,.
\EE

\noI Since $w^{p+1}$ is an even function and $v(0) = 0$, we find that
$v(y)$ is an odd function. Also, since $V_0^r$ is an even function, we
have by \eqref{nextorderu} that $U^{\prime\prime}$ is an even
function. The integral term on the right-hand side of
\eqref{solvability4} therefore evaluates to 0. Now with $v(\infty) =
-v(-\infty)$, we have from \eqref{solvability4}

\BE \label{x01}
 x_0^\prime = -\frac{q}{2(p+1)}K\frac{U_1^\prime(\infty) + 
U_1^\prime(-\infty)}{U_0} \,; \qquad K \equiv \frac{\intinf w^{p+1} \, dy}
{\intinf \left(w^\prime\right)^2 \, dy} \,.
\EE

\noI The quantities $U_1^\prime(\pm \infty) $ may be calculated from
the matching condition

\BEU
	U_1^\prime(\pm \infty) = u_{0x}(x_0^\pm) \,,
\EEU

\noI yielding from \eqref{x01}

\BE \label{x02}
 x_0^\prime = -\frac{q}{2(p+1)}K\frac{1}{G_{00}}\left\lbrack G_x(x_0^+;x_0) + 
G_x(x_0^-;x_0) \right\rbrack \,,
\EE

\noI where we have used \eqref{uoutsol} and \eqref{U0fin} for $u_0$
and $U_0$, respectively. Using \eqref{G} and \eqref{G00} in
\eqref{x02}, we have

\BE	\label{x03}
 x_0^\prime = -\frac{q}{2(p+1)\sqrt{D}}K \left\lbrack 
\tanh\left( \theta_0(1+x_0) \right) - 
\tanh\left( \theta_0(1-x_0) \right) \right\rbrack \,.
\EE

The quantity $K$ in \eqref{x03} is calculated in
\cite{nec2013explicitly}. We include the calculation here for
completeness. We first multiply \eqref{weqp} in \S \ref{GMfinline} by
$w^\prime$ and integrate to obtain

\BE \label{K1}
	\frac{1}{2}\left(w^\prime \right)^2 - 
\frac{1}{2}w^2 + \frac{1}{p+1}w^{p+1} = C \,,
\EE

\noI where $C = 0$ since $w, w^\prime \to 0$ as $|y| \to \infty$. Integrating \eqref{K1} over the entire real line yields

\BE \label{K2}
 1 - I_1 + \frac{2}{p+1}K = 0 \,; \qquad 
I_1 \equiv \frac{\intinf w^2 \, dy}{\intinf \left(w^\prime\right)^2 \, dy} \,.
\EE

\noI To obtain a second equation involving $I_1$ and $K$, we multiply
\eqref{weqp} by $w$, integrate by parts once on the $w
w^{\prime\prime}$ term and apply the decay condition of $w$ to find

\BE \label{K3}
	-1 - I_1 + K = 0 \,.
\EE

\noI Solving \eqref{K2} and \eqref{K3} simultaneously, we find

\BE \label{K}
	K = \frac{2(p+1)}{p-1} \,.
\EE

\noI Substituting \eqref{K} into \eqref{x03}, we obtain result
\eqref{x0ODE} of \S \ref{GMfinline}. Because of the slow
$\mathcal{O}(\varepsilon^2)$ drift of the spike, analysis of
$\mathcal{O}(1)$ time scale instabilities may be performed assuming a
``frozen'' spike centered at $x = x_0$. The analysis leading to an
explicitly solvable NLEP then proceeds as in Appendix \ref{appA} and
will not be included here. The reader may refer to
\cite{nec2013explicitly} for details.

\setcounter{equation}{0}
\section{Boundary spikes in the GS model and analysis of competition instability} \label{appC}

Here, we construct the two boundary spike equilibrium \eqref{GSeqbm}
of \eqref{GS} on the domain $x \in (-1,1)$. We then derive an NLEP and
solve it to obtain thresholds of competition (given in
\eqref{Ueminus}) and synchronous instabilities. To do so, we first
construct a one-spike equilibrium centered at $x = 1$ on $x \in
(0,2)$, taking only the interval $(0,1)$. We then apply a reflection
to obtain the two boundary-spike solution on the entire interval.

We let $x = \xi + 1$ so that $-1<\xi<1$. The spike is then centered at
$\xi = 0$. The construction of the one-spike equilibrium then follows
closely to that given in Appendix \ref{appA}. In the inner region with
stretched variable $\zeta = \xi/\varepsilon$ and $u(\xi) \sim
U_0(\zeta)$, $v(\xi) \sim V_0(\zeta)$, we find that $U_0$ is a
constant while $V_0(\zeta) = w(\zeta)/\sqrt{AU_0}$, with $w(\zeta)$
given in \eqref{eqbminf2}. In the outer region, $u = u_0(\xi)$
satisfies

\BE	\label{u0GSeq}
	Du_{0\xi\xi} + (1-u_0) = \frac{b}{A^{3/2}\sqrt{U_0}}\delta(\xi) \,, \qquad u_\xi(\pm 1) = 0 \,,
\EE

\noI where the weight of the delta function is calculated in the usual
way. Here, $b$ is defined in \eqref{eqbminf2}. The conditions
$u_\xi(\pm 1) = 0$ are correspond to even symmetry about $x = 0$ in
original coordinates. Note that the boundary conditions $u_x(\pm 1) =
0$ are satisfied by the constant inner solution for $u$. The solution
of \eqref{u0GSeq} may be written $u_0(\xi) = 1 + \tilde{u(\xi)}_0$,
where $\tilde{u}$ satisfies

\BE \label{utilde}
	\tilde{u}_0(\xi) = -\frac{b}{A^{3/2}\sqrt{U_0}}G(\xi;0) \,,
\EE

\noI where $G(\xi;0)$ satisfies

\BE \label{GGSeq}
	DG_{\xi\xi} - G = -\delta(\xi) \,, \qquad G_\xi(\pm 1;0) = 0 \,.
\EE

\noI The solution of \eqref{GGSeq} is

\begin{equation} \label{GGs}
G(\xi;0) = G_{00} \left\{ 
\begin{array}{lr}
\frac{\cosh\left(\theta_0(\xi+1) \right)}{\cosh\theta_0}
\,, & -1<\xi<0 \,, \\ 
\frac{\cosh\left(\theta_0(\xi-1) \right)}{\cosh\theta_0}
\,, & 0<\xi<1 \,,%
\end{array}
\right. \,; \qquad G_{00} =  \frac{1}{2\sqrt{D}\tanh\theta_0} \,,
\end{equation}

\noI where $\theta_0$ is defined in \eqref{G00}. The matching
condition $u_0(0) = U_0$ determines $U_0$, yielding $U_0 = 1 - b
G(0;0)/(A^{3/2}\sqrt{U})$, which is equivalent to \eqref{H} of \S
\ref{GScomp}. The one spike equilibrium on $\xi \in (-1,1)$ is thus
given by

\BE \label{onespike}
v_{e1}(\xi) = \frac{1}{\sqrt{AU_0}}w\left(\varepsilon^{-1} \xi \right) \,, 
\qquad u_{e1}(\xi) = 1 + \tilde{u}_0(\xi) \,,
\EE

\noI with $\tilde{u}_0$ given in \eqref{utilde}.

On $x \in (0,1)$, the two boundary-spike solution is given by the the
solution \eqref{onespike} on the interval $\xi \in (-1,0)$. That is,
on $x \in (0,1)$, $u$ and $v$ are given by

\BE \label{uvhalf}
 v = \frac{1}{\sqrt{AU_0}}w\left(\varepsilon^{-1}(x-1) \right) \,, 
\qquad u = 1 - \frac{b}{A^{3/2}\sqrt{U_0}} 
\frac{\cosh\left(\theta_0x\right)}{2\sqrt{D}\sinh\theta_0} \,.
\EE

\noI The solution on $x \in (-1,0)$ is an even reflection of
\eqref{uvhalf} about $x = 0$ so that $x \to -x$. Noting that
$w(\zeta)$ and $\cosh x$ are both even functions, we obtain
\eqref{GSeqbm} with $U_-$ replaced by $U_0$. Here, $U_0$ is determined
by the matching condition \eqref{H} and takes on the value $U_-$ or
$U_+$ depending on whether the top or bottom solution branch is being
considered.

To determine the stability of \eqref{GSeqbm}, we perturb the one-spike
equilibrium on $\xi \in (-1,1)$ as

\begin{equation}  \label{GSpert1}
v = v_{e1}(\xi) + e^{\lambda t}\phi \,, \qquad u = u_{e1}(\xi) + e^{\lambda t} \eta \,; \qquad \phi,\eta \ll 1 \,.
\end{equation}

\noI Here, $v_{e1}(\xi)$ and $u_{e1}(\xi)$ are the one-spike
equilibrium solutions for $v$ and $u$ on $\xi \in (-1,1)$ given by
\eqref{onespike}. Substituting \eqref{GSpert1} in \eqref{GSv} and
\eqref{GSu}, we obtain the linearized system of equations

\bes \label{GSlin}
\BE \label{GSlinphi}
 \lambda \phi = \varepsilon^2\phi_{\xi\xi} - \phi + 
  3Au_{e1}v_{e1}^2\phi + Av_{e1}^3\eta \,,
\EE
\BE	\label{GSlineta}
	\tau \lambda \eta = D\eta_{\xi\xi} - \eta - 
\frac{1}{\varepsilon}\left\lbrack 3u_{e1}v_{e1}^2 + 
v_{e1}^3 \eta \right\rbrack \,.
\EE
\ees

\noI The boundary conditions in $\xi$ for \eqref{GSlin} depend on the
mode of instability considered and is discussed below. In the inner
region with the stretched variable $\zeta = \xi/\varepsilon$, we find
from \eqref{GSlineta} that $\eta = \eta_0$ is a constant to leading
order. Note that this satisfies the no-flux conditions at $x = \pm1$
in the original coordinates. Writing $\phi = \Phi(\zeta)$, we find
that $\Phi$ satisfies

\BE \label{PhieqGS}
	L_0 \Phi + \frac{\eta_0}{\sqrt{A}U_0^{3/2}}w^3 = \lambda \Phi \,, 
\EE

\noI where the operator $L_0$ is defined in \eqref{L0}. In
\eqref{PhieqGS}, we have used \eqref{onespike} for $v_{e1}$ and the
leading order behavior $u_{e1} \sim U_0$ for $u_{e1}$ in the inner
region. The quantity $\eta_0$ must be obtained by solving the outer
equation for $\eta(\xi)$.

In the outer region for \eqref{GSlineta}, we proceed as in Appendix
\ref{appA} and represent the localized terms involving $\phi$ and
$v_{e1}$ as appropriately weighted delta functions. In this way, we
obtain the outer equation for $\eta$

\BE \label{etaeqGS}	
 D\eta_{\xi\xi} - (1+\tau\lambda)\eta = 
\left\lbrack \frac{b\eta_0}{\left(AU_0\right)^{3/2}} + 
\frac{3}{A}\intinf w^2 \Phi \, d\zeta  \right\rbrack \delta(\xi) \,.	
\EE

\noI The competition mode of instability, which leads to the growth of
one spike and the collapse of the other, is associated with an odd
eigenfunction. We thus impose that $\eta(\pm 1) = 0$ for the
competition mode, which corresponds to $\eta(0) = 0$ in the original
$x$ coordinate. The synchronous mode, which leads to the collapse of
both spikes, is associated with an even eigenfunction. This leads to
the symmetry condition $\eta_\xi(\pm 1) = 0$, which corresponds to
$\eta_x(0) = 0$ in the original $x$ coordinate. In imposing the
boundary conditions at $\xi = \pm 1$, we implicitly assume the
presence of image spikes centered at $\xi = \pm 2$.

For each mode, we define an associated Green's function with
appropriate boundary conditions

\BE \label{Gpm}
 DG_{\pm\xi\xi} - (1+\tau\lambda)G_\pm = -\delta(\xi) \,, 
\qquad G_{+\xi}(\pm 1;0) = 0 \,, \qquad G_-(\pm 1;0) = 0 \,,
\EE

\noI where $G_+$ ($G_-$) corresponds to the synchronous (competition)
mode. The solution of \eqref{etaeqGS} may then be written in terms of
$G_\pm$ as

\BE \label{etasolGS}
	\eta(\xi) = -\left\lbrack \frac{b\eta_0}{\left(AU_0\right)^{3/2}} + \frac{3}{A}\intinf w^2 \Phi \, d\zeta \right\rbrack G_\pm(\xi;0) \,.
\EE

\noI Finally, to find $\eta_0$, we apply the matching condition
$\eta(0) = \eta_0$ in \eqref{etasolGS} and calculate

\BE \label{eta0GS}
 \eta_0 = -\frac{\frac{3}{A} \intinf w^2\Phi \, d\zeta}
{\frac{1}{G_{\pm 00}} + \frac{b}{\left(AU_0\right)^{3/2}}} \,,
\EE

\noI where $G_{\pm 00} \equiv G_\pm(0;0)$.

Now we may substitute \eqref{eta0GS} for $\eta_0$ into \eqref{PhieqGS}
to obtain

\BE \label{NLEPGS1}
 L_0\Phi - 3w^3 \frac{\intinf w^2\Phi \, d\zeta}
{b + \frac{(AU_0)^{3/2}}{G_{\pm00}}} = \lambda \Phi \,.
\EE

\noI Using \eqref{H} in \S \ref{GScomp}, we may write $A^{3/2} =
bG_{00}/H(U_0)$ so that we obtain from \eqref{NLEPGS1} the NLEP

\BE \label{NLEPGS2}
	L_0 \Phi - \chi_\pm w^3 \intinf w^2\Phi \, d\zeta = \lambda\Phi \,,
\EE

\noI where $\chi_\pm$ is defined as

\BE \label{chiGS}
 \chi_\pm = \frac{3}{b\left(1 + 
\frac{G_{00}}{G_{\pm00}}\frac{U_0^{3/2}}{H(U_0)} \right)} \,.
\EE

\noI Here, $G_{00}$ is defined in \eqref{GGs}. It was shown in
Appendix \ref{appA} that the NLEP in \eqref{NLEPGS2} is explicitly
solvable, yielding

\BE \label{lambdaGS}
\lambda = 3 - \frac{3}{2}\chi_\pm \,.
\EE

To complete the derivation of $\lambda$, we require $G_{\pm00}$ in
\eqref{chiGS}. The solutions for $G_+(\xi;0)$ and $G_-(\xi;0)$ in
\eqref{Gpm} are given by

\BE \label{Gp}
		G_+(\xi;0) = G_{+00} \left\{ 
\begin{array}{lr}
\frac{\cosh\left(\theta_\lambda(1+\xi) \right)}{\cosh \left(\theta_\lambda \right)}
\,, & -1<\xi<0 \,, \\ 
\frac{\cosh\left(\theta_\lambda(1-\xi) \right)}{\cosh\left(\theta_\lambda \right)}
\,, & 0<\xi<1 \,,
\end{array}
\right. \,; \qquad G_{+00} = \frac{1}{2\sqrt{D}\sqrt{1+\tau\lambda}\tanh\theta_\lambda} \,,
\EE

\noI and

\BE \label{Gm}
	G_-(\xi;0) = G_{-00} \left\{ 
\begin{array}{lr}
\frac{\sinh\left(\theta_\lambda(1+\xi) \right)}{\sinh\left(\theta_\lambda \right)}
\,, & -1<\xi<0 \,, \\ 
\frac{\sinh\left(\theta_\lambda(1-\xi) \right)}{\sinh\left(\theta_\lambda \right)}
\,, & 0<\xi<1 \,,
\end{array}
\right. \,; \qquad G_{-00} = \frac{1}{2\sqrt{D}\sqrt{1+\tau\lambda}\coth\theta_\lambda} \,,
\EE

\noI where $\theta_\lambda$ is defined in \eqref{Glambdasol}. Note
that the parameter $\tau$ appears in the expressions only as
$\tau\lambda$. Since we consider only monotonic instabilities, which
occur as a single eigenvalue crosses into the right half-plane through
the origin, any increase or decrease in $\tau$ cannot trigger such an
instability. We may thus take $\tau = 0$ for simplicity while also
ensuring that no Hopf instabilities are present. With $\tau = 0$, we
have from \eqref{Gp} and \eqref{Gm}

\BE \label{Gratio}
 \frac{G_{00}}{G_{+00}} = 1 \,, \qquad \frac{G_{00}}{G_{-00}} = \coth^2\theta_0 \,.
\EE

\noI Finally, using \eqref{Gratio} in \eqref{chiGS} and
\eqref{lambdaGS}, we have the explicit expressions for the eigenvalues
corresponding to the synchronous ($\lambda_+$) and competition modes
($\lambda_-$)

\BE \label{lambdaGSfinal}
 \lambda_+ = 3 - \frac{9}{2\left\lbrack1 + \frac{U_0^{3/2}}{H(U_0)} 
\right\rbrack} \,, \qquad \lambda_- = 3 - \frac{9}{2\left\lbrack1 + 
\frac{U_0^{3/2}}{H(U_0)}\coth^2\theta_0 \right\rbrack} \,.
\EE

\noI Note that the expression for $\lambda_-$ in \eqref{lambdaGSfinal}
is the same as that given in \eqref{GSlambda} of \S
\ref{GScomp}. Setting $\lambda_- = 0$ yields the thresholds given in
\eqref{Ueminus}. Setting $\lambda_+ = 0$ in \eqref{lambdaGSfinal}, we
find that the stability threshold for the synchronous mode is $U_0 =
1/3$. Recalling that the upper branch corresponds to $0<U_0<1/3$ while
the lower branch corresponds to $1/3 < U_0 < 1$, we find that the
threshold for the synchronous mode occurs at the saddle point, which
was stated in \S \ref{GScomp}. A simple calculation shows that the
upper branch is always stable to the synchronous mode while the lower
branch is always unstable.

\subsubsection*{Acknowledgements}

T.~Kolokolnikov and M.~J.~Ward gratefully acknowledge the grant support of 
NSERC. J.~C.~Tzou was supported by an AARMS Postdoctoral Fellowship.

\bibliographystyle{elsart}
\bibliography{delay}

\end{document}